\documentclass{article}

\usepackage[utf8]{inputenc}
\usepackage[top=50mm, bottom=50mm, left=50mm, right=50mm]{geometry}
\usepackage{lineno}
\usepackage{amssymb}
\usepackage{amsmath}
\usepackage{amsthm}
\usepackage{epsfig}
\usepackage{graphicx}
\usepackage{graphics}
\usepackage{float}
\usepackage{subcaption}
\usepackage{multirow}
\usepackage{color}
\usepackage{lineno}
\usepackage{fullpage}
\usepackage[normalem]{ulem} 
\usepackage{makeidx}
\usepackage{xspace}
\usepackage{wrapfig}
\usepackage{url}
\usepackage{authblk}
\usepackage{hyperref}
\usepackage{pdfpages}
\usepackage{soul}
\usepackage[table]{xcolor}

\newcommand{\virgolette}[1]{``#1''}

\everymath{\displaystyle}

\title{SEIHRDV: a multi-age multi-group epidemiological model and its validation on the COVID-19 epidemics in Italy}

\author[1,*]{{Luca Dede'}}
\author[1]{{Nicola Parolini}}
\author[1,2]{Alfio Quarteroni}
\author[3]{Giulia Villani}
\author[4]{{Giovanni Ziarelli}}
\affil[1]{MOX, Department of Mathematics, Politecnico di Milano}
\affil[2]{Institute of Mathematics, \'Ecole Polytechnique F\'ed\'erale De Lausanne}
\affil[3]{Department of Mathematics, University of Rome ``La Sapienza"}
\affil[4]{Department of Mathematics, Università degli Studi di Milano}

\affil[*]{Corresponding author, \texttt{luca.dede@polimi.it}}

\date{January 3, 2025}

\begin{document}

\maketitle

\begin{abstract}
    We propose a novel epidemiological model, {referred to as} SEIHRDV, for the numerical simulation of the COVID-19 epidemic, which we validate using data from Italy {starting in September 2020}. SEIHRDV features the following compartments: Susceptible (S), Exposed (E), Infectious (I), Healing (H), Recovered (R), Deceased (D) and Vaccinated (V). The model is {age-stratified}, as it considers the population split into $15$ age groups. {Moreover, }it {takes into account} $7$ different contexts of exposition to the infection (family, home, school, work, transport, leisure, other contexts){, which impact on the transmission mechanism}. Thanks to these features, {the} model can address the analysis of the epidemics and the efficacy of non-pharmaceutical interventions, as well as possible vaccination strategies and the introduction of the Green Pass, a containment measure introduced in Italy in 2021. {By leveraging on the SEIHRDV model, we successfully analyzed epidemic trends during the COVID-19 outbreak from September 2020 to July 2021. The model proved instrumental in conducting comprehensive what-if studies and scenario analyses tailored to Italy and its regions. Furthermore, SEIHRDV facilitated accurate forecasting of the future potential trajectory of the epidemic, providing critical information for informed decision making and public health strategies.}
\end{abstract}

\section{Introduction}
Coronavirus pandemic was a public health emergency that has affected the whole World since its outbreak in December 2019. {Up to December 6, 2023, more than $6.9$ million deaths have been registered, and over $770$ million cases have been detected according to the World Health Organization}. In this context, various mathematical models have been proposed to describe the course of the epidemic and forecast its progression, aiming to assist authorities in selecting optimal control strategies to mitigate the spread of the virus. {The primary objective is to prevent} the collapse of the healthcare system {and ultimately enhance the public health response \cite{AA}}.

{Most epidemiological models are derived from the compartmental framework originally proposed by Kermack and McKendrick \cite{AG}, known as the classic SIR model, which partitions the population into susceptible, infected, and recovered individuals; see also \cite{AN}.
In recent years, numerous compartmental epidemiological models specifically designed to describe the COVID-19 pandemic have been proposed (see, e.g., \cite{BB, OO, PP, AB, AC, AD, AF, AI, AJ, AM, AO, AQ}). Generally, these models divide the population into subsets, referred to as compartments, that represent different stages of disease progression. The associated mathematical formulations are often deterministic systems of Ordinary Differential Equations (ODEs) governing the mobility of individuals among compartments.
In addition, uncertainty in parameters can be incorporated by reformulating the problem as a set of stochastic differential equations \cite{lekone2006statistical, zimmer2017likelihood}.
Beyond compartmental models, various data-driven approaches have emerged. These include agent-based models \cite{hoertel2020stochastic, kerr2021covasim, lasser2022assessing} and machine learning algorithms \cite{olumoyin2021data, bertaglia2022asymptotic, shaier2022data, ziarelli2024learning, millevoi2023physics, ziarelli2024model}, which offer complementary perspectives for studying the dynamics of infectious diseases.}

Building on the foundation of compartmental epidemiological models, the standard SIR model can be generalized to incorporate specific features of the disease under investigation, such as its virological and medical characteristics. Most of these models operate under the assumption of a homogeneous epidemic, disregarding distinctions in factors such as age groups, exposure contexts, and geographical variations that can influence transmission dynamics \cite{CC}. While this simplification may reduce the accuracy of forecasts, homogeneous models are computationally efficient, straightforward to calibrate, and often yield results that are interpretable and actionable for policy-making. 
{Since the beginning of the pandemic, we have observed that the impact of COVID-19 infection varies with the age of infected individuals, with clear evidence showing that older people experience more severe outcomes than younger individuals. Additionally, the risk of infection varies depending on the context in which susceptible individuals come into contact with infected persons. 
Vaccinations, which were widely administered starting in 2021, have significantly impacted the transmission mechanisms of the virus and reduced the severity of symptoms, thereby mitigating the effects of COVID-19 \cite{HH}.}

In this work, we introduce a novel age-stratified ODE-based compartmental model, named SEIHRDV, specifically designed to describe the spread of the Coronavirus epidemic in Italy. This model was applied to analyze the period from September 2020 to July 2021, accounting for varying exposure contexts where infection occurred with differing probabilities, as well as the effects of the vaccination campaign (see e.g., \cite{AP}).
The model leverages on \textit{Tables of Contacts}, also known as Mossong tables \cite{DD}, which detail the average daily number of contacts between members of different age groups across various exposure contexts. We consider the following contexts: home, school, work, transport, leisure, and others. As a key innovation, we refined the original \textit{home} category in the Mossong tables by distinguishing between {intra-household} contacts (\textit{family}) and {household-visitor} contacts (\textit{home}). This distinction is significant for capturing the impact of Non-Pharmaceutical Interventions (NPIs) that imposed specific restrictions on interactions between relatives and friends visiting different households, as was the case in Italy during certain periods of the first year of the pandemic \cite{AE, AR}.
This enhanced model enables detailed evaluations of the effects of various NPIs with differing levels of restrictions on social behaviors across $15$ age groups and $7$ exposure contexts, supporting both retrospective analyses and future scenario projections.

In this paper, we retrospectively assess the effectiveness of restrictions associated with the introduction of the Green Pass, an NPI implemented in Italy in 2021 to maximize vaccination coverage, particularly in public exposure contexts. By incorporating the Green Pass into the model, we aim to simulate a realistic scenario, evaluate its efficacy in reducing infection rates across different contexts, and analyze its impact on the health outcomes of vaccinated individuals.\\

The paper is organized as follows: in Section \ref{mathematical_model}, we describe the multi-age/multi-context structure of the mathematical model, introducing the different parameters and the {way we derived the} tables of contacts. Then we describe the calibration process, and how we take into account the vaccination campaign and the introduction of Green Pass restrictions. In Section \ref{results}, we show {an extensive campaign of numerical results}, {set} in Italy and {two Italian administrative regions}: Lombardy and Lazio. In Section \ref{conclusions}, we draw our conclusions and we discuss {possible improvements and possible future extensions of the SEIHRDV.}

\section{Mathematical model}
\label{mathematical_model}
Starting from the standard SEIR model \cite{BME}, we develop an ODE-based compartmental model tailored to COVID-19. In Section \ref{subsec:basicSEIHRDV}, we present the homogeneous SEIHRDV model as a foundational framework. In Section \ref{seihrdv_ma_mc}, we extend this model to incorporate multi-age and multi-context dynamics. Section \ref{tables_of_contacts} details the implementation of these multi-age and multi-context elements. Finally, in Section \ref{subsec:calValGP}, we outline the calibration procedure, including the integration of vaccination data and considerations related to the Green Pass policy.

\subsection{{SEIHRDV model}}
\label{subsec:basicSEIHRDV}
When a \textit{Susceptible} individual (compartment $S$) is infected but still in the incubation period, transition to the \textit{Exposed} state (compartment $E$) occurs. After the incubation period, an Exposed individual becomes \textit{Infected} (compartment $I$) and can transmit the infection. Before recovery, infected individuals enter the \textit{Healing} state (compartment $H$), during which they remain positive but are no longer infectious. The progression of the disease ultimately results in either full recovery (\textit{Recovered} compartment $R$) or death (\textit{Dead} compartment $D$).
To account for the vaccination campaign, we introduce a \textit{Vaccinated} compartment ($V$), which accounts for individuals who have received at least the first dose of the vaccine. In this work, we neglect the effects of second-dose vaccinations, as our analysis conclude in the summer of 2021, whereas widespread second-dose administration in Italy only began in September 2021. A possible extension of the model, incorporating the complete vaccination cycle, could be inspired by \cite{ziarelli2023optimized}.

\begin{figure}[t!]
\centering
\includegraphics[width=0.5\textwidth]{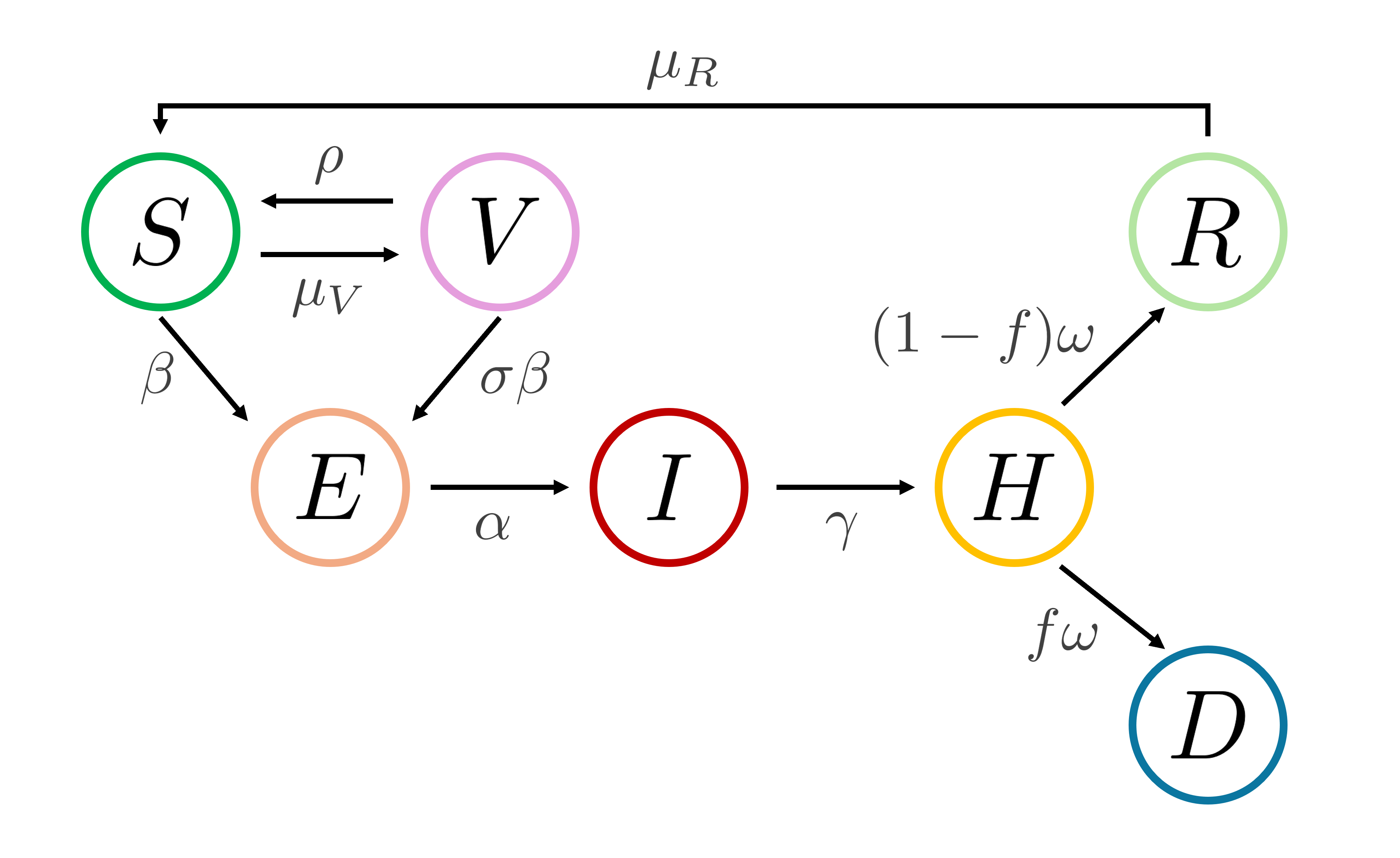}
\caption{{Sketch of} SEIHRDV model ($S$ = Susceptible, $E$ = Exposed, $I$ = Infectious, $H$ = Healing, $R$ = Recovered, $D$ = Deceased, $V$ = Vaccinated).}
\label{fig:SEIHRDV}
\end{figure}

The SEIHRDV model {(cf. Figure \ref{fig:SEIHRDV})} is designed to describe all the stages an individual may undergo, from being susceptible to entering the healing phase. {The corresponding mathematical model is given by:}

%On the mathematical point of view, this model is described by the following system of Ordinary Differential Equations {with initial conditions:}
\begin{equation}\label{eq:SEIHRDV}
    \begin{array}{l}
    \displaystyle
\frac{dS}{dt} = - c\, \beta  \, \frac{S\, I}{N} - d \, \frac{S}{S+R} + \mu_{R} \, R+ \mu_{V} \, V, \\[3mm]
\displaystyle \frac{dE}{dt} = c\, \beta \frac{(S+\sigma V)\, I}{N} -\alpha E, \\[3mm]
\displaystyle \frac{dI}{dt} = \alpha \, E - \gamma \, I, \\[3mm]
\displaystyle \frac{dH}{dt} = \gamma \, I - \omega \, H \qquad \qquad \qquad \qquad \qquad \qquad \qquad \qquad \qquad t \in (0,T], \\[3mm]
\displaystyle \frac{dR}{dt} = (1-f(S,V))\, \omega \, H - \mu_{R} \, R, \\[3mm]
\displaystyle \frac{dD}{dt} = f(S,V)\omega H, \\[3mm]
\displaystyle \frac{dV}{dt} = - \sigma \, c \, \beta \, \frac{V\, I}{N} + d \frac{S}{S+R} - \mu_{V} \, V, \\[3mm]
S(0) = S_{0}, \quad E(0) = E_{0}, \quad I(0) = I_{0}, \quad H(0) = H_{0}, \\[3mm]
R(0) = R_{0}, \quad D(0) = D_{0}, \quad V(0) = V_{0}\\[3mm]

\end{array}
\end{equation}
{where $N$ denotes the total number of individuals of the population that is given by the sum of the number of individuals belonging to one of the $7$ compartments, namely $N = S + E + I + H + R + D + V$.} {Albeit {each compartment changes in time}, we assume that the total number of individuals $N$ in the population remains constant over time.}

The SEIHRDV model is characterized by the following parameters:
\begin{itemize}
    \item {the product $(c\,  \beta)$ stands as} the transmission rate due to contacts between a susceptible individual and an infected individual, where $\beta:(0,T]\rightarrow \mathbb{R}$ is a {time-dependent function that captures the variation in contacts, which also depends on the implementation of possible NPIs}, while $c:(0,T]\rightarrow \mathbb{R}$ {embodies changes in the transmission rate, which we will extract from data during the calibration process (cf. Section \ref{calibration_and_updates})};
    \item $\alpha > 0$ denotes the rate at which an exposed individual becomes infectious and is defined as the inverse of the average incubation time;
    \item $\gamma > 0$ denotes the rate at which an infectious individual becomes {unable to infect, but still infected. This rate is derived from the average infectious time};
    \item $\omega > 0$ is a rate derived from the average removal time for deceased, and it is multiplied by $f$ or $1-f$;
    \item $\sigma \in [0,1]$ denotes the {inefficacy rate} of the vaccine in transmissibility of the disease \cite{JJ};
    \item $f:[\tau,T] \rightarrow [0,1]$ is the fatality function, defined as
    \begin{equation}
        \displaystyle f(t) = f(S(t),V(t); \tau) = \hat{f} \, \frac{S(t - \tau) +\theta\, \sigma\, V(t - \tau)}{S(t - \tau)+\sigma \, V(t - \tau)},
       \label{eq_fifr}
    \end{equation}
    \noindent where $\hat{f}$ is the \textit{infectious fatality rate} (\cite{LL}) taken from an Imperial College report of October 2020 \cite{GG}, and $\tau$ is the average {elapsing time} between the first exposition and death, that we set at $25$ days {according to clinical studies \cite{zhou2020clinical}}. We introduce the parameter $\theta \in [0,1]$ which denotes the rate of inefficacy of the vaccine in the reduction of the severity of the disease. {In this way, $f$ is obtained by rescaling the original $\hat{f}$ with the ratio between the sum of individuals susceptible and vaccinated in which the vaccine is not effective either in reducing the transmissibility of the virus or in reducing the severity of the disease, and} all the individuals who can potentially contract the virus;
    \item $d \in \mathbb{N}$ is the number of the average of first doses administered daily, {estimated from the data made available from Dipartimento di Protezione Civile Italiana\footnote{\url{https://github.com/italia/covid19-opendata-vaccini}}};
    \item $\mu_{R} > 0$ and $\mu_{V} > 0$ denote the {waning immunity rates for Recovered and Vaccinated individuals respectively, i.e. the} rate at which {Recovered individuals and Vaccinated individuals, respectively, come back to the susceptible compartment}.
    
\end{itemize}
In our model, the parameters $\alpha$, $\gamma$, $\omega$, $\theta$, $\hat{f}$, $\mu_{R}$ and $\mu_{V}$ are kept fixed in time, while $d$, $\beta$ and $\sigma$ are instead functions of time.
{The values of these parameters have been deduced from established studies \cite{HH} or directly from available data. Following the methodology in \cite{AA2}, we estimate the initialization of compartments that are not directly accessible from the data provided by the Dipartimento di Protezione Civile Italiana\footnote{\url{https://github.com/pcm-dpc/COVID-19}}, such as the Exposed and Infected individuals. This estimation is obtained thanks to algebraic relationships among measurable quantities, including the CFR (Case Fatality Rate), IFR (Infection Fatality Rate) \cite{GG}, and the recorded number of deceased individuals.}
{The number of delivered doses $d$ in \eqref{eq:SEIHRDV} is rescaled by the factor $\frac{S}{S+R}$ in order to take into account that a portion of doses are administered to both detected and undetected recovered individuals.}

\newpage

\subsection{The multi-age and multi-context SEIHRDV model}
\label{seihrdv_ma_mc}

\begin{figure}[t!]
\centering
\includegraphics[width=0.7\textwidth]{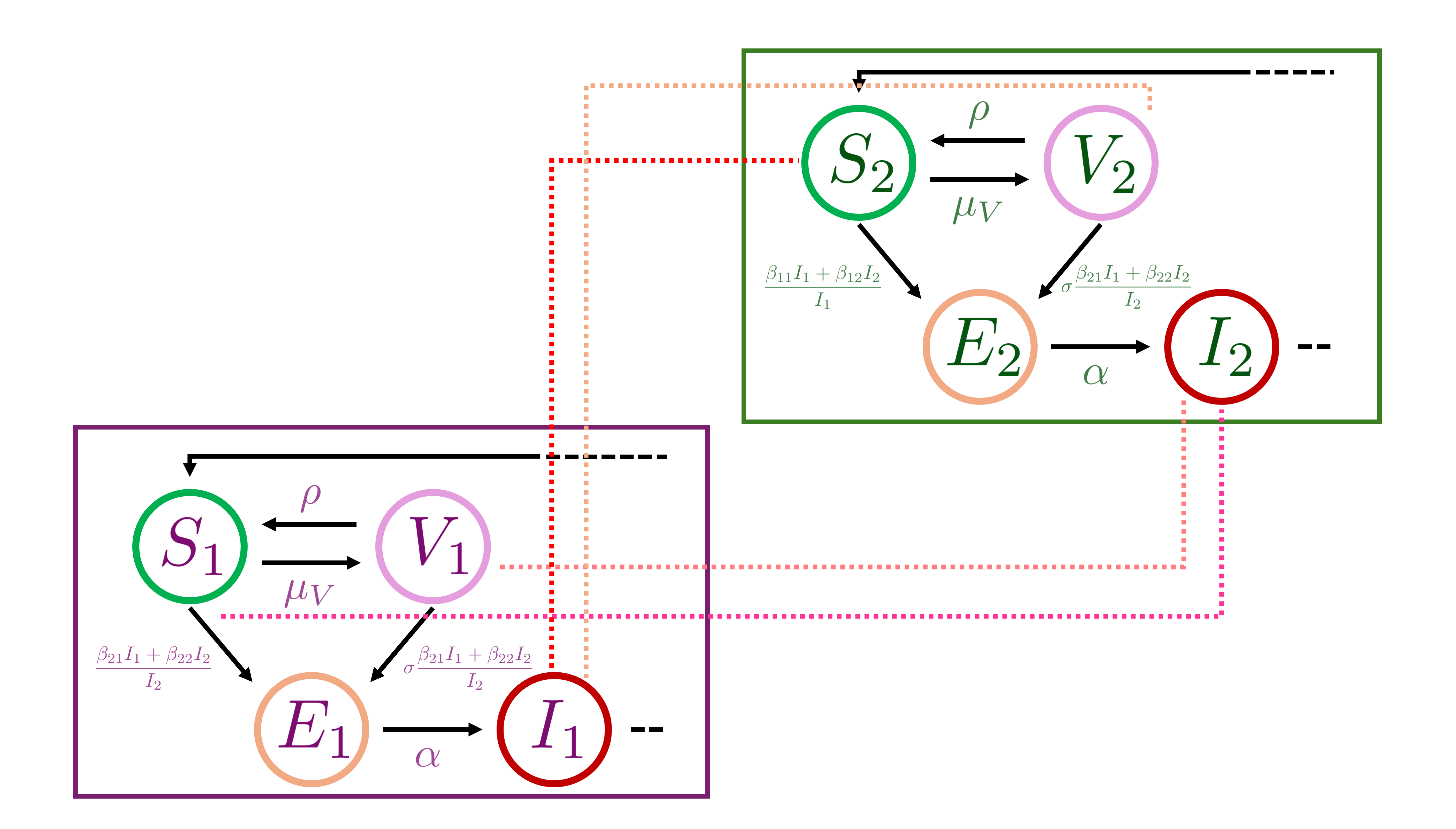}
\caption{{Sketch of the SEIHRV model in a single exposure context and two ages group.}}
\label{fig:S1S2}
\end{figure}

{In this section we present the SEIHRDV model.}
We consider: 
\begin{itemize}
    \item $N_{a} = 15$ age groups, {by assigning to each of the following age groups the corresponding SEIHRDV variables: }$\{ 0\div4, 5\div9, 10\div14, 15\div19, 20\div24, 25\div29, 30\div34, 35\div39, 40\div44, 45\div49, 50\div54, 55\div59, 60\div64, 65\div69, 70+ \}$;
    \item $K = 7$ contexts of exposition: family, home, school, work, transport, leisure and others.
\end{itemize} 

\begin{figure}[t!]
\centering
\includegraphics[width=0.7\textwidth]{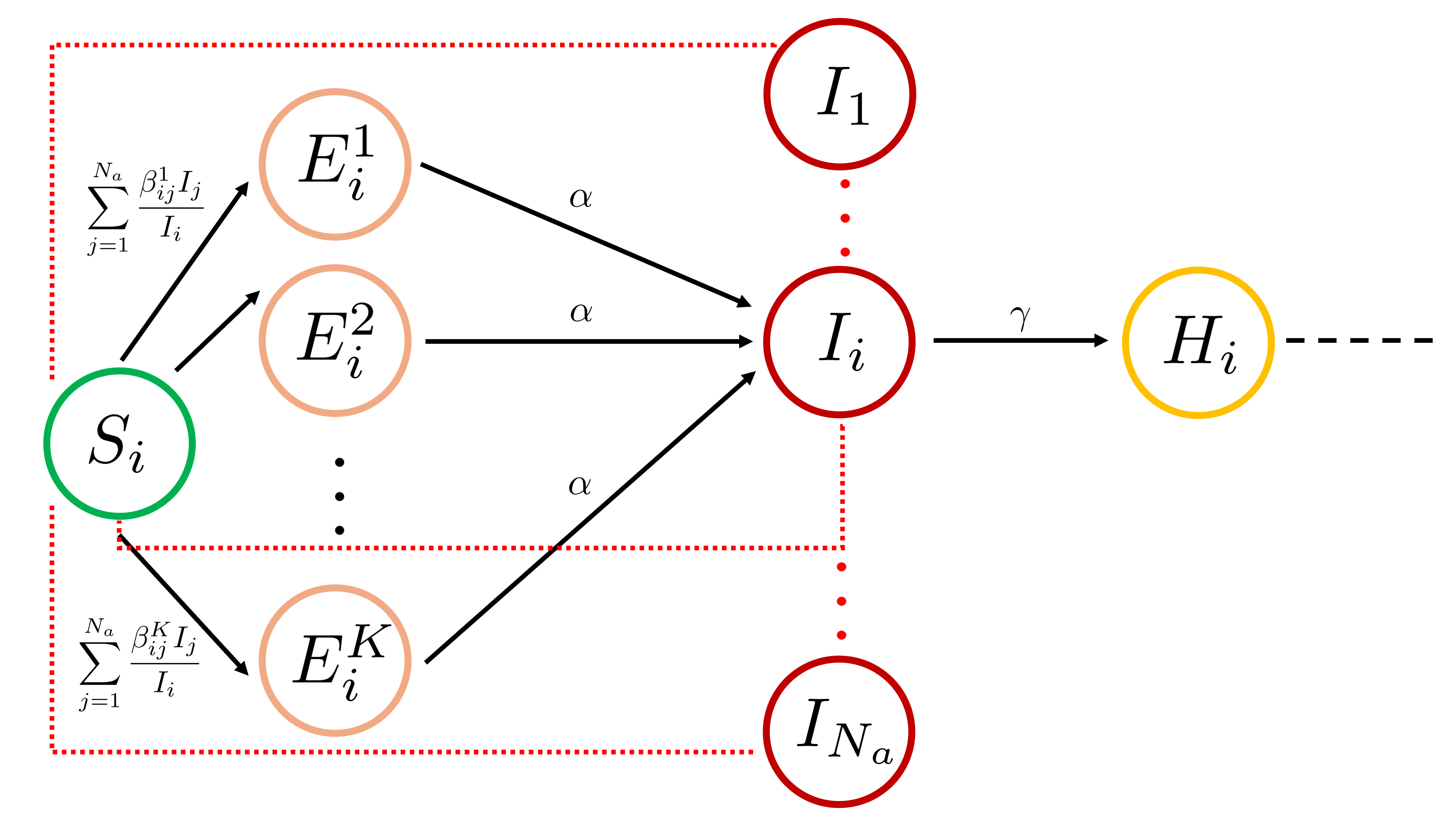}
\caption{Interactions among the compartments in SEIHRDV multi-age and multi-group model, with $K=7$ contexts of exposition and $N_a=15$ age groups.}
\label{fig:E1E2}
\end{figure}

The interactions among the compartments of the various age groups (\cite{NN}) in the different exposure context are {sketched} in Figure \ref{fig:S1S2}. {For the sake of simplicity, we limit ourselves to two age groups, named \textit{group 1} and \textit{group 2}, and only one exposure context for each age group.} The corresponding compartments are denoted with $S_{i}$, $E_{i}$, $I_{i}$, $H_{i}$, $R_{i}$, $D_{i}$, and $V_{i}$, with $i \in {1,2}$. {The disease spreads as far as} an individual from $S_{1}$ can have a contact with either an infectious individual from the same age group, $I_{1}$, or an individual from $I_{2}$. Therefore the transmission rate $\beta$ of the virus is replaced by the expression $\frac{\beta_{11}\, I_{1}+\beta_{12}\, I_{2}}{I_{1}}$, that takes into account both the possible contacts. {The same holds for an individual in $S_2$ with the appropriate modifications.} 

Extending this argument for all the $N_a=15$ age groups, and considering the $K=7$ contexts of exposition, we obtain the complete SEIHRDV multi-age and multi-group model (Figure \ref{fig:E1E2}): an individual of age group $i$ can have contacts with infectious individuals belonging to the $N_a$ age groups in the $K$ contexts of exposition. In particular, we denote with $E_{i}^{k}$ the number of individuals in age group $i$ who become exposed in the context $k$, and with $\beta_{ij}^{k}$ the transmission rate of $I_{j}$ in contact with $S_{i}$ in the context of exposition $k$. 
{In the multi-age and multi-context SEIHRDV, $\beta$ can be seen as a third order tensor with components $\beta_{ij}^{k}$}. 

The equations describing the multi-age and multi-context model {read as:}
\begin{equation}
    \begin{array}{l}
    \displaystyle
\frac{dS_{i}}{dt} =  - \Bigg(\sum_{k=1}^{K} \Bigg(\sum_{j=1}^{N_{a}}c \, \beta_{ij}^{k} I_{j} \Bigg) \Bigg)\frac{S_{i}}{N_{i}} - d_{i} \, \frac{S_{i}}{S_{i}+R_{i}} + \mu_{R}\, R_{i} + \mu_{V}\, V_{i}, \quad \quad \\[4mm]

\displaystyle \frac{dE_{i}^{k}}{dt} = \Bigg(\sum_{j=1}^{N_{a}}c\, \beta_{ij}^{k}I_{j} \Bigg) \frac{(S_{i} + \sigma\,  V_{i})}{N_{i}} - \alpha \, E_{i}^{k},  \\[4mm]
\displaystyle \frac{dI_{i}}{dt} = \alpha \Bigg( \sum_{k=1}^{K} E_{i}^{k} \Bigg) -\gamma I_{i}, \qquad \qquad \qquad k=1,\dots, K, \\[4mm]

\displaystyle \frac{dH_{i}}{dt} = \gamma I_{i} - \omega H_{i}, \qquad \qquad \qquad \qquad \qquad \qquad \qquad \qquad \qquad \qquad \qquad \qquad \qquad \qquad t \in (0,T]\\[4mm]

\displaystyle \frac{dR_{i}}{dt} = (1-f(S_{i},V_{i}))\omega H_{i} - \mu_{R} R_{i}, \\[4mm]
\displaystyle \frac{dD_{i}}{dt} = f(S_{i},V_{i})\omega H_{i}, \\[4mm]

\displaystyle \frac{dV_{i}}{dt} = - \Bigg(\sum_{k=1}^{K} \Bigg(\sum_{j=1}^{N_{a}}c \beta_{ij}^{k} I_{j} \Bigg) \Bigg)\frac{\sigma V_{i}}{N_{i}} + d_{i} \frac{S_{i}}{S_{i}+R_{i}} - \mu_{V} V_{i}, \\[4mm]

S_{i}(0) = S_{i_{0}}, \quad E^k_{i}(0) = E^k_{i_{0}}, \quad I_{i}(0) = I_{i_{0}}, \quad H_{i}(0) = H_{i_{0}}, \quad R_{i}(0) = R_{i_{0}}, \quad D_{i}(0) = D_{i_{0}}, \quad V_{i}(0) = V_{i_{0}}

 \\[4mm]
\forall \; {i = 1,...,N_{a}.}
 \end{array}
\end{equation}

\subsection{Tables of contacts and $\mathbf{\boldsymbol{\beta}-matrices}$}  %perché devo chiamarli beta-arrays?
\label{tables_of_contacts}

To estimate the transmissibility rate $\beta$ {for different age-groups in different exposure contexts}, we use the tables of contacts, based on those proposed by the POLYMOD study \cite{DD}. {The social contact matrices reported in \cite{DD} for different countries estimate} the average number of contacts that individuals from an age group have in the following contexts: home, school, work, transport, leisure and other. In particular, { we use the table referring to the Italian case that was also considered by the Italian Scientific Technical Committee (CTS) in the early phase of the COVID-19 epidemics (see \cite{EE})}. The CTS was established in Italy on February 2020 to provide advice and support for coordination activities to overcome the epidemiological emergency due to the spread of the Coronavirus; see also \cite{AH}. {In addition to the $5$ exposure contexts outside the household, as outlined in \cite{DD}, we account for a total of $K=7$ distinct exposure contexts. Specifically, based on data from the Italian National Statistics Institute (ISTAT) regarding population and household composition, we disaggregate household contacts into two categories: those occurring exclusively among family members (\textit{"family"}) and those involving both family and non-family members (\textit{"home"}). This distinction enables the consideration of different restrictions implemented in Italy during short periods in 2021. Notably, in specific administrative regions, visits to family members were occasionally prohibited.}. {The actual social contact matrix used in the present work is reported in Table \ref{table:1}.}

\begin{table}[t]
\centering

\begin{tabular}{|c|c|c|c|c|c|c|c|c|}
\hline
% &  & k =  & 1 & 2 & 3 & 4 & 5 & 6 & 7 = K \\
\multicolumn{2}{|c|}{} & \multicolumn{7}{c|}{\textbf{Exposure context}} \\
\hline
\textbf{Ages}   & \textbf{Tot} & \textbf{Family} & \textbf{Home} & \textbf{School} & \textbf{Work} & \textbf{Transport} & \textbf{Leisure} & \textbf{Other}  \\
\hline
\cellcolor{yellow} \textbf{0$\div$4}    & \cellcolor{lime}16.54 & 2.30  & 2.19  & 5.27   & 0.00    & 0.98      & 3.06    & 2.75 \\
\hline
\cellcolor{yellow}\textbf{5$\div$9}    & \cellcolor{lime}20.49 & 2.27  & 2.34  & 8.87   & 0.00    & 1.12      & 4.53    & 1.37 \\
\hline
\cellcolor{yellow}\textbf{10$\div$14}  & \cellcolor{lime}27.38 & 2.21  & 2.22 & 11.98  & 0.20  & 1.35      & 5.62    & 3.80  \\
\hline
\cellcolor{yellow}\textbf{15$\div$19}  & \cellcolor{lime}29.28 & 2.05  & 2.54 & 13.22  & 0.05 & 1.74      & 6.83    & 2.87 \\
\hline
\cellcolor{yellow}\textbf{20$\div$24}  & \cellcolor{lime}22.15 & 1.49  & 2.02  & 1.17   & 4.49 & 0.96      & 7.23    & 4.80  \\
\hline
\cellcolor{yellow}\textbf{25$\div$29}  & \cellcolor{lime}21.00 & 1.04  & 2.43  & 2.23   & 5.21 & 1.13      & 6.30    & 2.66 \\
\hline
\cellcolor{yellow}\textbf{30$\div$34}  & \cellcolor{lime}18.03 & 1.26  & 2.29  & 0.85   & 3.92 & 0.76      & 5.24    & 3.72 \\
\hline
\cellcolor{yellow}\textbf{35$\div$39}  & \cellcolor{lime}21.25 & 1.75  & 2.63  & 0.68   & 7.78 & 1.05      & 3.92    & 3.45 \\
\hline
\cellcolor{yellow}\textbf{40$\div$44}  & \cellcolor{lime}22.35 & 1.63  & 2.25  & 2.53   & 7.00    & 0.67      & 4.48    & 3.79 \\
\hline
\cellcolor{yellow}\textbf{45$\div$49}  & \cellcolor{lime}19.27 & 1.50  & 1.49  & 2.61   & 8.24 & 0.88      & 1.93    & 2.64 \\
\hline
\cellcolor{yellow}\textbf{50$\div$54}  & \cellcolor{lime}22.30 & 1.38  & 1.37  & 5.54   & 8.05 & 0.52      & 2.02    & 3.41 \\
\hline
\cellcolor{yellow}\textbf{55$\div$59}  & \cellcolor{lime}18.27 & 1.11  & 1.77  & 1.41   & 4.60  & 0.68      & 3.62    & 5.06 \\
\hline
\cellcolor{yellow}\textbf{60$\div$64}  & \cellcolor{lime}18.43 & 0.91  & 2.37  & 1.07   & 6.05 & 0.87      & 3.53    & 3.63 \\
\hline
\cellcolor{yellow}\textbf{65$\div$69}  & \cellcolor{lime}12.74 & 0.71  & 2.39  & 0.55   & 0.48 & 0.95      & 3.33    & 4.33 \\
\hline
\cellcolor{yellow}\textbf{70+}    & \cellcolor{lime}10.55 & 0.71  & 2.53  & 0.06   & 1.04 & 0.22      & 4.22    & 1.77 \\
\hline
\end{tabular}
\caption{Average number of contacts by age groups, total and disaggregated by social context in which the contact takes place, adding  \virgolette{family} context.}
\label{table:1}
\end{table}

\begin{figure}[]
\centering
\includegraphics[width=0.35\textwidth]{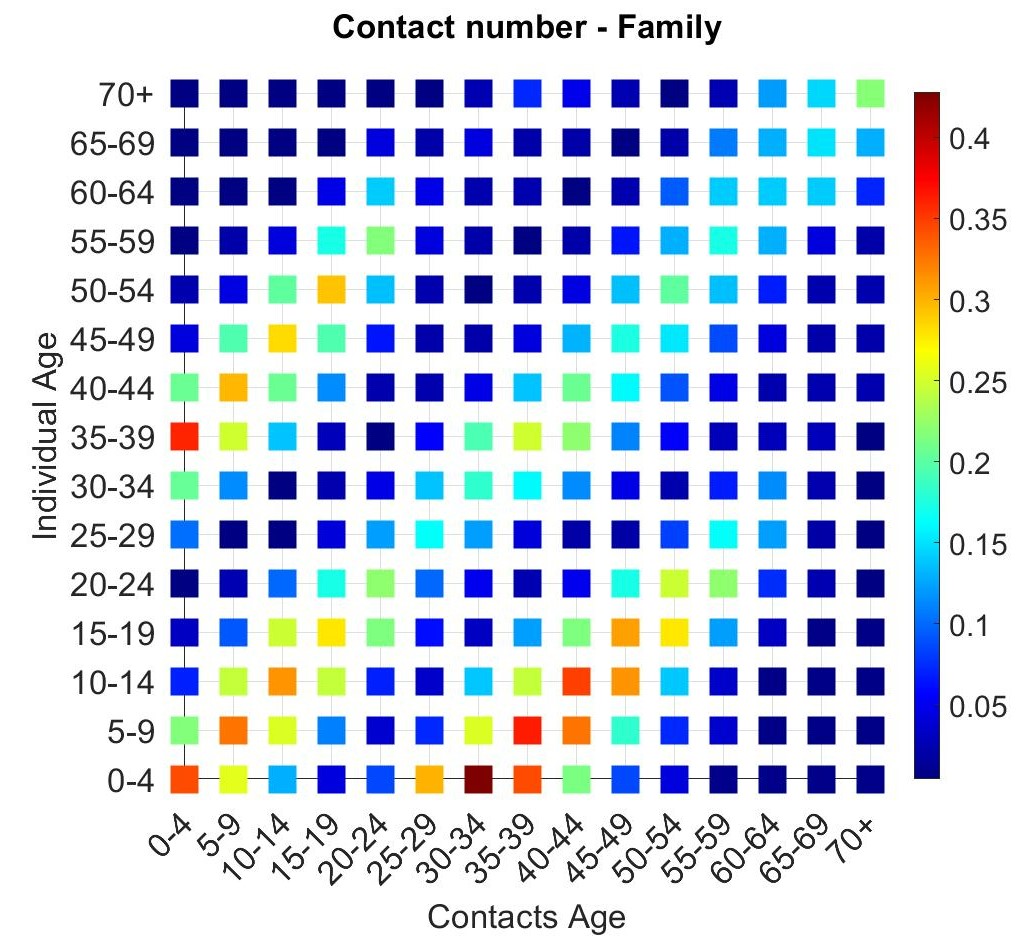}
\includegraphics[width=0.35\textwidth]{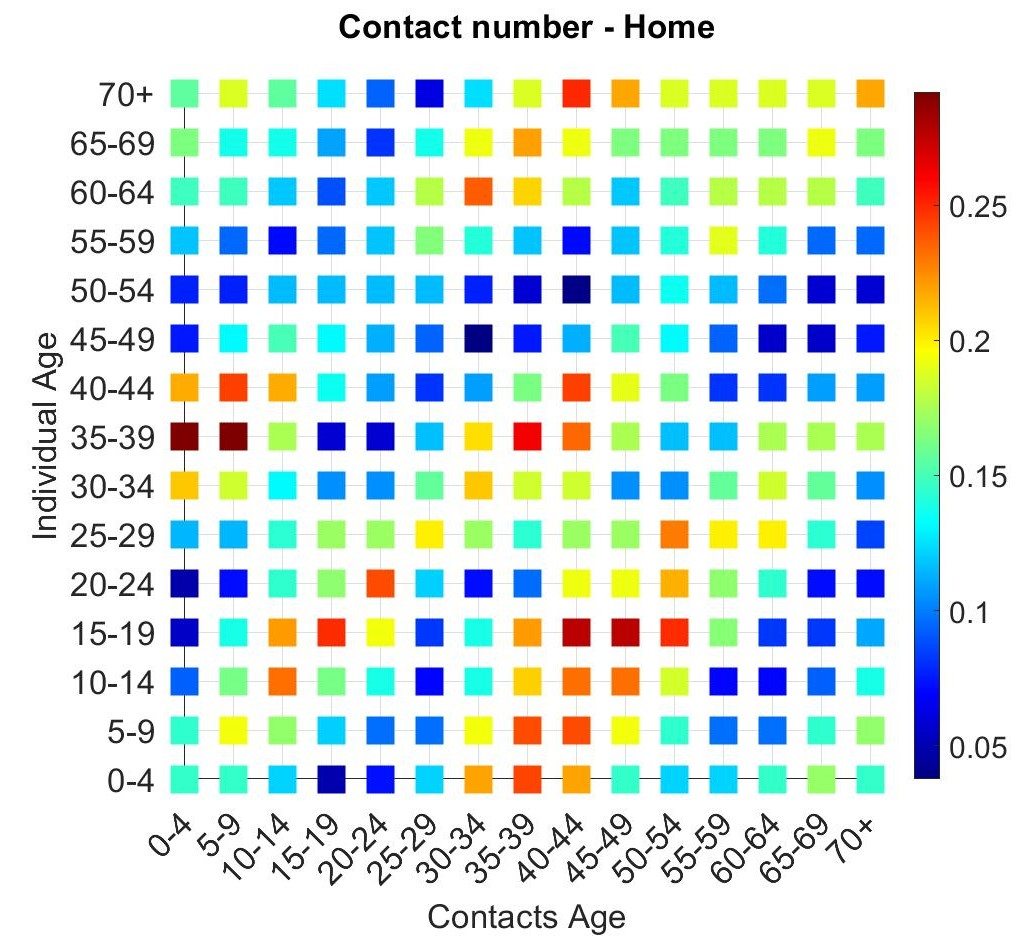}\\
\includegraphics[width=0.35\textwidth]{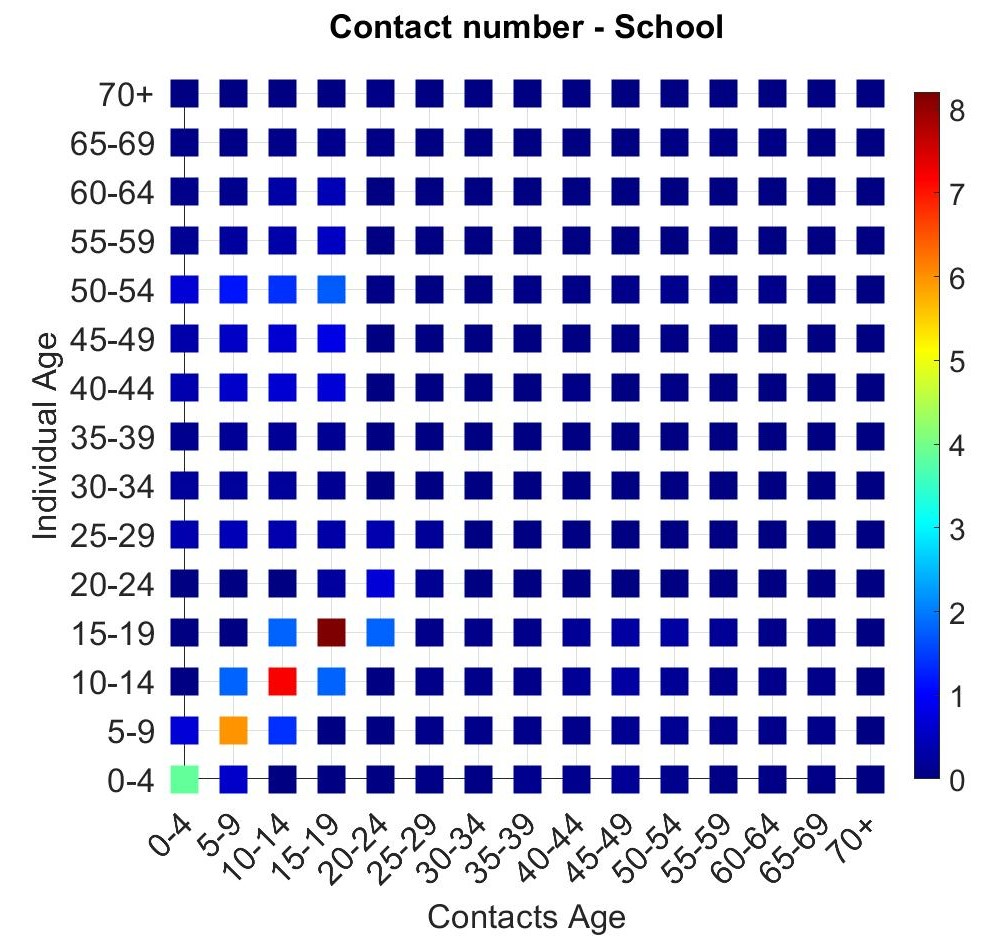}
\includegraphics[width=0.35\textwidth]{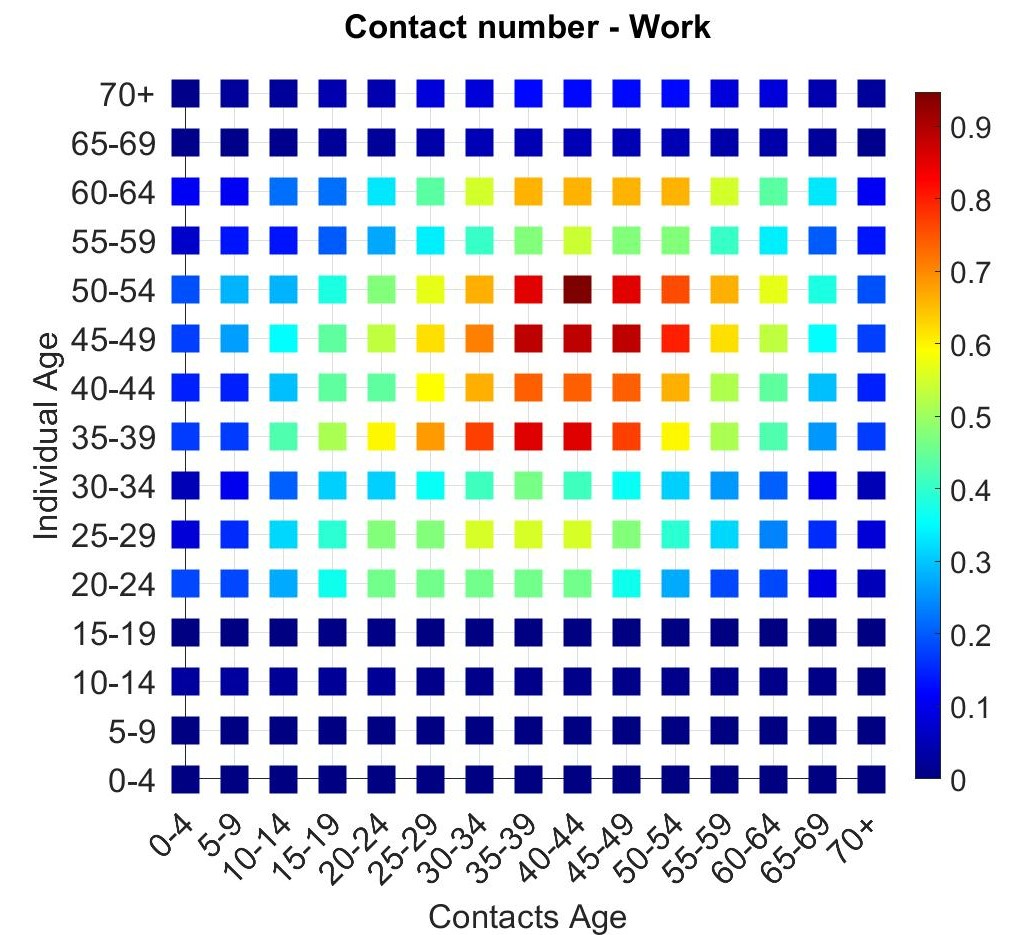}\\
\includegraphics[width=0.35\textwidth]{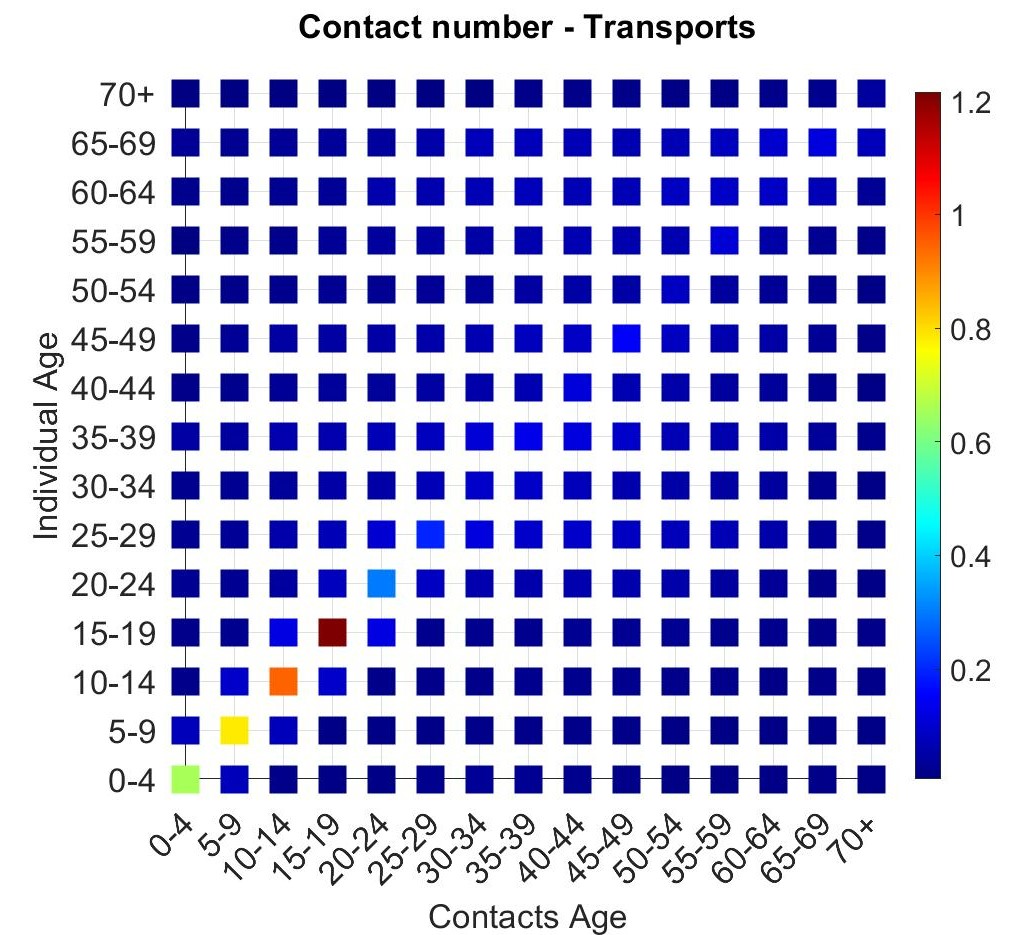}
\includegraphics[width=0.35\textwidth]{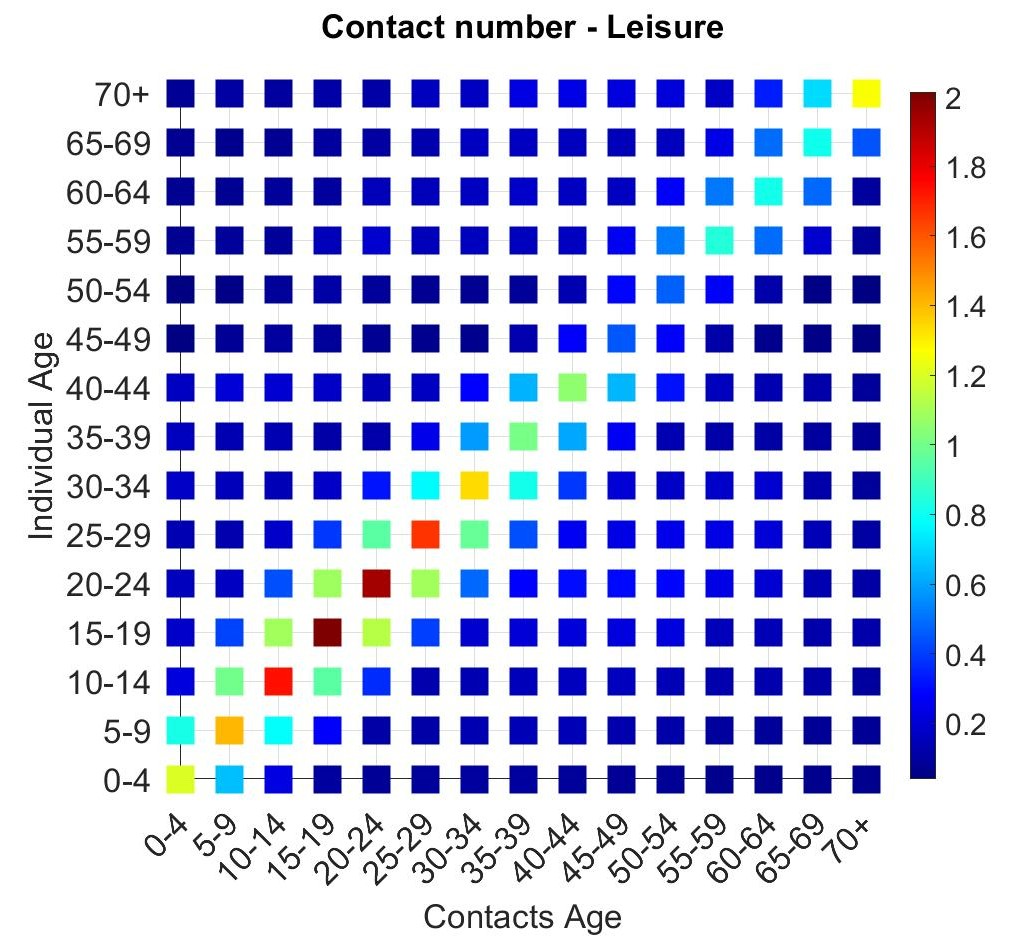}\\
\includegraphics[width=0.35\textwidth]{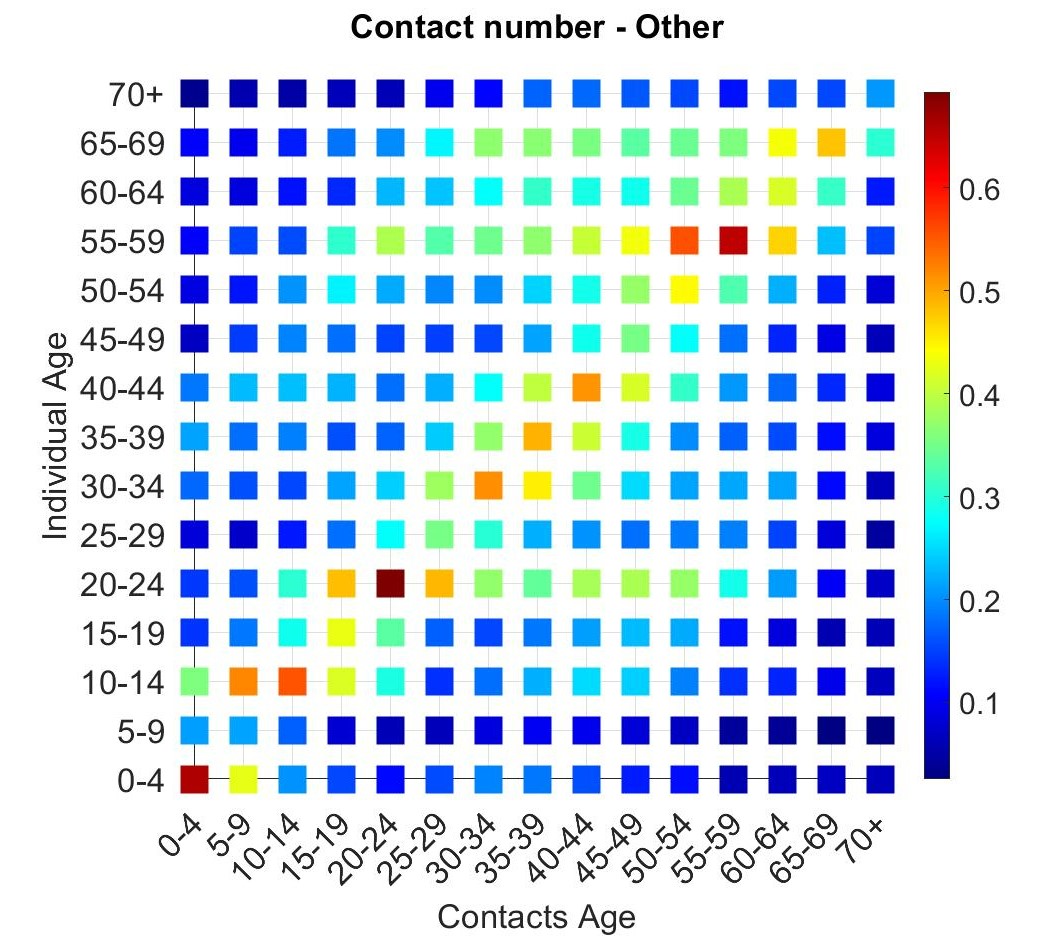}
\caption{POLYMOD matrices $M^{k}_{i,j}$, $k=1,...,7 = K$, $i,j = 1,...,15 = N_a$: total number of contacts that each individual has with individuals from the 15 age groups, in the seven contexts of exposition. }
\label{fig:context_matrix_contact}
\end{figure}

\begin{figure}[t!]
\centering
\includegraphics[width=0.45\textwidth]{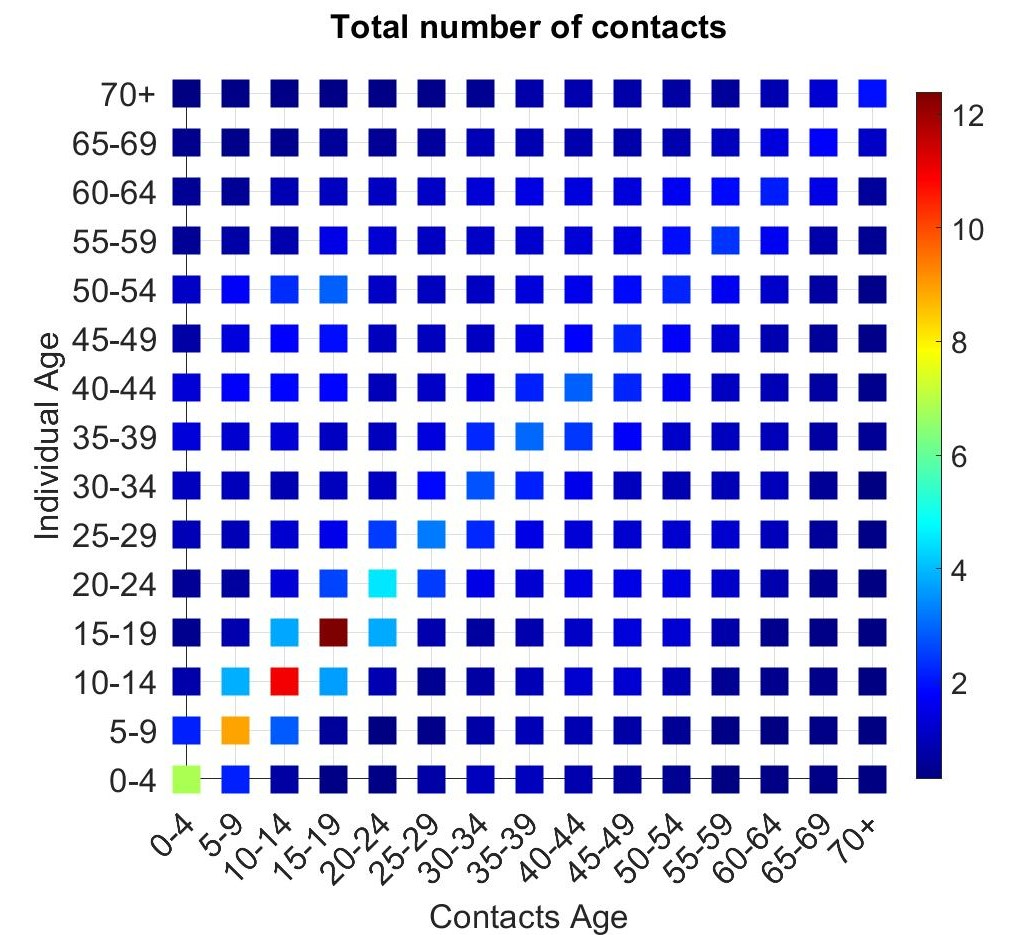}
\includegraphics[width=0.45\textwidth]{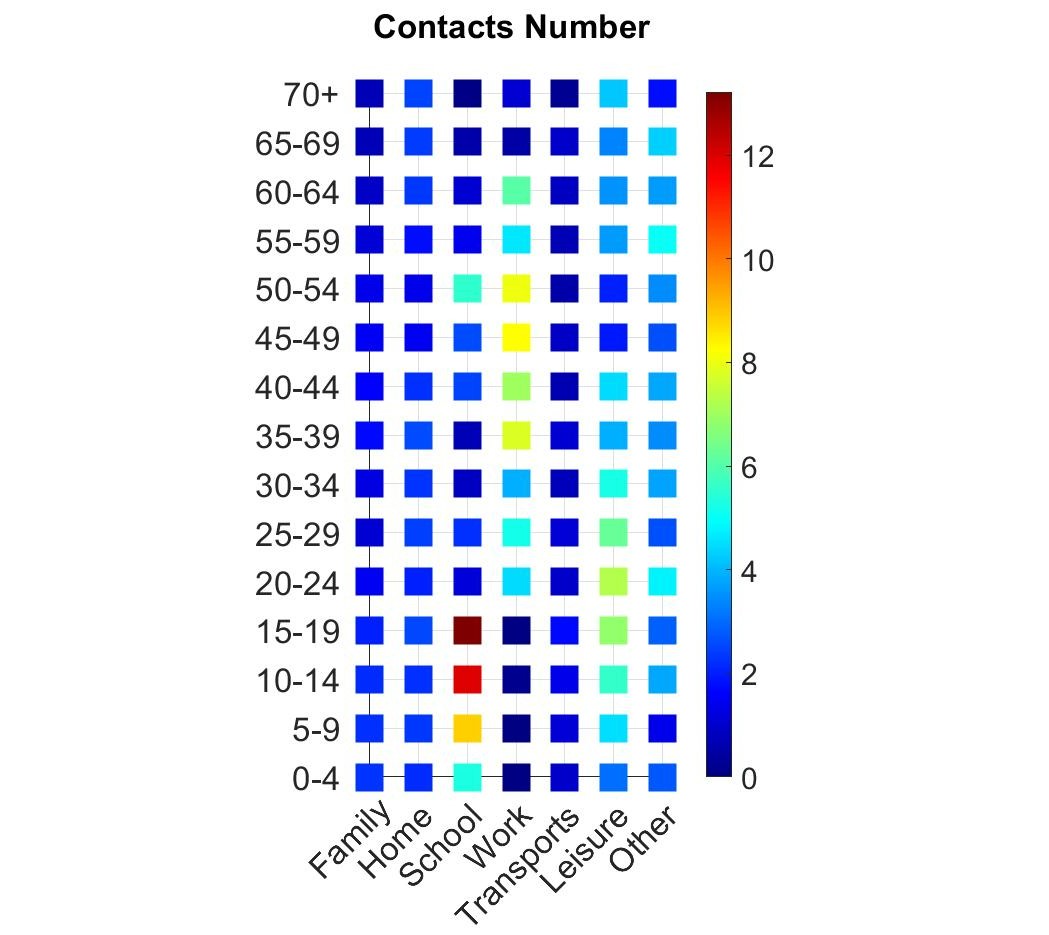}
\caption{Total number of contacts that each individual has with individuals from the 15 age groups at left, for each exposure context at right. }
\label{fig:matrix_tot_contact}
\end{figure}

{On the basis of the social contact data provided in \cite{DD}, it is possible to estimate, for each context of exposition, the daily contacts of each individual distributed in the different age groups (cf. Figure \ref{fig:context_matrix_contact}), as well as the total number of daily contacts that each individual of a given age group has with individuals in different age groups and in different contexts of exposition  (cf. Figure \ref{fig:matrix_tot_contact}).}

These contact matrices have been used to determine the $\beta-matrices$ that express the transmissibility rate for each age class in each exposure contexts. {However, the number of contacts is not the only factor that should be taken into account. Indeed, as suggested in \cite{EE}, age-dependent susceptibility to the infection should also be considered, that we denote $risk_A$ (Table  \ref{table:2}), as well as different levels of risk to contract the virus in the different contexts of exposition \cite{FF}, denoted as $risk_E$ (Table \ref{table:3}).}

\begin{table}[t!]
    \centering
    \begin{tabular}{|c|c|c|c|c|c|c|c|c|c|c|c|c|c|c|}
        \hline
        \textbf{Age groups}  & 0$\div$4  & 5$\div$9 & 10$\div$14 & 15$\div$19 & 20$\div$24 & 25$\div$29 & 30$\div$34 & 35$\div$39 \\
        \hline
        $\mathbf{risk_A}$ & 0.34 & 0.34 & 0.34 & 1 & 1 & 1 & 1  & 1  \\
        \hline
        \hline
        \hline
        \textbf{Age groups}  & 40$\div$44 & 45$\div$49 & 50$\div$54 & 55$\div$59 & 60$\div$64 & 65$\div$69 & 70+ &\\
        \hline 
        $\mathbf{risk_A}$  & 1 & 1 & 1 & 1 & 1 & 1.47 & 1.47 &  \\
        \hline
    \end{tabular}
    \caption{Susceptibility by age group}
    \label{table:2}
\end{table}

\begin{table}[t!]
    \centering
    \begin{tabular}{|c|c|c|c|c|c|c|c|}
        \hline
        \textbf{Level of the risk} & Low  & Medium-Low & Medium & Medium-High & High & Very High  \\
        \hline
        $\mathbf{risk_E}$ & 0.25  & 0.5 & 1 & 1.5 & 2 & 2.5  \\
        \hline
    \end{tabular}
    \caption{Levels of risk associated to the contexts of exposition}
    \label{table:3}
\end{table}

The risk associated to each age group is based on a CTS report, \cite{EE}, which states that children under 14 have a $66$\% lower than average risk of contract the infection, while adults over the age of 65 have a $47$\% above average risk of infection. 
{On the other hand, the risk factor associated with each exposure context depends on the relative contribution of the various sectors (sub-contexts) that constitute it. Specifically, we assign a risk level to each sector by considering four key factors: the presence of closed spaces, the duration of interactions, the density of the crowd, and activities involving forced exhalation (such as sneezing, yelling, singing, and coughing).}
{For the following contexts of exposition, we consider different sectors as follow:
\begin{itemize}    
    \item \textbf{school context}: it is split into six sectors (nursery, kindergarten, elementary school, middle school, high school and university);
    \item \textbf{work context}: it is divided into seven sectors (essential, healthcare, manufacturing, trade, construction, restaurant/hotel, other);
    \item \textbf{leisure context}: is divided into six sectors (outdoor restaurants, indoor restaurants, outdoor pub/disco, indoor pub/disco, indoor sport, outdoor sport);
    \item \textbf{\virgolette{other} context}: is divided into three sectors (essential retail, other not essential retail, mass and religious events).
\end{itemize}
The home context takes into account contacts that occur in households, but not among co-habitants, which are instead referred to the family context. No split is considered for the transport context.}

Then, a scenario can be defined by assigning values in $[0,1]$ for each sector of the various contexts, to indicate their opening/closing considered according to the regulations in force ($0$ if the sector is totally closed, $1$ if it is totally opened, values in $(0,1)$ if there are other restrictions). 
To obtain the $\beta-matrix$ that expresses the effective risk of transmission for each age group in contact with other age groups in the different contexts of exposition, we multiply the \textit{k-th} contact matrix by a function depending on the risk parameters defined above and on the chosen scenario, which expresses the level of opening/closure of the sub-contexts: 
\begin{equation}
    \beta_{i,j}^{k} =  M_{i,j}^{k} \ h_{k}(risk_{E}^{S_k}, risk_{A}, scenario_k^l)   \quad  \text{for all sectors } l=1,\ldots,S_k,  \quad  \text{for all context } k=1,\ldots,K,
\end{equation}
where $i$ is the age of the individual, $j$ is the age of the contact, $M^{k}$ is the contact matrix of the context $k$, $S_k$ is the number of sectors of the $k^{th}$ context of exposition and $h_{k}$ is a piecewise constant context-dependent function, in which for each age group we calculate a value expressing the global risk associated to the specific context $k$. {The latter function} depends on:
\begin{itemize}
    \item $risk_{E}^{S_k}$, that is the value of risk assigned to each of the $S_k$ sectors of the context $k$ between low, medium-low, medium, medium-high, high and very high risk (Table \ref{table:3});
    \item ${risk_{A}}_{i}$, that is the vector containing the risk for each age group (Table \ref{table:2});
    \item $scenario_k^l$, that is the level of opening/closing (in $[0,1]$) of each sector $l$ of the exposure context $k$, as we defined before.
\end{itemize}  
%We obtain a $\beta-matrix$ of dimension $N_a^2$ for each of the seven context of exposition.\\

Starting from November 2020, in Italy, containment measures differentiated by region are introduced, due to the fact that the spread of the virus was different on the peninsula. This differentiation is expressed in colors: \textit{yellow} for the regions with a transmissibility index less than $1.25$, \textit{orange} for regions with transmissibility index between $1.25$ and $1.5$, where the associated NPIs have more restrictions, and \textit{red} for regions with transmissibility index greater than $1.5$. The NPIs associated to red regions are the more restrictive{, corresponding to almost complete lockdowns}. These measures are issued with an order from the Minister of Health, in agreement with the president of the region concerned and the CTS, and must have a minimum duration of $15$ days, {since their impact on the transmission mechanism is delayed, as estimated, e.g., in \cite{summan2022timing}}. The conditions for the regions to switch the associated color have changed during the evolution of the pandemic, and \textit{white} zones have been also introduced in January 2021, {corresponding to the almost complete dismissal of mobility restrictions}.

\begin{figure}[]
\centering
\includegraphics[width=0.4\textwidth]{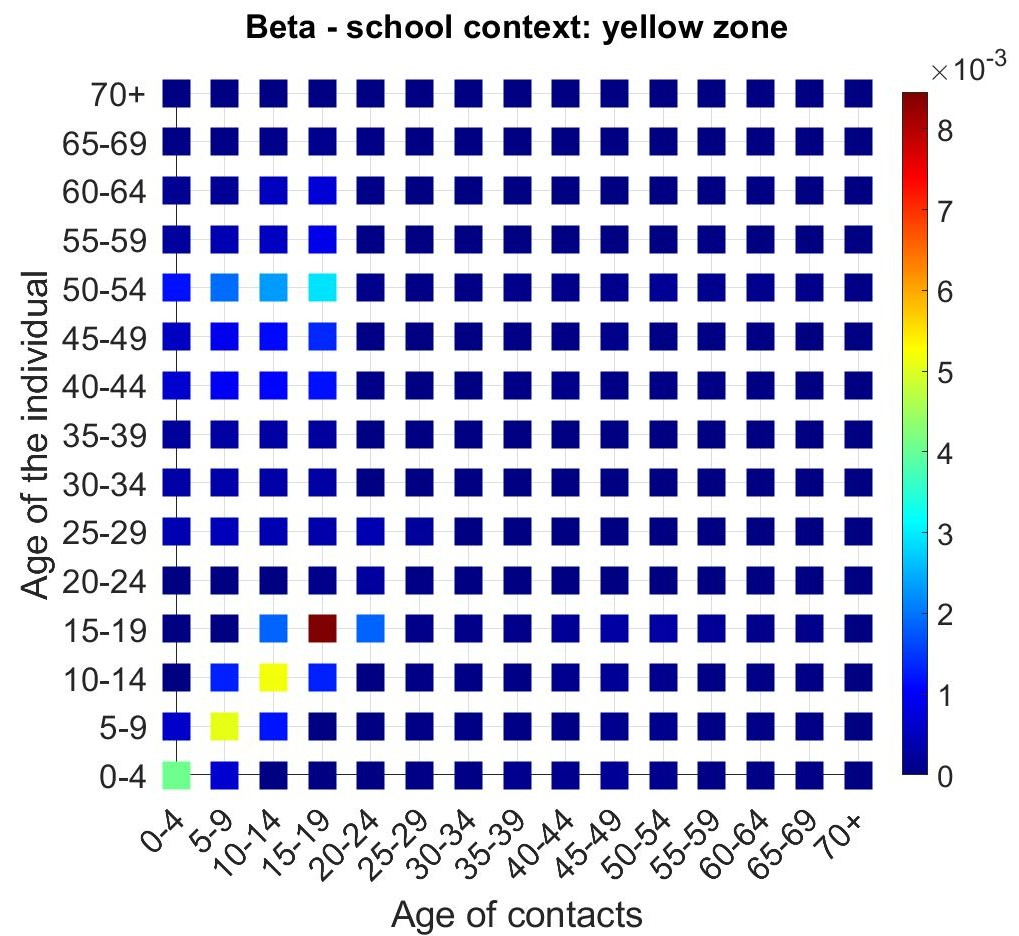}
\includegraphics[width=0.4\textwidth]{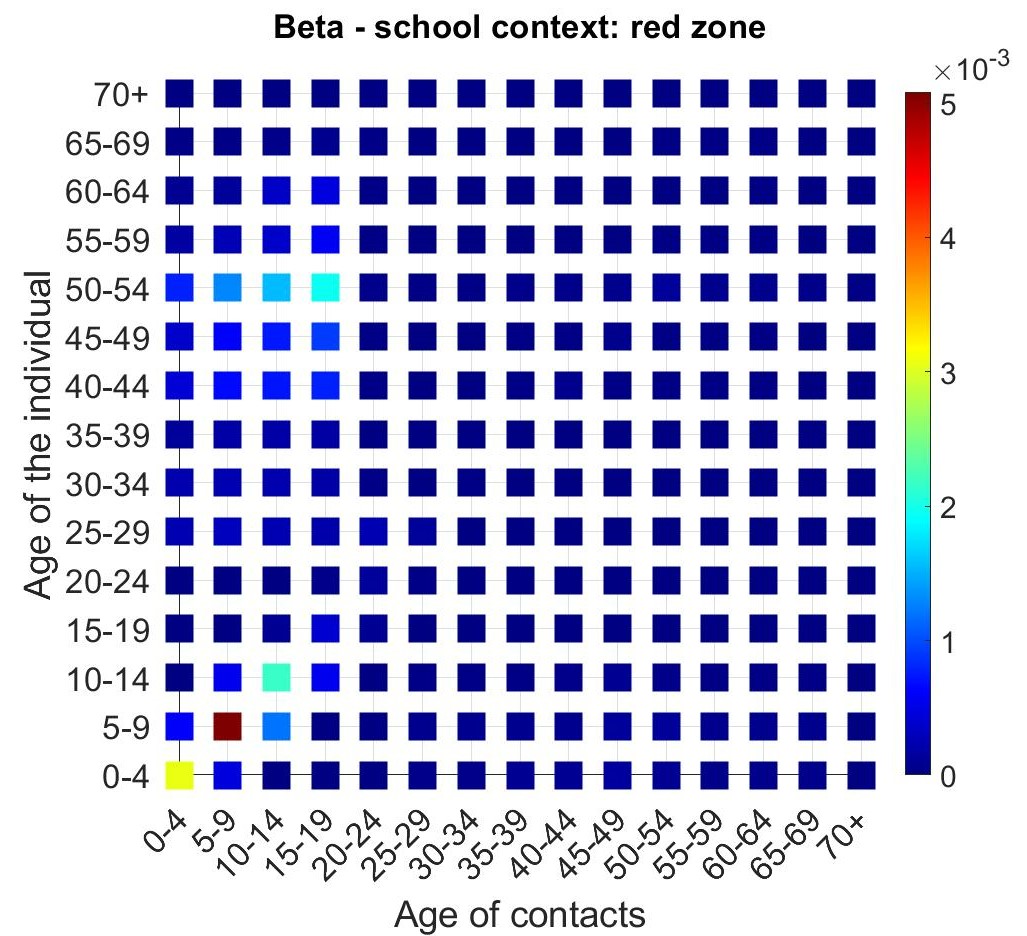}
\caption{$\beta-matrices$ in school context, for yellow zone scenario at left, for red zone scenario at right. }
\label{fig:beta_matrix_yellow_red}
\end{figure}

In Figure \ref{fig:beta_matrix_yellow_red} we can see the $\beta-matrices$ of school context associated to two different scenarios: yellow zone and red zone. In yellow zone the restrictions included schools totally open up to the third year of middle school and high schools open to $50\%$, while in red zone we had schools open up to the first year of middle school, while second and third year of middle school, high schools and universities were totally closed.

With different containment measures in force for each region, to define the scenario we have to make different considerations depending on whether we are simulating on a national or regional basis. At regional level it is sufficient to reconstruct the temporal history of the regional restrictions, instead at national level it is necessary to take into account the different containment measures in force in each region, {and to reconstructed an average measure for the restrictions}. To do that, for each week, we analyzed the regional restrictions in force (which regions where in yellow zone, which in orange, which in red, which in white zone), and we set the final scenario weighing all these different scenarios on the population of each region. In fact regions with small populations have less weight on Italian situation than large regions.

{In the national case, we consider an array $\mathbf{n}$ in which each component corresponds to the number of inhabitants of an Italian region.}
Moreover, we consider arrays for each color zone (for example red zone, orange zone, yellow zone and green zone) {having one entry for each region, in which we have $1$ if the corresponding region (in the same order of the previous array) has government restriction of that color, $0$ if there are other color restriction. For example, calling \textbf{r} the array of binary values corresponding to red zone, \textbf{o} the respective array for the orange zone, \textbf{y} for the yellow zone and \textbf{w} for the white one.}
Multiplying the inhabitants array $\mathbf{n}$ with each color zone array we obtain the number of the Italian population under that color zone restriction:
\begin{equation*}
    r_n = \sum_{i=1}^{N} \mathbf{r}_i \mathbf{n}_{i}, 
\qquad
    o_n = \sum_{i=1}^{N} \mathbf{o}_i \mathbf{n}_{i},
\qquad 
    y_n = \sum_{i=1}^{N} \mathbf{y}_i \mathbf{n}_{i},
\qquad \text{and} \qquad
    w_n = \sum_{i=1}^{N} \mathbf{w}_i \mathbf{n}_{i},
\end{equation*}
where $r_n$ is the number of Italian people in red Zone restrictions, $o_n$ is the number of Italian people in orange zone restrictions, $y_n$ is the number of Italian people in yellow zone restrictions and $w_n$ is the number of Italian people in White Zone restrictions.

{Finally, we derive the national scenario by averaging} as follows:
\begin{equation*}
    scenario_{k}^{l} = \frac{r_n s_{r_{k}}^{l} + o_n s_{o_{k}}^{l} + y_n s_{y_{k}}^{l} + w_n s_{w_{k}}^{l}}{r_n + o_n + y_n + w_n}, \quad \quad l\in {1,...,S_{k}}, \quad k \in {1,...,K}
\end{equation*}
where $scenario_{k}^{l}$ is the value in $[0,1]$ assigned to the $l^{th}$ sector of the $k^{th}$ context of exposition of the scenario, $s_{r_{k}}^{l}$.
Instead, $s_{o_{k}}^{l}$, $s_{y_{k}}^{l}$ and $s_{w_{k}}^{l}$ are the values in $[0,1]$ assigned to the $l^{th}$ sector of the $k^{th}$ context of exposition of the red zone scenario, the orange zone scenario, the yellow zone scenario and the white zone scenario respectively. 

%We assigned value $1$ to home scenario because there are no more limitation inviting people at home. Schools are totally closed except for the university, which we have assumed to be $10\%$ frequented by school staff and some student who uses classrooms reserved for personal study.
%In work scenario we have rescaled the values to take into consideration the fact that in August many sectors work less, except for restaurant and hotel sectors. Furthermore, we had no limitation in outdoor restaurants and pub, but discos were closed so at the voice "outdoor pub/disco" we considered $0.70$. Theaters, museums, concerts and shows have resumed in the summer but with limited capacity, on average at $50\%$. \\
%This is how we set the scenario, which therefore depends strictly on the containment measures in force. \\
%For each context of exposition we define a function that calculates values that enclose environmental risk, the age risk and scenario considered.  
%Now to define $\beta- matrices$, for each context of exposition, we multiply the contact matrix for the values obtained. These matrices give us the transmission rate for the different age groups and exposure contexts.\\ 

\subsection{{Data generation and calibration of the parameters}}
\label{subsec:calValGP}
{In this section, we describe the calibration procedure (see Section \ref{calibration_and_updates}) for determining model parameters that could not be directly derived from clinical or population cohort studies. Furthermore, we detail the processing of vaccination data (cf. Section \ref{vaccination}) and describe the adjustments made to the model to incorporate the social restrictions introduced under the \textit{Green Pass} initiative (cf. Section \ref{greenpass}).}

\subsubsection{Calibration}
\label{calibration_and_updates}
Model calibration {poses significant challenges in epidemic modeling, as inaccurate reconstructions can critically compromise the quality of scenario analyses and forecasts. Special attention must be given to parameters governing the transmission mechanism, particularly the function $c(t)$ in our model, as it plays a pivotal role in determining the model's reliability \cite{ziarelli2024model}.

For this purpose, we undergo a Least-Square calibration procedure based on} reported deceased data from Dipartimento di Protezione Civile Italiana (DPC).
{Indeed, the compartment of deceased individuals is the only one with a direct counterpart in DPC data and is generally the least affected by monitoring uncertainties.
In our cases, both $c(t)$ and $\beta(t)$ (cf. Section  \ref{subsec:basicSEIHRDV}) are piecewise constant functions.
The values of $c(t) = \sum_{i}^{N_{int}} c_i I_{(t_i, t_{i+1}]}(t)$ are calibrated separately for each time subinterval corresponding to a specific scenario (e.g., a particular restriction level).}
{The iterative calibration stage is carried out according to the following steps}:
\begin{enumerate}
    \item {Associate} a $\beta-matrix$ for each subinterval {corresponding to the prescribed scenario};
    \item {Solve the following optimization problem:
    \begin{equation}
        \min_{\{c_i\}_i} J(c) = \sum_{t=0}^{T} \frac{(D(t) - \hat{D}(t))^{2}}{\left(\hat{D}(t)\right)^2}, 
    \end{equation}
    where $\hat{D}$ is the total number of deceased individuals provided by the DPC in [0,T]
    }  
\end{enumerate}

\begin{figure}[t!]
\centering
\includegraphics[width=0.9\textwidth]{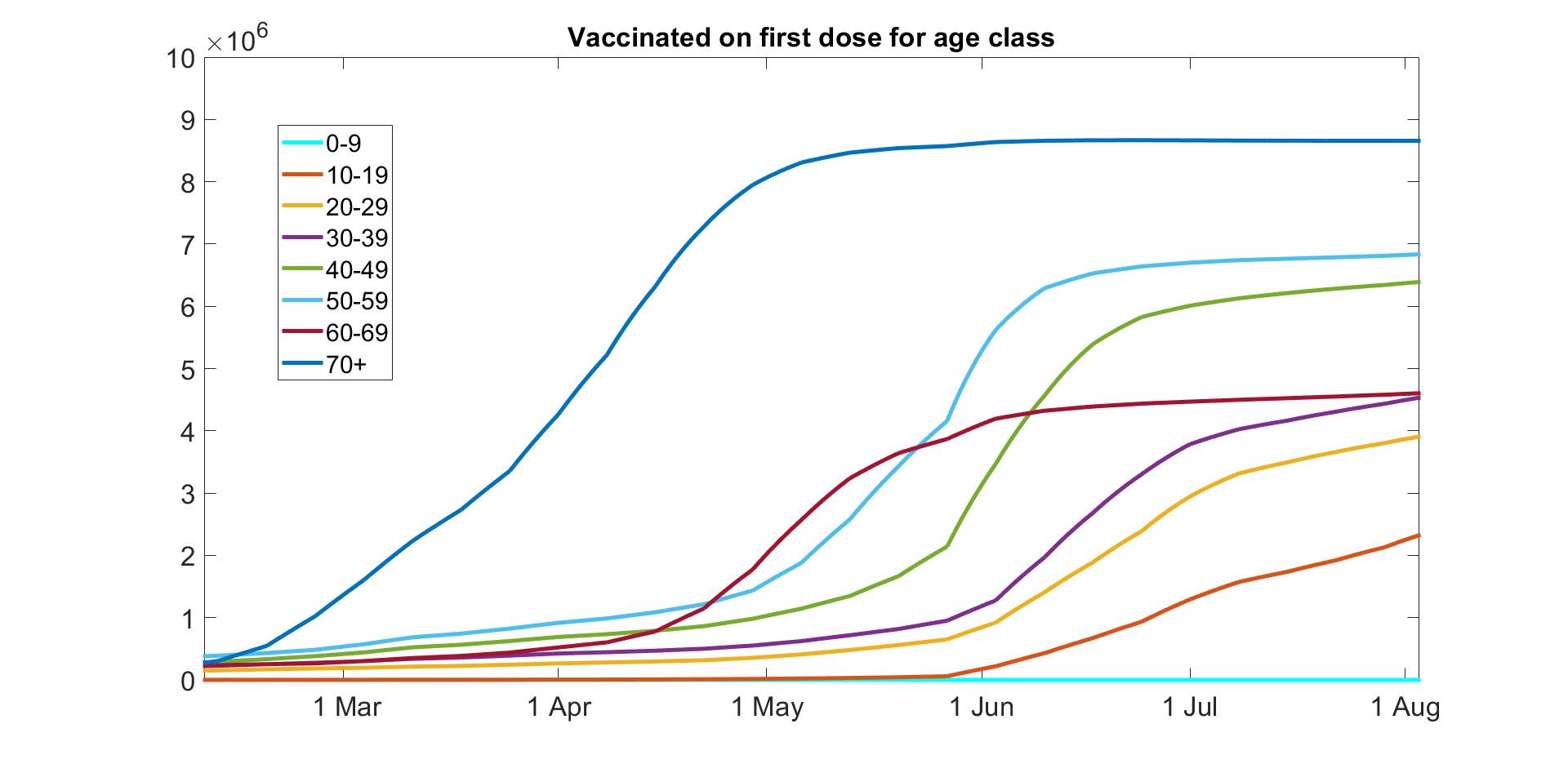}
\caption{Vaccinated in first dose for {DPC} age groups. }
\label{fig:vaccinati}
\end{figure}

\subsubsection{Vaccination campaign}
\label{vaccination}
{The Italian COVID-19 vaccination campaign included the administration of two doses and a booster (third) dose in the late periods of the outbreak. These dynamics are incorporated into our model using vaccination data provided by DPC, enabling us to track the participation of individuals in the vaccination campaign.} At the beginning of the vaccination campaign, vaccinations were initially {delivered} by age group. In particular, these data {are split} into eight age groups $\{0\div9, 10\div19, 20\div29, 30\div39, 40\div49, 50\div59, 60\div69, 70+\}$, even though in our model we consider fifteen age groups $\{0\div4, 5\div9, 10\div14, 15\div19, 20\div24, 25\div29, 30\div34, 35\div39, 40\div44, 45\div49, 50\div54, 55\div59, 60\div64, 65\div69, 70+\}$.
{Thus,} to initialize our model we {make the following} assumption: calling $A_i$ the age groups provided by data, with $i=1,...,8$, and $B_j$ the age groups used in our model, with $j=1,...,15$, the number of the initial vaccinated individuals of the age groups $B_j$ is given by
\begin{equation*}
    |B_j| = \frac{|A_{k}|}{2}, \quad j=1,\dots, 14,
\end{equation*}
where 
\begin{equation}
 k = \begin{cases}
        \biggl \lfloor \frac{j}{2}\biggr \rfloor + 1 \quad \text{if} \quad j\mod{2} = 1\\[0.3cm]
        
        \biggl \lfloor \frac{j}{2}\biggr \rfloor \quad \text{if} \quad j\mod{2} = 0
     \end{cases}
    \label{eq:dinamica}
\end{equation}
and $|B_{15}| = |A_8|$. 
In this way, we are able to exploit the SEIHRDV multi-age and multi-group for reproducing the curve of vaccinated people evolving over time for each age group.
{In Figure \ref{fig:vaccinati} we report this evolution in the 8 DPC age-classes, finding agreement with the actual administrations \textit{per-age}}.

{In our model, we explicitly include only individuals vaccinated with at least one dose in the compartment $V$. The efficacy parameter $\sigma$, as estimated from clinical trials \cite{HH}, was an appropriate choice when the number of individuals with only the first dose surpassed that of fully vaccinated individuals. However, by late 2021, approximately $78$\% of the Italian population had received at least one dose, with nearly $74$\% fully vaccinated. Furthermore, vaccine effectiveness in reducing COVID-19 transmissibility is estimated to decline approximately six months after full vaccination, necessitating the introduction of the third, or booster, dose. Consequently, for these later periods, we compute $\sigma$ as a weighted parameter, determined as follows:}  
\begin{equation}
    \sigma(t) = \frac{\sigma_{1} \, V_{D_1}(t) + \sigma_{2}\, ( V_{D_2}(t) + V_{D_4}(t)) + \sigma_{3} \, V_{D_3}(t)}{V_{D_1}(t) + V_{D_2}(t) + V_{D_3}(t) + V_{D_4}(t)},
\end{equation}
with
\begin{itemize}
    \item $\sigma_{1},$ rate of ineffectiveness in reducing virus transmissibility for those who received only the first dose of the vaccine;
    \item $\sigma_{2},$ rate of ineffectiveness in reducing virus transmissibility for those who have been fully vaccinated (double dose of vaccine) for less than six months;
    \item $\sigma_{3},$ rate of ineffectiveness in reducing virus transmissibility for those who have been fully vaccinated (double dose of vaccine) for more than six months;
    \item $V_{D_1}(t)$ number of individuals who received only the first dose of vaccine on day $t$ (DPC); 
    \item $V_{D_2}(t)$ number of individuals who are fully vaccinated (with two doses) for less than six months on day $t$ (DPC);
    \item $V_{D_3}(t)$ number of individuals who are fully vaccinated (with two doses) for more than six months on day $t$ (DPC);
    \item $V_{D_4}(t)$ number of individuals who are vaccinated with the booster dose on day $t$ (DPC).
\end{itemize}
We assume that individuals receiving the booster dose regain the vaccine efficacy that achieved in the first few months after full vaccination.
In our model we do not distinguish on the various vaccines used, hence the ineffectiveness values are estimated from average values of different vaccines data collected in the literature. In particular we {estimate} $\mathbf{\sigma_{1}} = 0.30$, $\mathbf{\sigma_{2}} = 0.12$ and $\mathbf{\sigma_{3}} = 0.50$.

{Finally, to determine the value of $\theta$, introduced in the model in (\ref{eq_fifr}), we utilize data from the ISS as reported in \cite{II}. Specifically, we analyze the ratio of unvaccinated individuals who died relative to unvaccinated COVID-19 diagnoses and the ratio of vaccinated individuals who died relative to vaccinated COVID-19 diagnoses. From this analysis, we set $\theta = 0.8$, which implies that for vaccinated individuals diagnosed with COVID-19, the vaccine reduces mortality by $20$\%.}

\subsubsection{Green Pass}
\label{greenpass}
{On June 17, 2021, Italy introduced the Green Pass restriction for the first time. The Green Pass, a digital COVID certificate proposed by the European Commission, aimed to facilitate the safe and unrestricted movement of European Union citizens during the COVID-19 pandemic. In Italy, the Green Pass certified one of the following conditions: receipt of a COVID-19 vaccination (issued after each vaccine dose), a negative result from a rapid antigen test within the previous $48$ hours or a molecular test within the previous $72$ hours, or recovery from COVID-19 within the last $6$ months.}

We {modify} the model to take into account the restrictions due to the Green Pass. 
\begin{figure}[t!]
\centering
\includegraphics[width=0.5\textwidth]{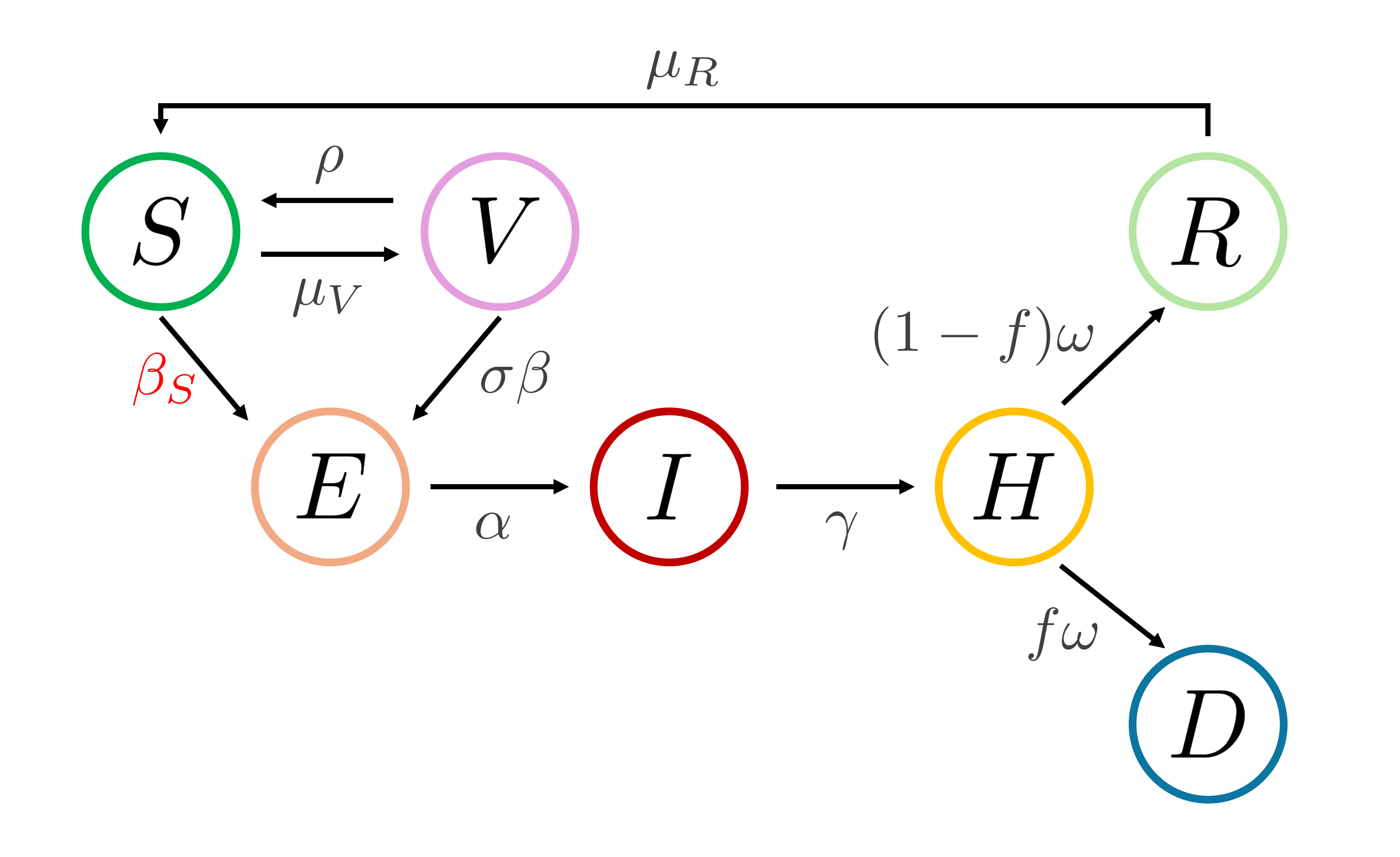}
\caption{SEIHRDV model adapted to take into account Green Pass restrictions.}
\label{fig:beta_greenpass}
\end{figure}
For this purpose, we {introduce} a second rate of transmissibility, $\beta_{S}$, which expresses the probability that a susceptible individual will contract the virus after a contact with an infectious, considering the limitations due to not having the Green Pass. For simplicity we are considering as Green Pass holders the individuals who have received at least one dose of vaccine. So the individuals in $V$, which are Green Pass holders, are not subject to the same restrictions. For them, we consider the transmissibility rate $\beta $ defined earlier. If the Green Pass restrictions were lifted, then $\beta_{S}=\beta$.
Therefore, the equations become:
\begin{equation}
    \begin{array}{l}
    \displaystyle
\frac{dS_{i}}{dt} =  - \Bigg(\sum_{k=1}^{K} \Bigg(\sum_{j=1}^{M}c(t) \, \textcolor{red}{\beta_{S}}_{ij}^{k} I_{j} \Bigg) \Bigg)\frac{S_{i}}{N_{i}} - d_{i} \, \frac{S_{i}}{S_{i}+R_{i}} + \mu_{R}\, R_{i} + \mu_{V}\,  V_{i}, \\[4mm]

\displaystyle \frac{dE_{i}^{k}}{dt} = \Bigg(\sum_{j=1}^{M}c(t)\, \textcolor{red}{\beta_{S}}_{ij}^{k}\, I_{j} \Bigg) \frac{S_{i}}{N_{i}} + \Bigg(\sum_{j=1}^{M}c(t)\, \beta_{ij}^{k}\, I_{j} \Bigg) \frac{\sigma \, V_{i}}{N_{i}} - \alpha \, E_{i}^{k}, \qquad \qquad \qquad \qquad k=1,\dots, K, \\[4mm]
\displaystyle \frac{dI_{i}}{dt} = \alpha \, \Bigg( \sum_{k=1}^{K} E_{i}^{k} \Bigg) -\gamma \, I_{i}, \\[4mm]

\displaystyle \frac{dH_{i}}{dt} = \gamma \, I_{i} - \omega \, H_{i}, \qquad \qquad \qquad \qquad \qquad \qquad \qquad \qquad \qquad \qquad  \qquad \qquad \qquad \qquad \qquad  t\in (0,T],\\[4mm]

\displaystyle \frac{dR_{i}}{dt} = (1-f(S_{i},V_{i}))\, \omega \, H_{i} - \mu_{R}\,  R_{i}, \\[4mm]

\displaystyle \frac{dD_{i}}{dt} = f(S_{i},V_{i})\, \omega \, H_{i}, \\[4mm]

\displaystyle \frac{dV_{i}}{dt} = - \Bigg(\sum_{k=1}^{K} \Bigg(\sum_{j=1}^{M}c(t) \, \beta_{ij}^{k} \, I_{j} \Bigg) \Bigg)\frac{\sigma\,  V_{i}}{N_{i}} + d_{i} \, \frac{S_{i}}{S_{i}+R_{i}} - \mu_{V} \, V_{i}, \\[4mm]

S_{i}(0) = S_{i_{0}}, \quad E^k_{i}(0) = E^k_{i_{0}}, \quad I_{i}(0) = I_{i_{0}}, \quad H_{i}(0) = H_{i_{0}}, \quad R_{i}(0) = R_{i_{0}}, \quad D_{i}(0) = D_{i_{0}}, \quad V_{i}(0) = V_{i_{0}},\\[2mm]
\forall i = 1,...,N_{a}.\\
 \end{array}
\end{equation}

\section{Results}
\label{results}
We report {numerical} results obtained in {different scenarios for both the global and regional Italian levels}. Our aim is to {capture how the COVID-19 developed in the different contexts of exposition among the different age-groups and to assess the validity of the proposed novel compartmental model}. For each period selected, an accurate research was carried out to set the scenario parameters, taking into account the containment measures in force and their variations in the respective time-frame. 

{In particular, in Section~\ref{calibration_sept_july} we present the results set at the Italian national level from September 2020 to July 2021, in Section~\ref{subsec:regionalNR} we analyze the same period in Lombardy and Lazio, whilst in Section~\ref{subsec:natGP} we report the results for the whole 
Country up to December 2021, in order to evaluate the impact of the Green Pass measure.}

\subsection{September 2020 - July 2021: Italy}
\label{calibration_sept_july}
{At the beginning of the period of interest} very few NPI restrictions were in force, and in particular schools were open. With the increase in infections, regional restrictions were introduced starting from November, with the classification in 3 colors according to the risk: yellow, orange and red. 

%We set up the scenario taking into account the color changes of the individual regions week by week, determining the corresponding $\beta$ - matrices. Then we defined the function $f_{\beta}$ as in (\ref{eq:f_beta}): our aim was to estimate the function $c(t)$ calibrating on the deceased individuals. The calibration is done using DPC data of total deceased individual from 1 September 2020 to 31 July 2021. We also set up the initial conditions. We chose to start the calibration in September to get an overview of the second and third wave trends.

For the first calibration we set $\theta = 1$ and $\mu_{R}=0$, so we consider that, once the positive diagnosis is received, vaccines are totally ineffective in reducing deaths, and that recovered individuals {become immune}. Note that if we consider $\theta = 1$, the function $f(S,V; \tau)$ defined in (\ref{eq_fifr}) is equal to the \textit{Infectious Fatality Rate}.

\begin{figure}[t!]
\centering
\includegraphics[width=0.4\textwidth]{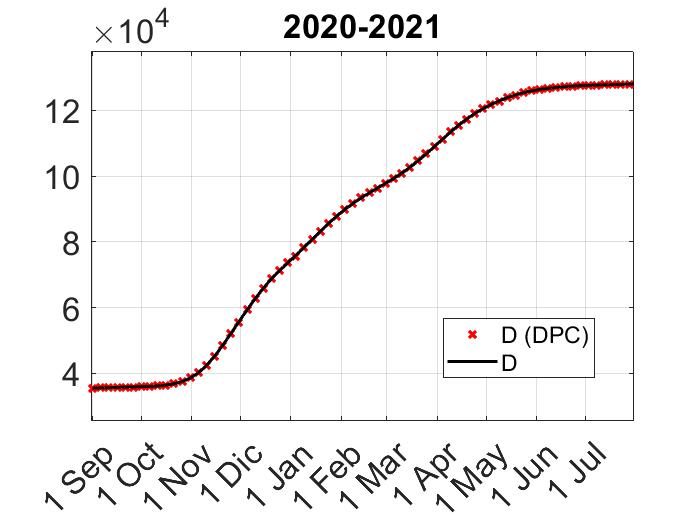}
\includegraphics[width=0.4\textwidth]{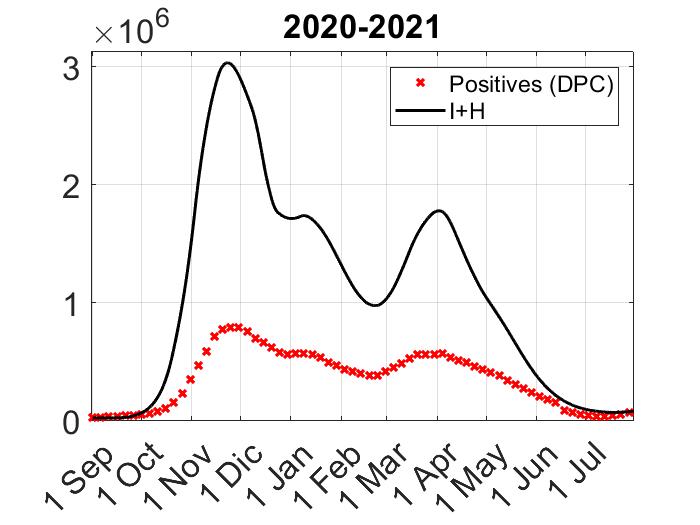}
\caption{Calibration results in Italy: the calibration on reported deceased (left), the sum of Infectious and Healing estimated by the model compared with reported positives (right). }
\label{fig:calibration_results}
\end{figure}

As we can see in Figure \ref{fig:calibration_results}, {the amount of deceased matches with actual monitoring}, showing a reasonable behavior of the curve of positive individuals. Indeed, it is possible to see that $I + H$ curve follows the trend of the positives given by DPC. The gap between the two curves is due to {monitoring uncertainty}: SEIHRDV model does not distinguish detected from undetected individuals. 

\begin{figure}[t!]
\centering
\includegraphics[width=0.24\textwidth]{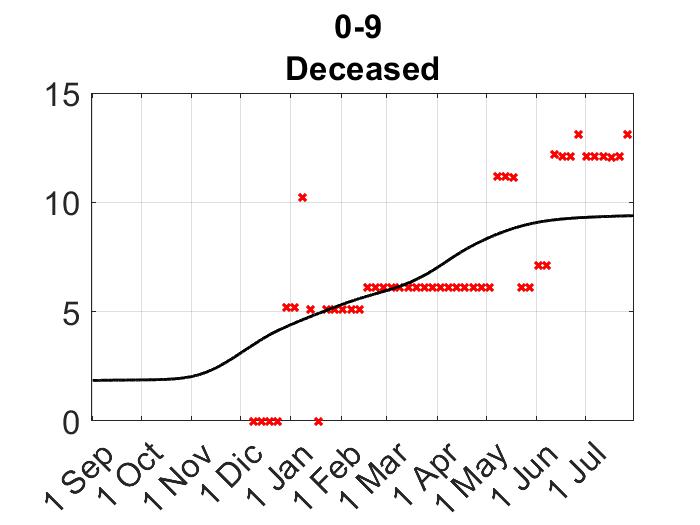}
\includegraphics[width=0.24\textwidth]{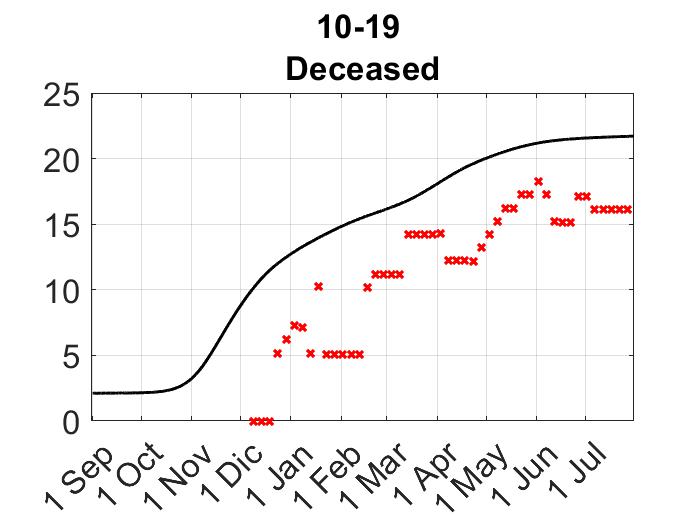}
\includegraphics[width=0.24\textwidth]{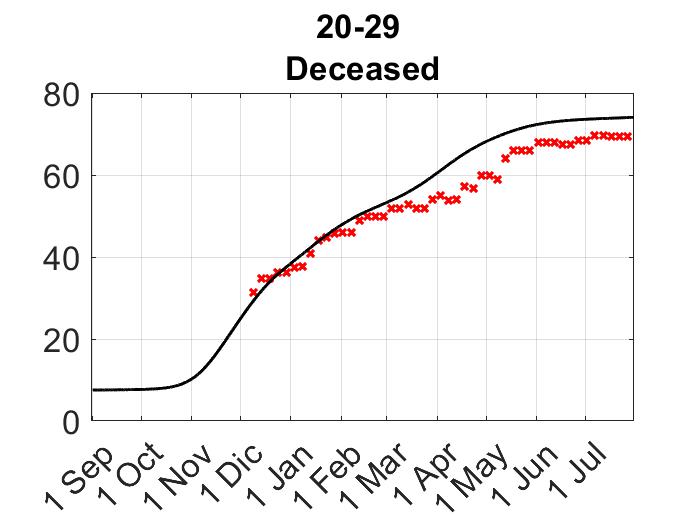}
\includegraphics[width=0.24\textwidth]{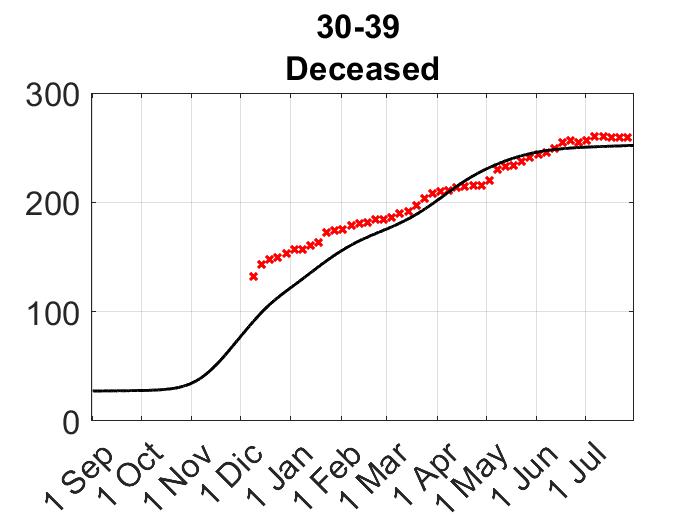}
\includegraphics[width=0.24\textwidth]{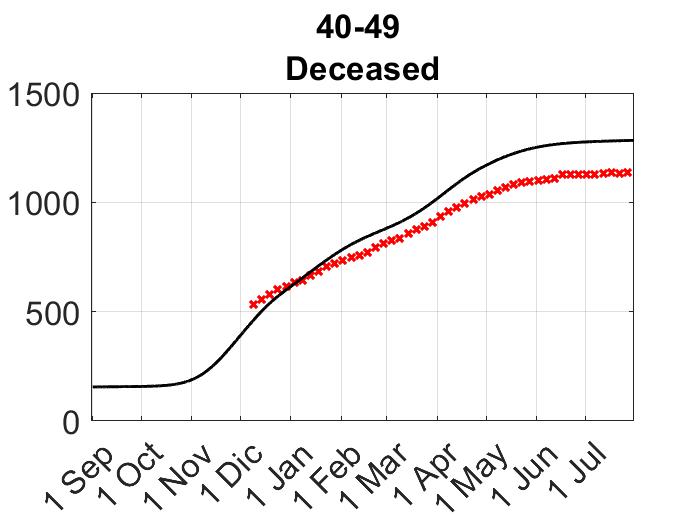}
\includegraphics[width=0.24\textwidth]{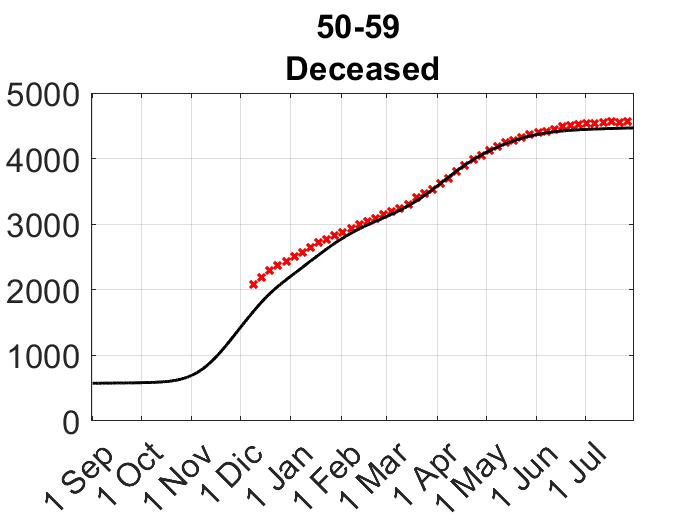}
\includegraphics[width=0.24\textwidth]{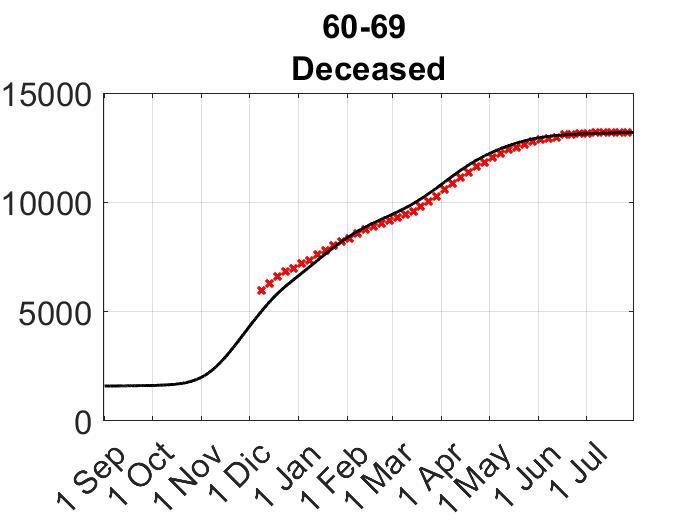}
\includegraphics[width=0.24\textwidth]{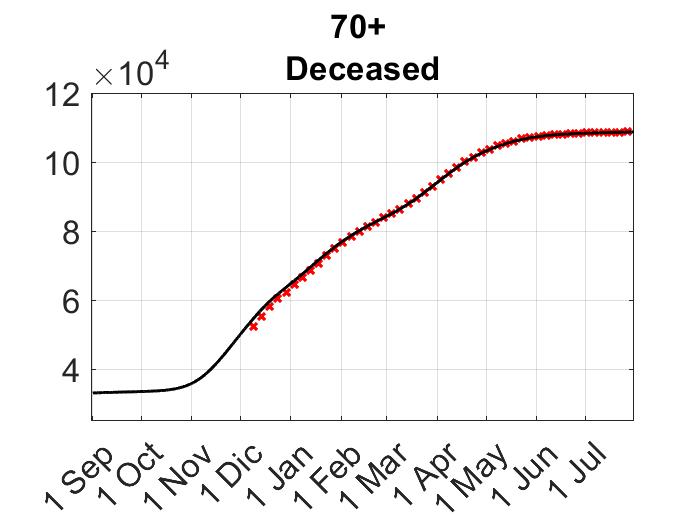}
\caption{Deceased in Italy from September 2020 to July 2021 estimated by SEIHRDV multi-age and multi-context model divided by age group, compared with data (in red).}
\label{fig:deceased_age}
\end{figure}

\begin{figure}[t!]
\centering
\includegraphics[width=0.24\textwidth]{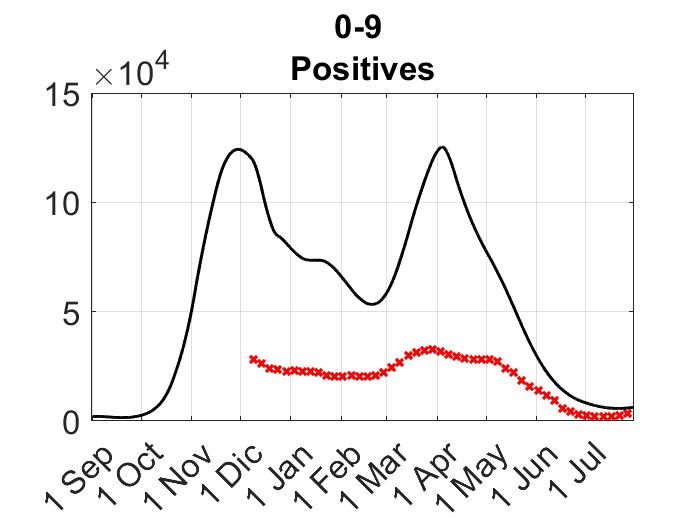}
\includegraphics[width=0.24\textwidth]{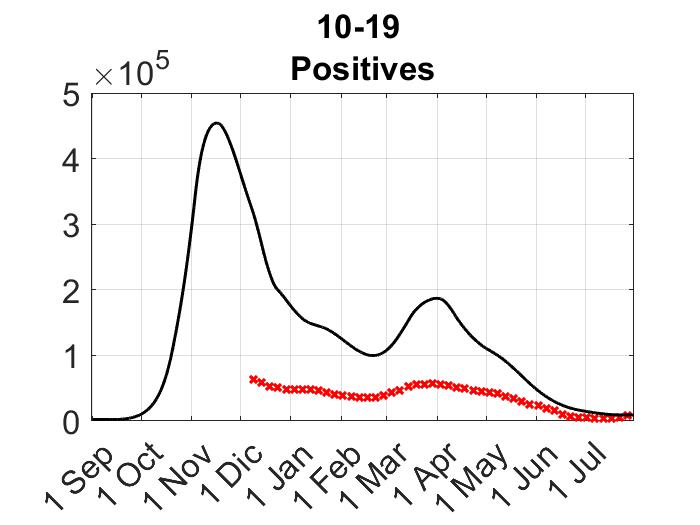}
\includegraphics[width=0.24\textwidth]{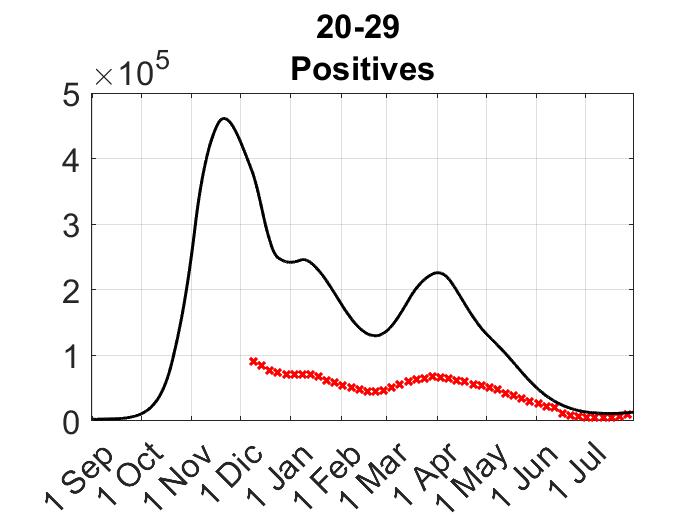}
\includegraphics[width=0.24\textwidth]{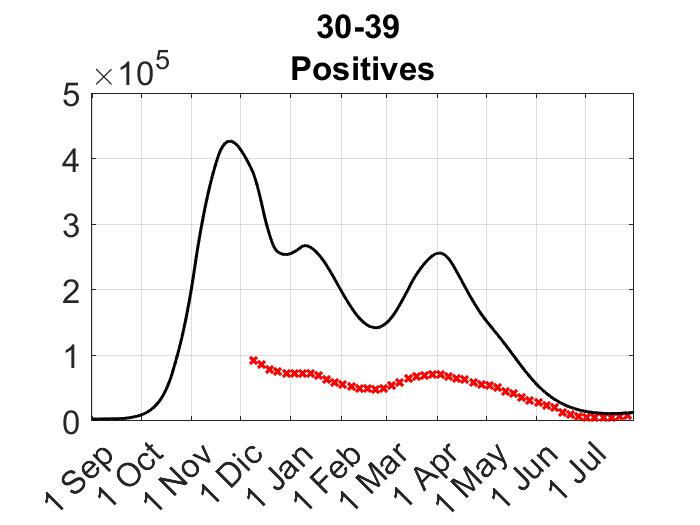}
\includegraphics[width=0.24\textwidth]{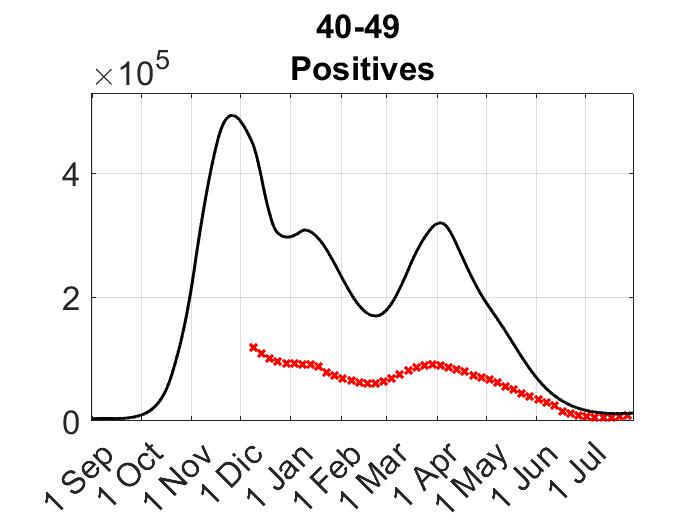}
\includegraphics[width=0.24\textwidth]{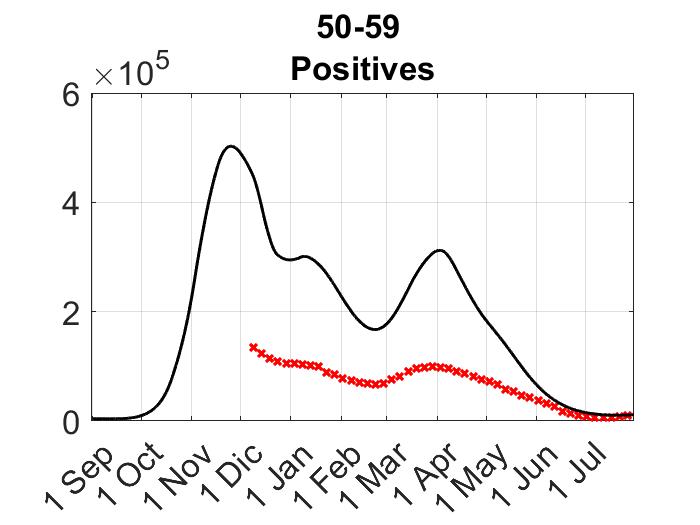}
\includegraphics[width=0.24\textwidth]{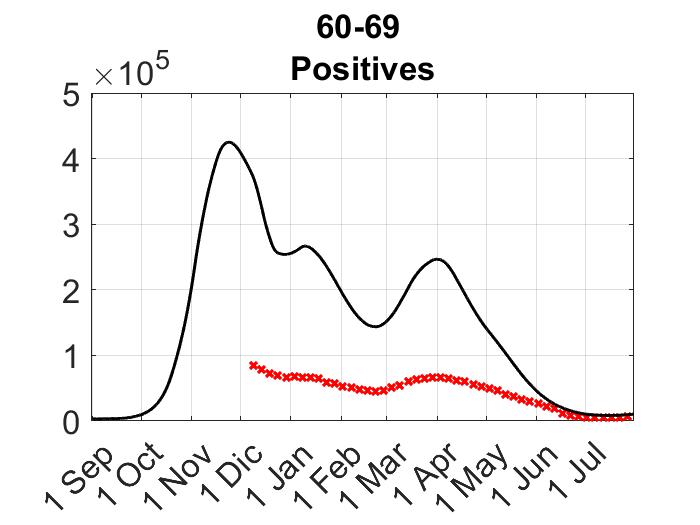}
\includegraphics[width=0.24\textwidth]{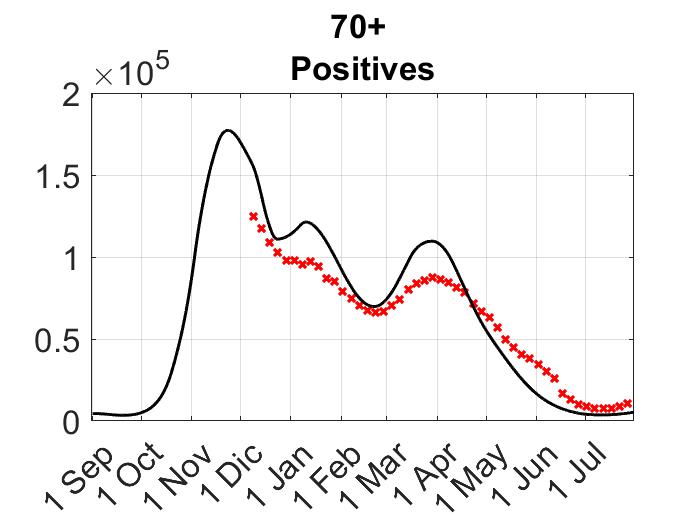}
\caption{Positives ($I+H$) in Italy from September 2020 to July 2021 estimated by SEIHRDV multi-age and multi-context model divided by age group, compared with data (in red).}
\label{fig:positives_age}
\end{figure}

\begin{figure}[t!]
\centering
\includegraphics[width=1\textwidth]{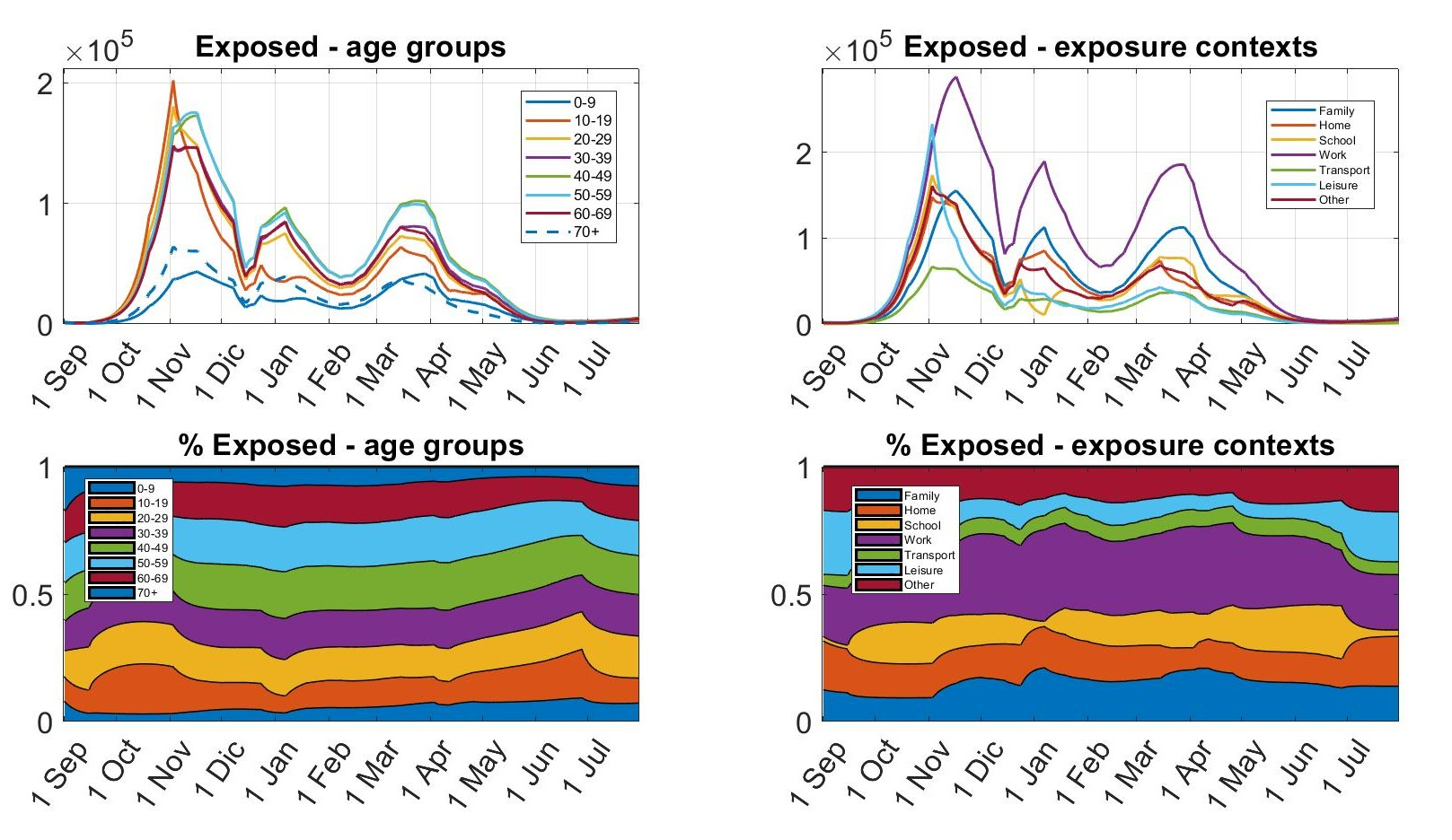}
\caption{Exposed in Italy from September 2020 to July 2021: at the top, evolution of the exposed individuals divided by age group (a) and context of exposition (b); at the bottom, distribution of age groups (c) and contexts of exposition (d) in the exposed compartment.}
\label{fig:exposed}
\end{figure}

{In Figure \ref{fig:deceased_age}} we compare deceased individual divided by age groups{, since we are} able to reconstruct the IFR for each age class. 
{Even in this case, the stratification of deceased matches with available data.}
In Figure \ref{fig:positives_age} the curves of positives individuals (i.e. Infectious and Healing individuals) estimated by SEIHRDV multi-age and multi-context model follow the trend of the data. As in Figure \ref{fig:calibration_results}, the gap between the curves and the data {is motivated by} the {limitations in monitoring}. In particular, in the 70+ age group we observe that estimated curve and data are similar: this is probably due to the fact that older individuals are more often symptomatic than other age groups, and thus they are more easily detected.

{The peculiarity} of the SEIHRDV multi-age and multi-context is that we can observe how exposed individuals are distributed in each age group and context of exposition. In Figure \ref{fig:exposed} {we show the age-repartition (a) and contexts of exposition (b) of exposed individuals}, while at the bottom we have the distribution of age groups (c) and context of exposition (d) in the exposed compartment.
{In this way, we can interpret the numerical results in light of the specific NPIs placed in act and evaluate the impact of specific interventions.
For instance,} in the middle of September 2020 schools were totally open. This led to an increase {of cases} in the younger age groups until the beginning of November, when some restrictions where applied with the introduction of colored zones. Since then, the most exposed age groups were 40-49 and 50-59. {For what concerns the }contexts of exposition we can see that, in the first part of the growth of the curve, \virgolette{leisure} was the context with the greatest exposures. Then, with the introduction of the containment measures of November, the context in which most infections occur was the \virgolette{work} sector. In the same way, the {infections in the} \virgolette{school} context, which was growing, began to decline and then remained more or less constant (with an exception for Christmas holiday). School-related restrictions particularly affected part of middle school students and high school students. 

\begin{figure}[t!]
\centering
\includegraphics[width=0.6\textwidth]{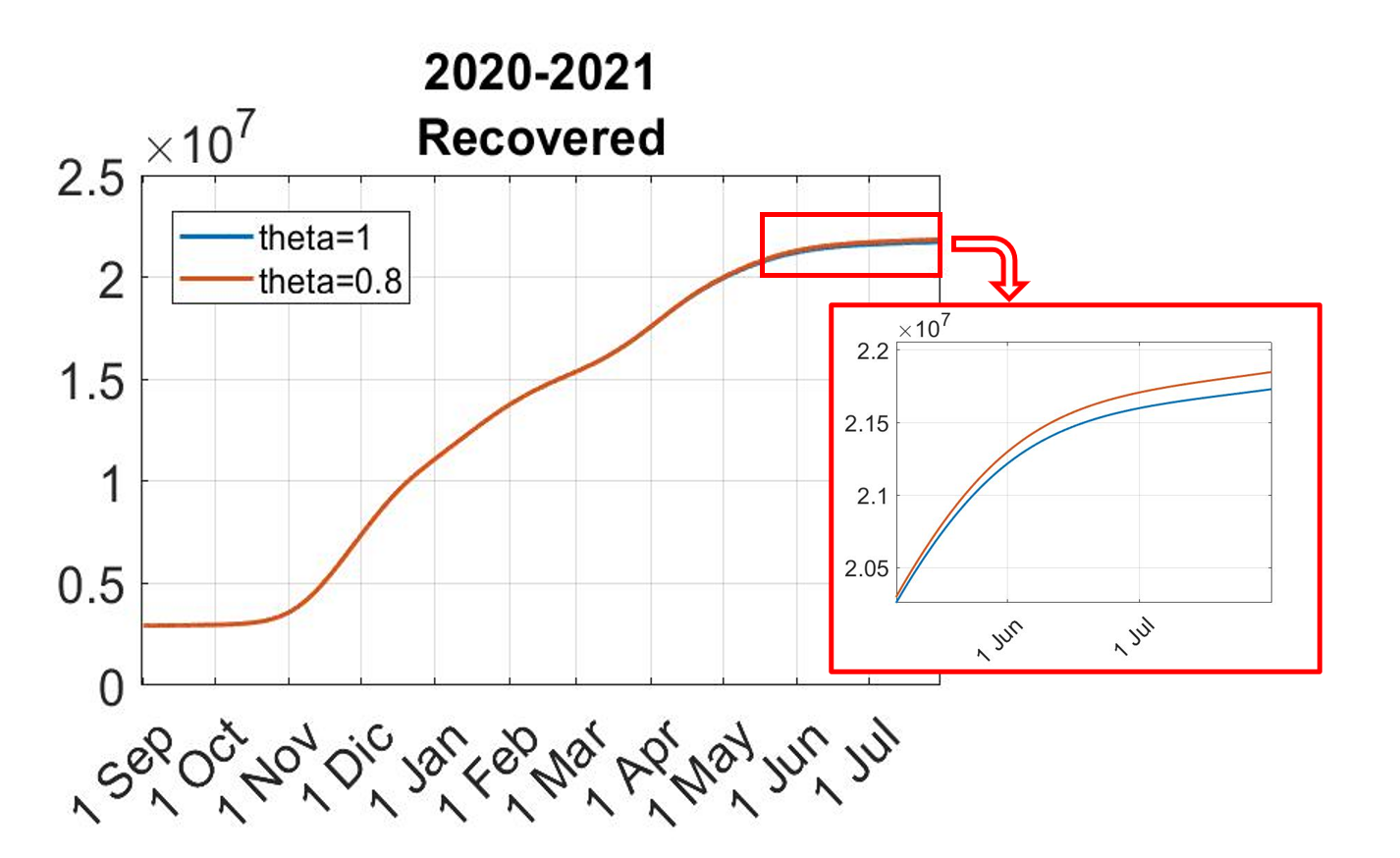}
\caption{Curves of Recovered with $\theta = 1$ and $\theta = 0.8 $ compared in the whole time period. }
\label{fig:recovered_comparison}
\end{figure}

{For the sake of completeness, we repeated the numerical tests by} imposing $\theta = 0.8$. We estimated this value of $\theta$, that is the rate of inefficacy of vaccination in the reduction of death, from ISS data reported in \cite{II}. We obtained similar results. In particular we can see a difference in the Recovered compartment: if individuals die fewer, we will have more recovered (Figure \ref{fig:recovered_comparison}). 

%\begin{figure}[]
%\centering
%\includegraphics[width=0.4\textwidth]{fig/Italy/confronto theta/confronto_guariti.jpg}
%\includegraphics[width=0.4\textwidth]{fig/Italy/confronto theta/confronto_guariti_zoom.jpg}
%\caption{Recovered: curves of Recovered with $theta = 1$ and $theta = 0.8 $ compared in the whole time period at the left, zoom on the last period at the right. }
%\label{fig:exposed}
%\end{figure}

% \newpage
\subsection{September 2020 - July 2021: regions Lombardy and Lazio}
\label{subsec:regionalNR}
The trend of the national epidemic curve is influenced by the trends of the individual regions. In fact, the situation may be different in each region, depending on average age of individuals, daily habits, hospital situation, climatic situation. For this reason, the NPI in the chosen time period are adopted on a regional scale.
Hence, we analyzed what happened in two of the most populous Italian region: Lombardy and Lazio. 
{As we did in Section \ref{calibration_sept_july}}, we set up the scenario, the initial conditions by imposing $\theta = 1$ and $\mu_{R}=0$.

\subsubsection{Lombardy}
\begin{figure}[t!]
\centering
\includegraphics[width=0.4\textwidth]{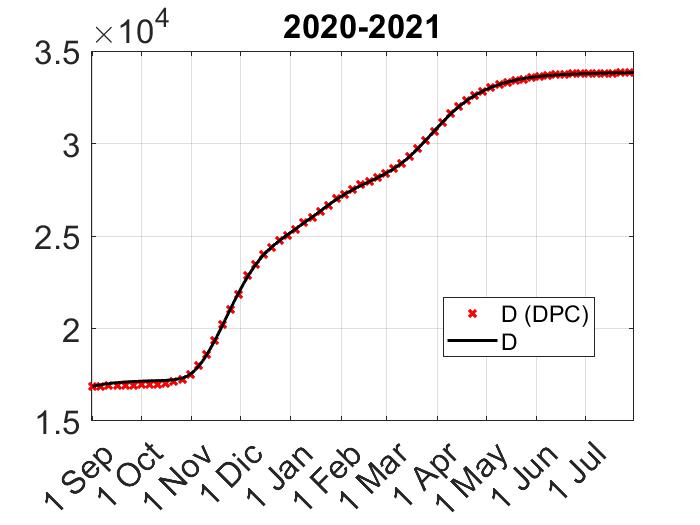}
\includegraphics[width=0.4\textwidth]{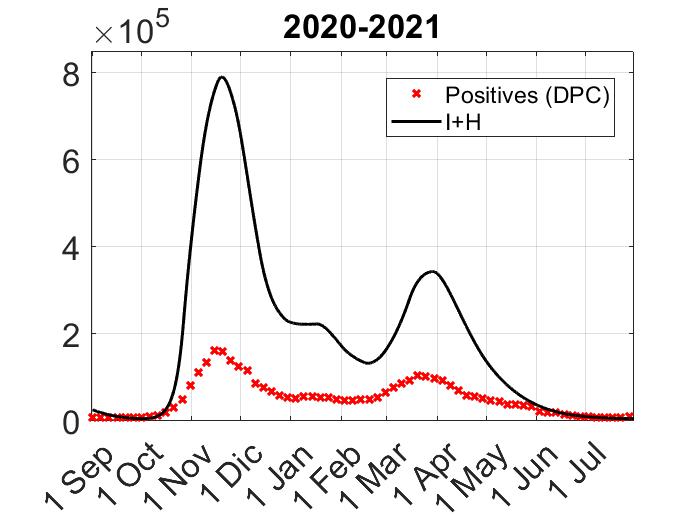}
\caption{Calibration results in Lombardy: the calibration on reported deceased at the left, the sum of Infectious and Healing estimated by the model compared with reported positives at the right. }
\label{fig:calibration_results_lombardy}
\end{figure}

\begin{figure}[h]
\centering
\includegraphics[width=0.24\textwidth]{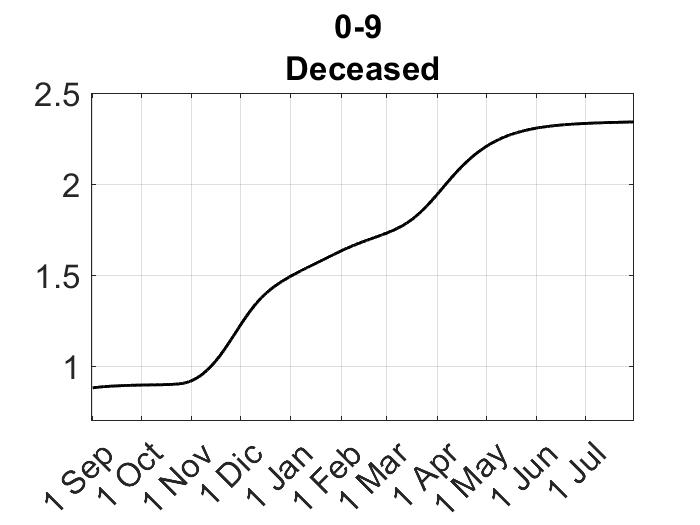}
\includegraphics[width=0.24\textwidth]{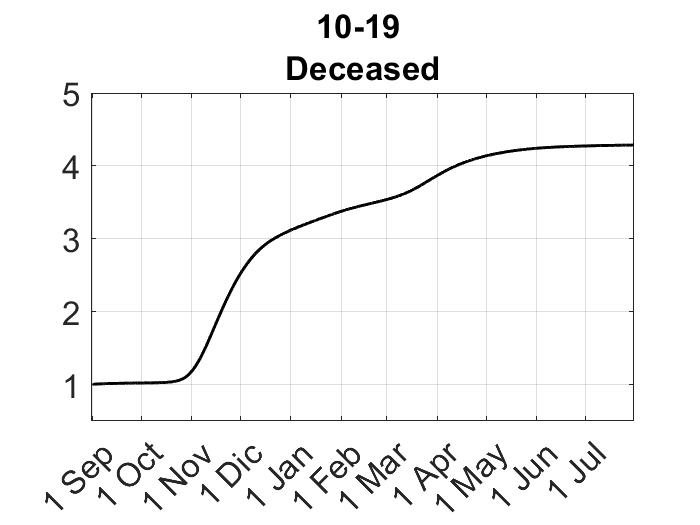}
\includegraphics[width=0.24\textwidth]{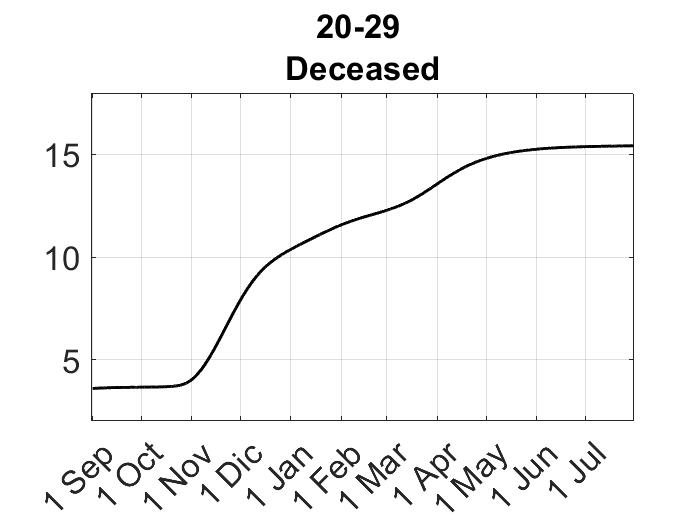}
\includegraphics[width=0.24\textwidth]{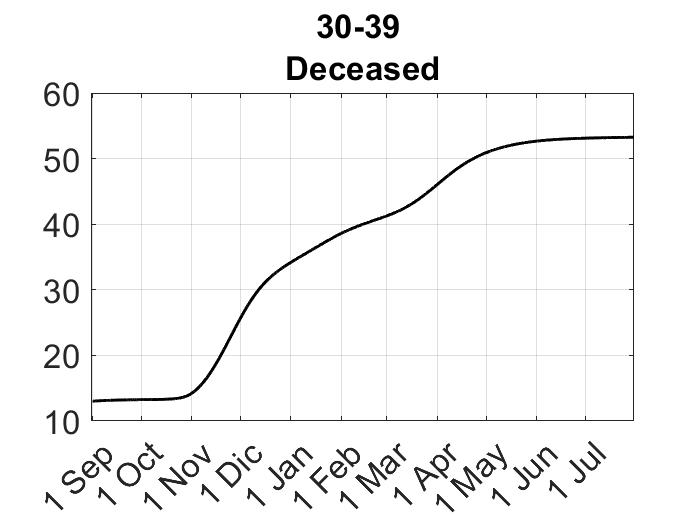}
\includegraphics[width=0.24\textwidth]{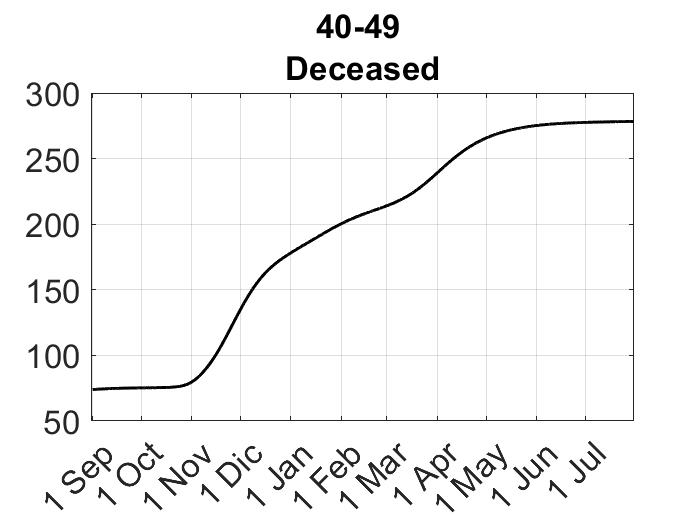}
\includegraphics[width=0.24\textwidth]{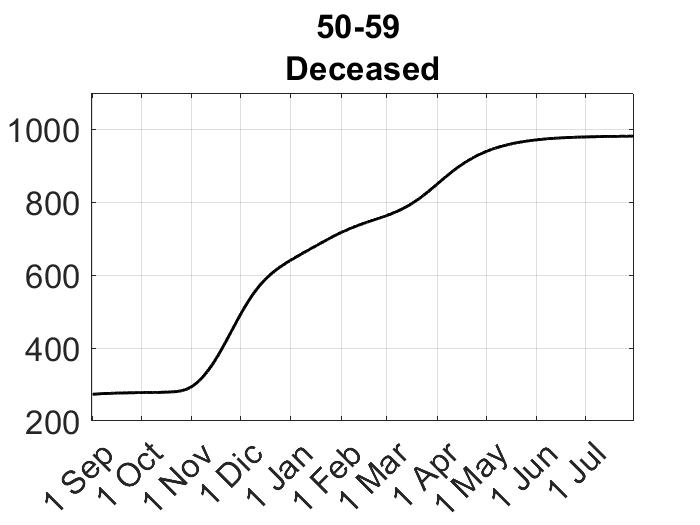}
\includegraphics[width=0.24\textwidth]{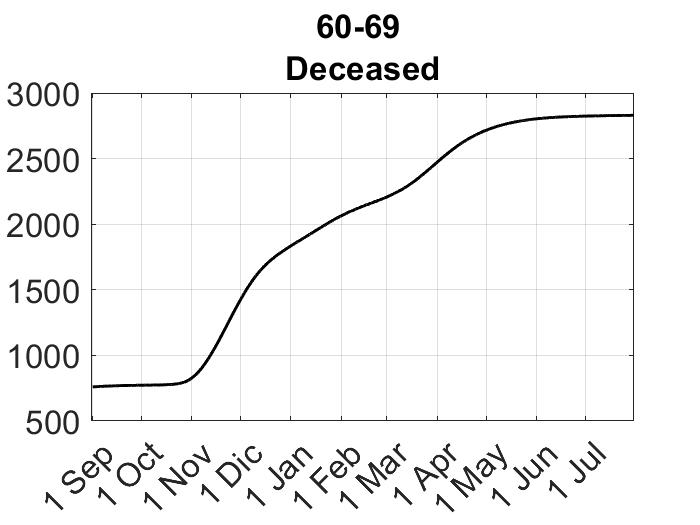}
\includegraphics[width=0.24\textwidth]{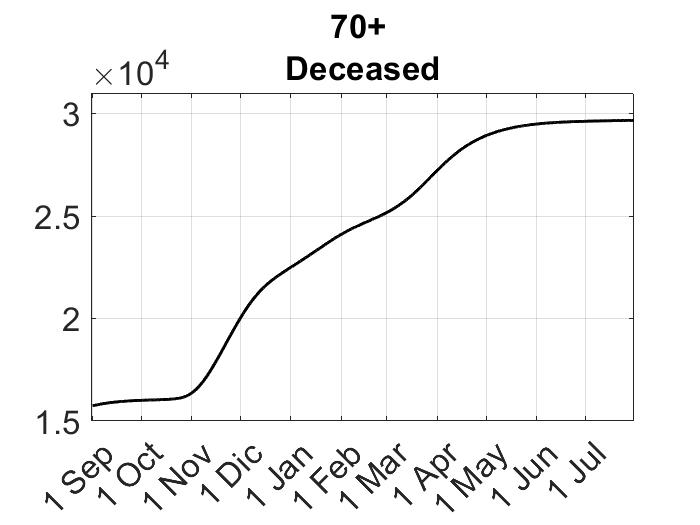} 

\caption{Deceased in Lombardy estimated by SEIHRDV multi-age/multi-context model divided by age group.}
\label{fig:deceased_age_lombardy}
\end{figure}

\begin{figure}[H]
\centering
\includegraphics[width=0.24\textwidth]{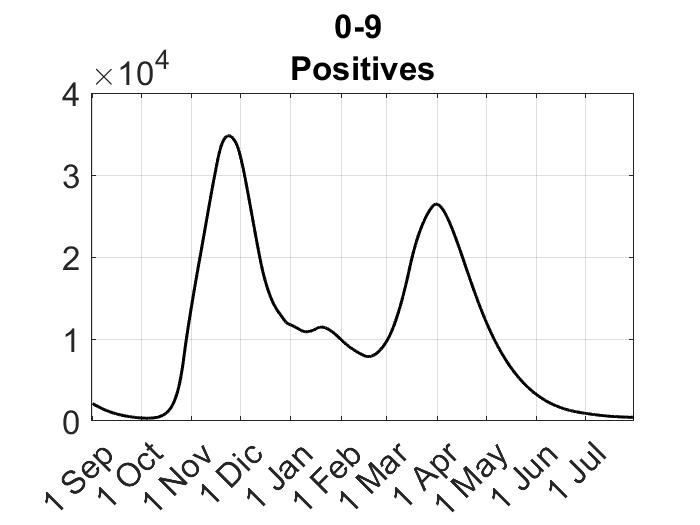}
\includegraphics[width=0.24\textwidth]{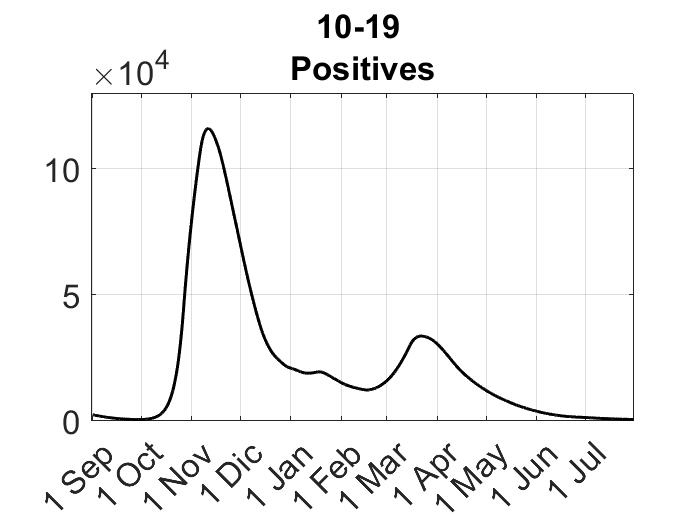}
\includegraphics[width=0.24\textwidth]{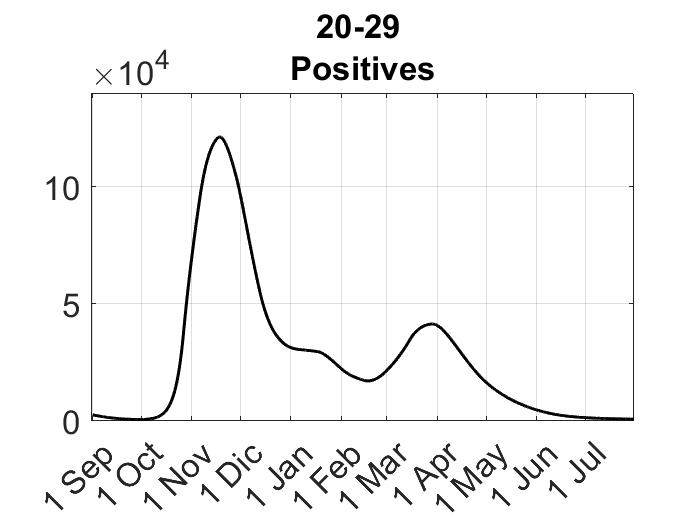}
\includegraphics[width=0.24\textwidth]{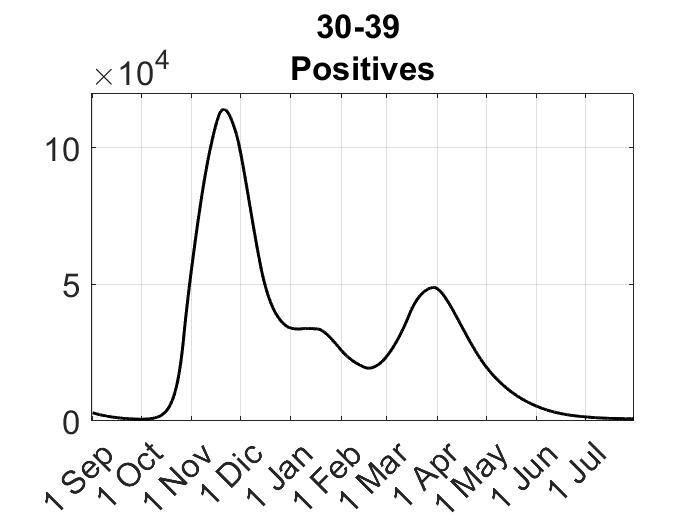}
\includegraphics[width=0.24\textwidth]{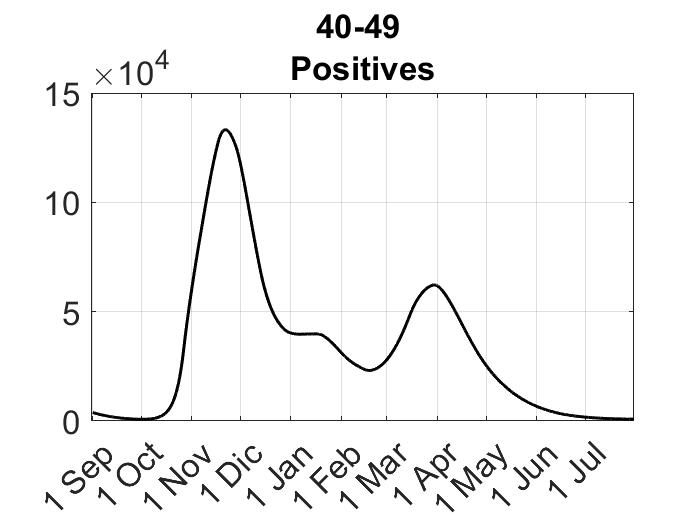}
\includegraphics[width=0.24\textwidth]{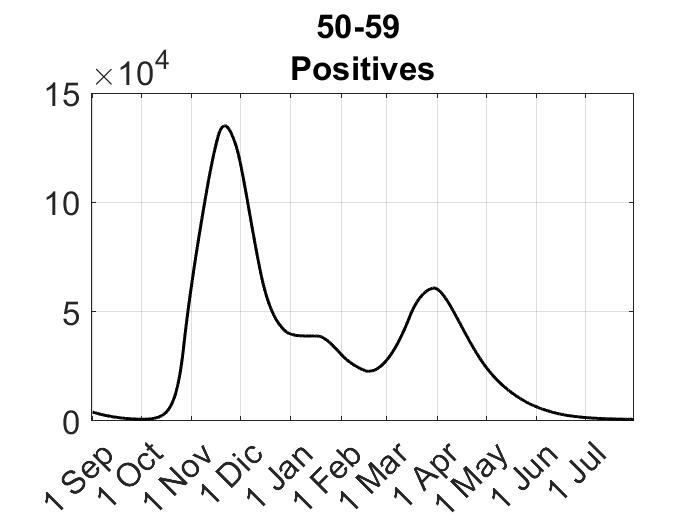}
\includegraphics[width=0.24\textwidth]{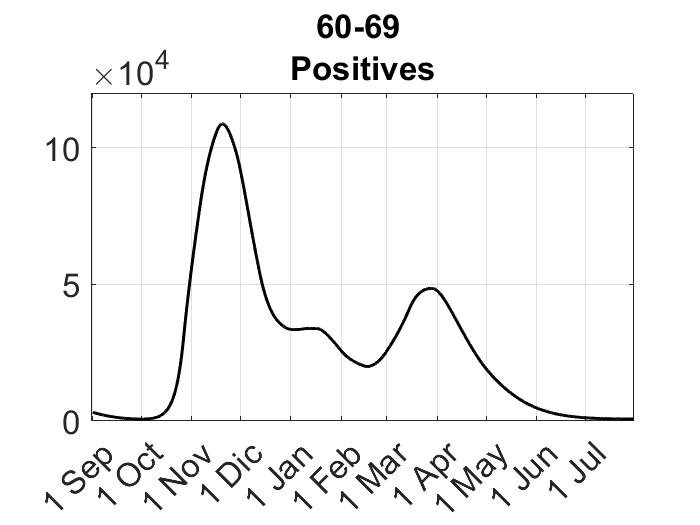}
\includegraphics[width=0.24\textwidth]{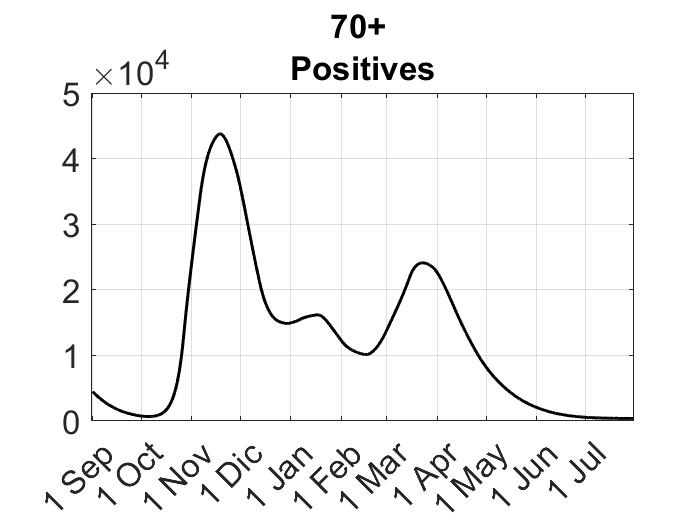}
\caption{Positives ($I+H$) in Lombardy estimated by SEIHRDV multi-age/multi-context model divided by age group.}
\label{fig:positives_age_lombardy}
\end{figure}

As we can see in Figure \ref{fig:calibration_results_lombardy}, {the amount of deceased individuals matches with} actual data, showing also in Lombardy a reasonable behavior of positive individuals. The $I + H$ curve follows the trend of the positives given by DPC. Even in this case, the gap between the two curves is due  to the accuracy of the detection.
{In Figures \ref{fig:deceased_age_lombardy} and \ref{fig:positives_age_lombardy} we report deceased individuals and positives individuals divided by age groups}. 

%\vspace{-0.5cm}
%\vspace{-0.4cm}
\vspace{-0.8cm}

\begin{figure}[h]
\centering
\includegraphics[width=1\textwidth]{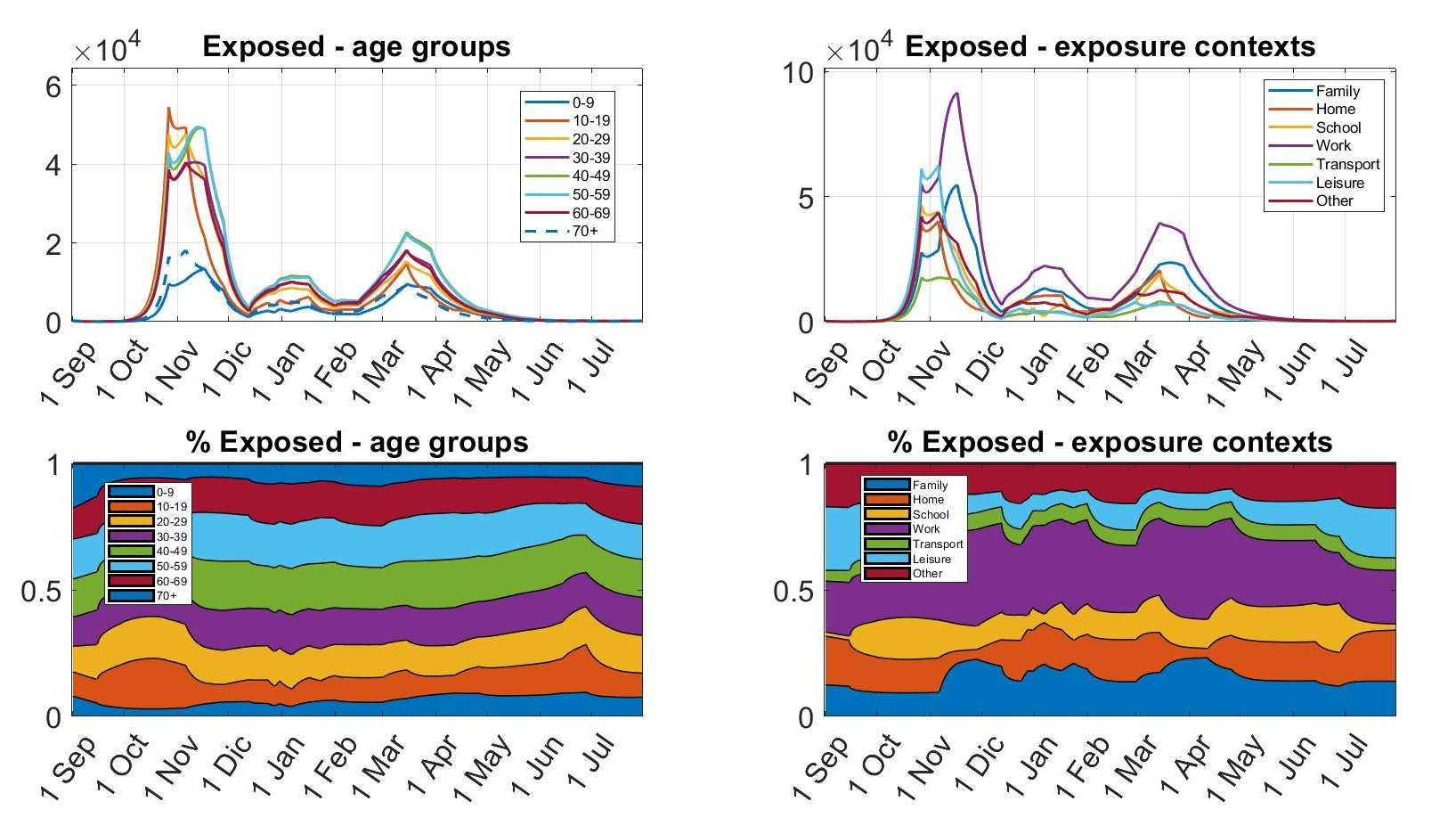}
\caption{Exposed in Lombardy: at the top, evolution of the exposed individuals divided by age group (a) and context of exposition (b); at the bottom, distribution of age groups (c) and contexts of exposition (d) in the exposed compartment. }
\label{fig:exposed_lombardy}
\end{figure}

With SEIHRDV multi-age and multi-context {we are able to track the evolution of infections in} each age group and each context of exposition. {In Figure \ref{fig:exposed_lombardy} we report the amount of exposed divided by age groups and context of exposition, while at the bottom we can see how the exposed (and therefore infections) are divided into age groups and contexts of exposition}.
{As we expected}, the curves in Figure \ref{fig:exposed_lombardy} {follow different trends with respect to} the national curve of the exposed individuals (Figure \ref{fig:exposed}).  

\subsubsection{Lazio}  %modifica da qui e carica immagini
\begin{figure}[t!]
\centering
\includegraphics[width=0.4\textwidth]{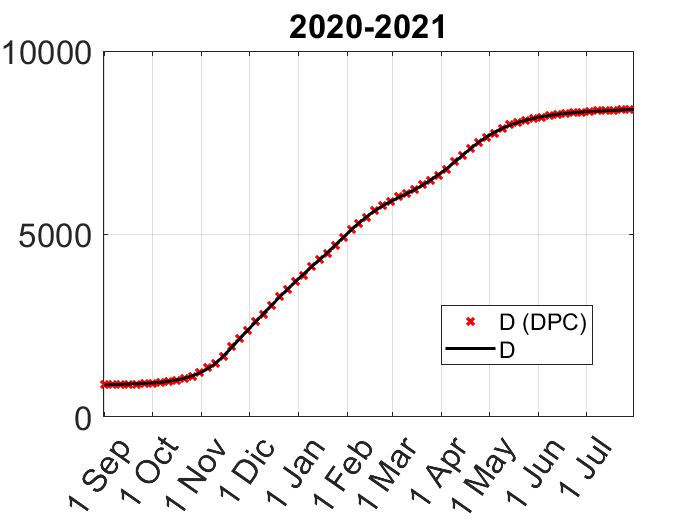}
\includegraphics[width=0.4\textwidth]{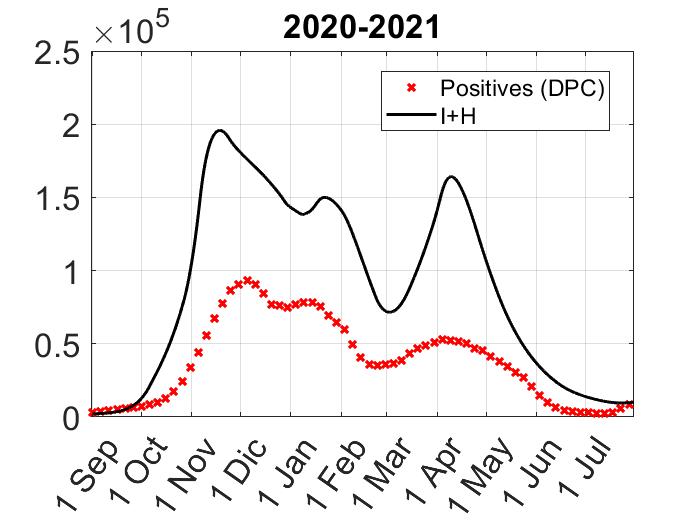}
\caption{Calibration results in Lazio: the calibration on reported deceased at the left, the sum of Infectious and Healing estimated by the model compared with reported positives at the right. }
\label{fig:calibration_results_lazio}
\end{figure}

\begin{figure}[t!]
\centering
\includegraphics[width=0.24\textwidth]{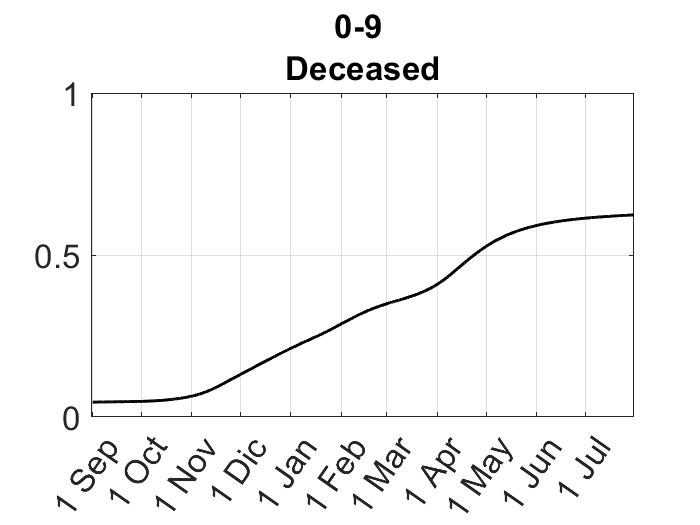}
\includegraphics[width=0.24\textwidth]{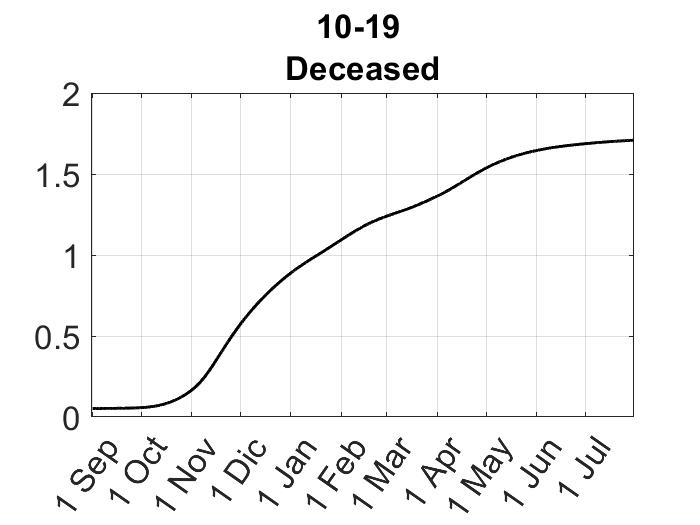}
\includegraphics[width=0.24\textwidth]{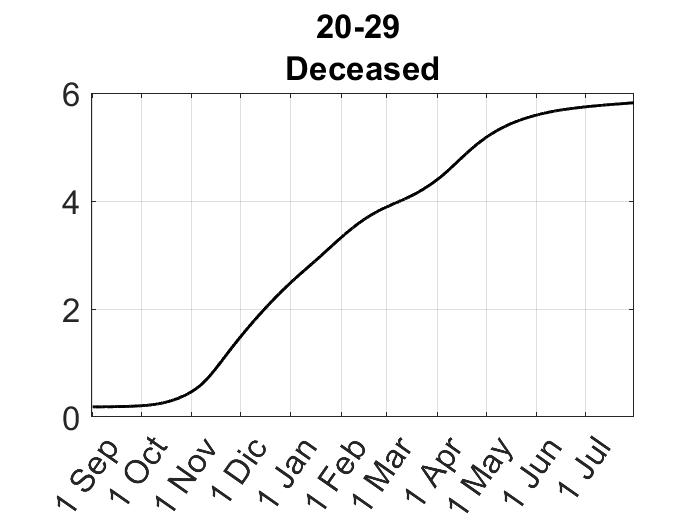}
\includegraphics[width=0.24\textwidth]{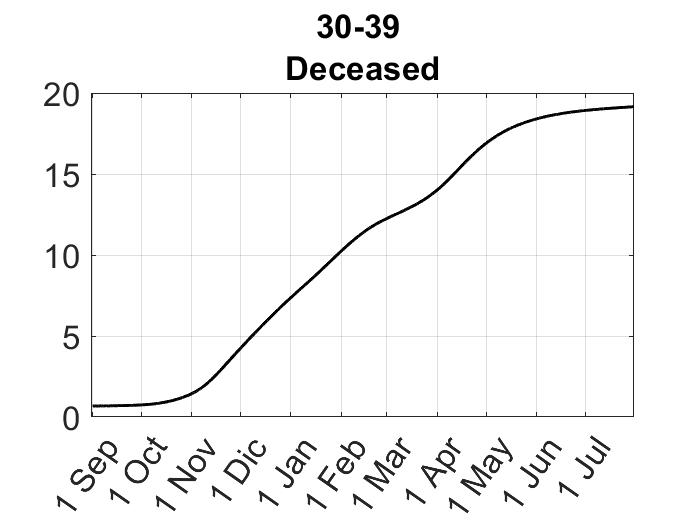}
\includegraphics[width=0.24\textwidth]{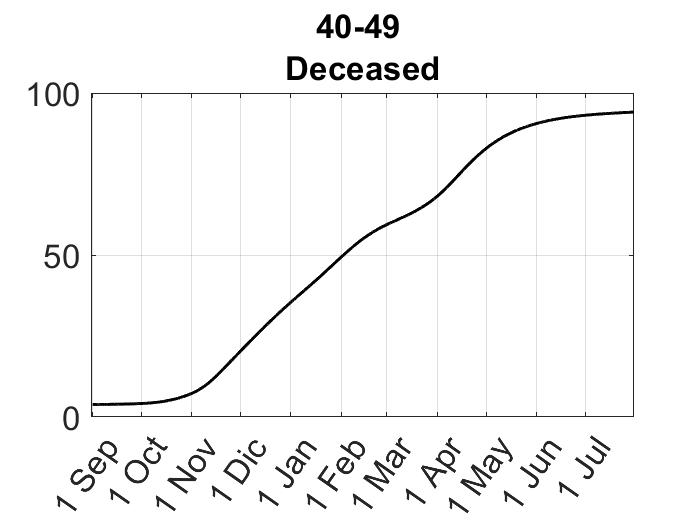}
\includegraphics[width=0.24\textwidth]{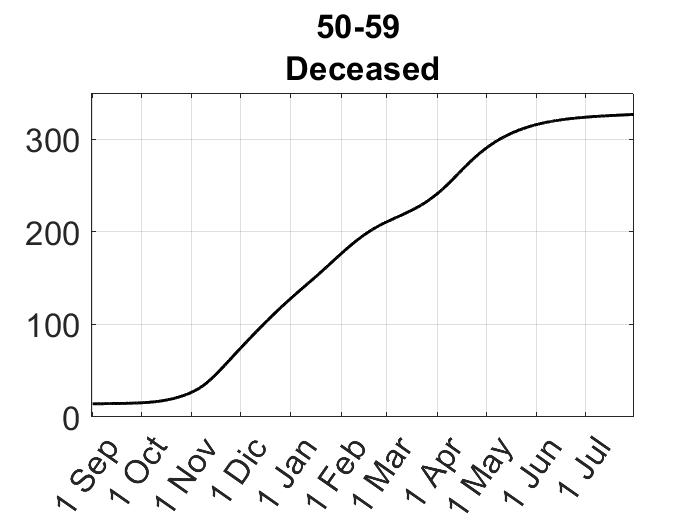}
\includegraphics[width=0.24\textwidth]{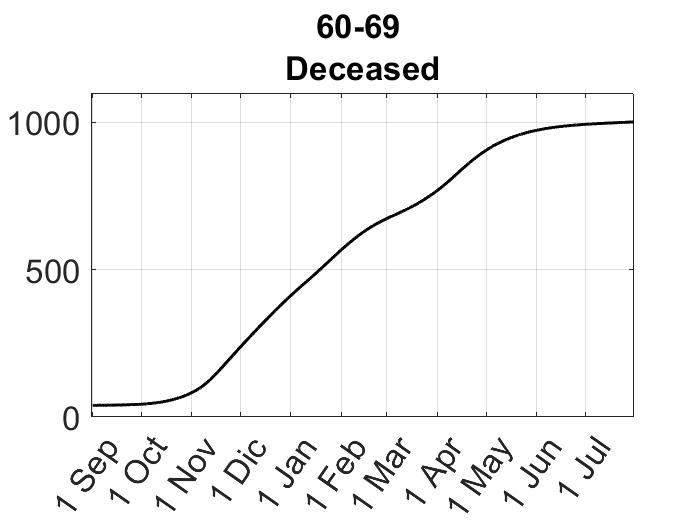}
\includegraphics[width=0.24\textwidth]{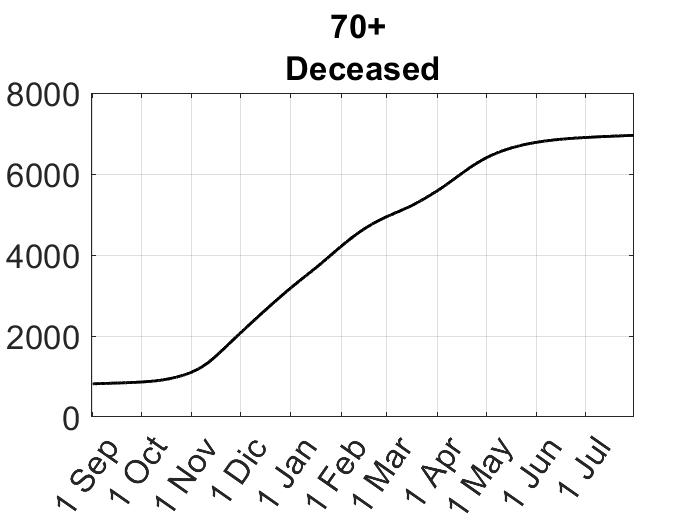} 
\caption{Deceased in Lazio estimated by SEIHRDV multi-age/multi-context model divided by age group.}
\label{fig:deceased_age_lazio}
\end{figure}

\begin{figure}[t!]
\centering
\includegraphics[width=0.24\textwidth]{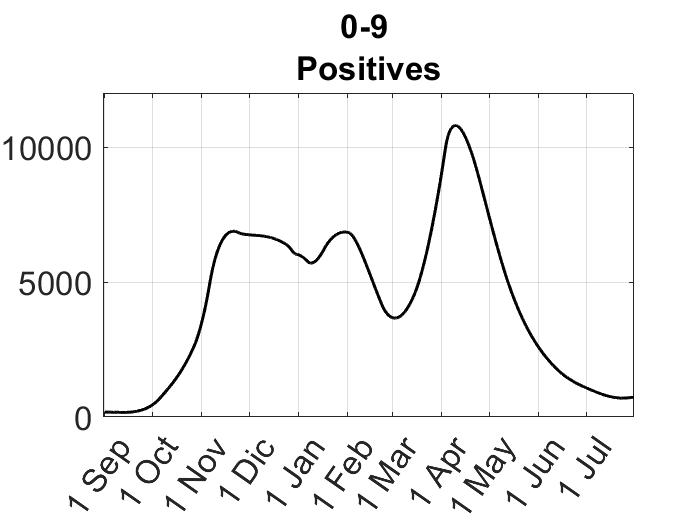}
\includegraphics[width=0.24\textwidth]{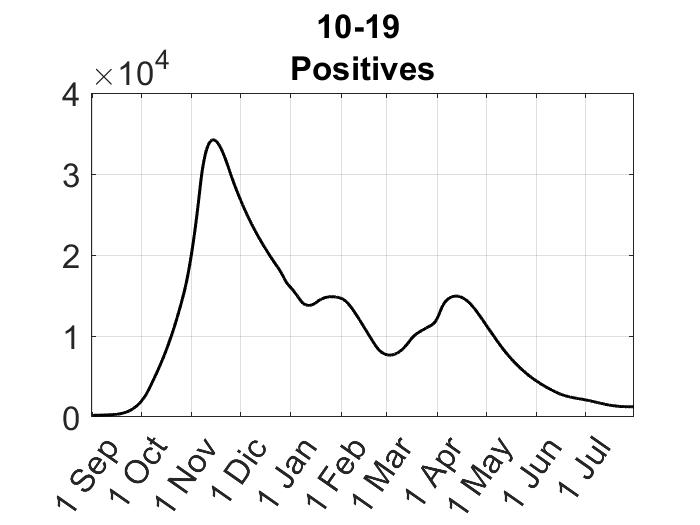}
\includegraphics[width=0.24\textwidth]{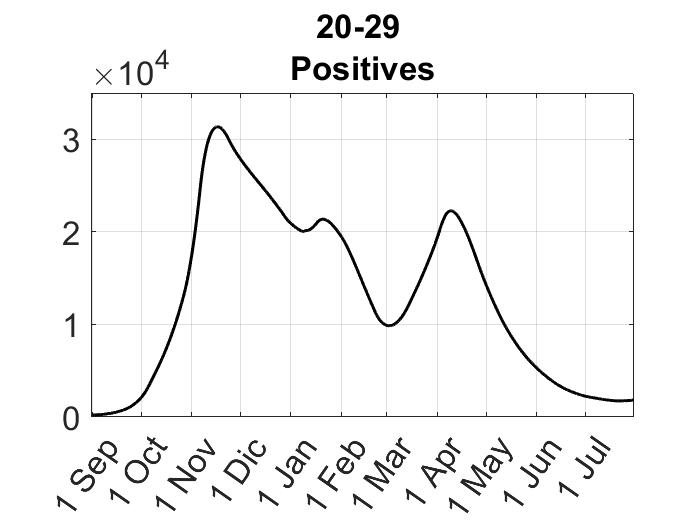}
\includegraphics[width=0.24\textwidth]{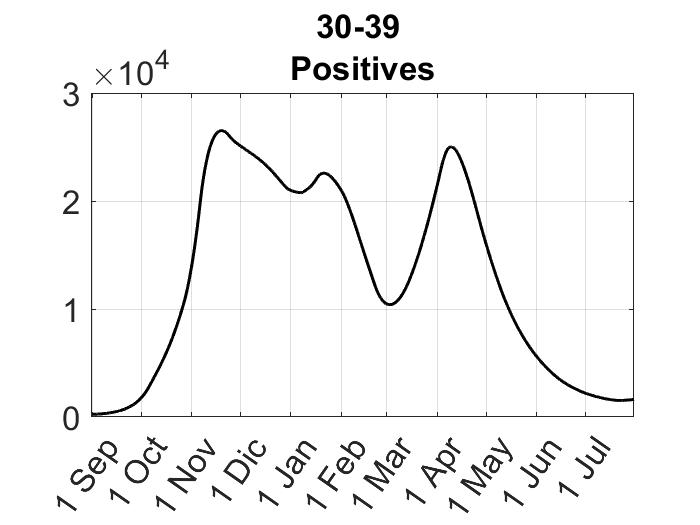}
\includegraphics[width=0.24\textwidth]{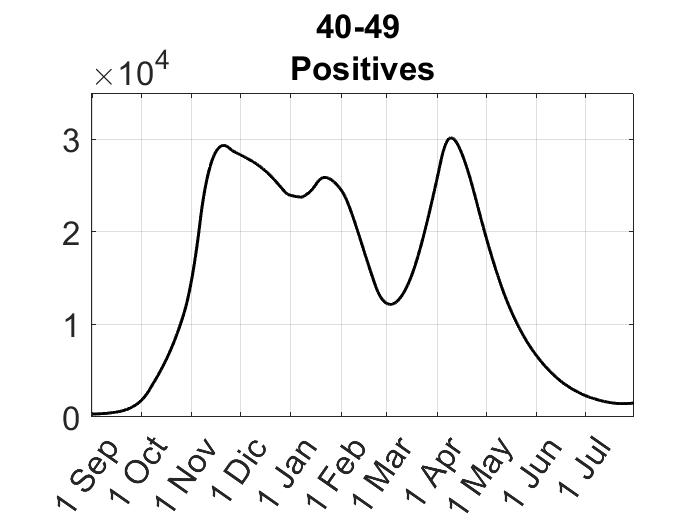}
\includegraphics[width=0.24\textwidth]{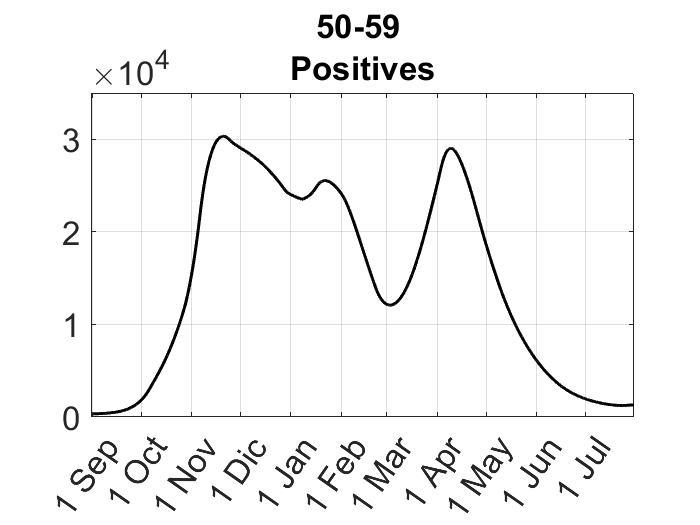}
\includegraphics[width=0.24\textwidth]{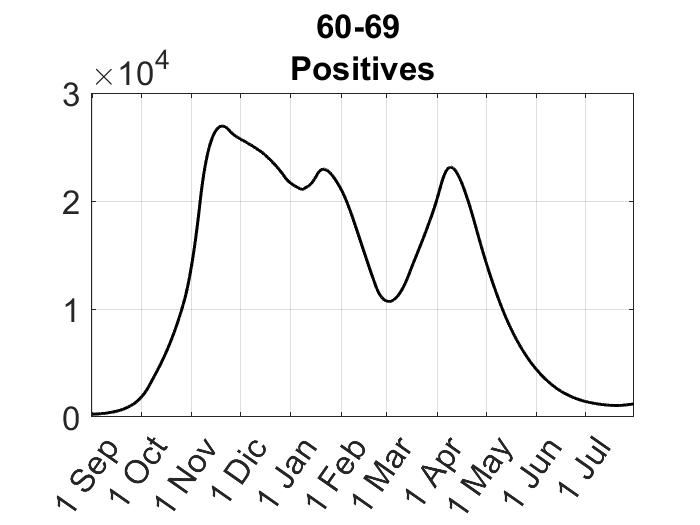}
\includegraphics[width=0.24\textwidth]{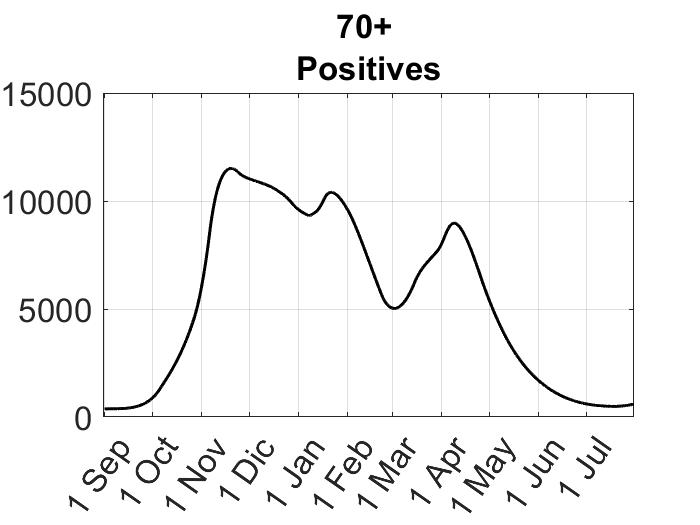}
\caption{Positives ($I+H$) in Lazio estimated by SEIHRDV multi-age/multi-context model divided by age group.}
\label{fig:positives_age_lazio}
\end{figure}

In Figure \ref{fig:calibration_results_lazio} {we observe that even in this case, deceased individuals match the actual data}. The curve of positives in the same figure follows the trend of the positives given by {DPC, with the usual gap ascribed to monitoring uncertainty.}
{In Figure \ref{fig:deceased_age_lazio} and \ref{fig:positives_age_lazio} we report deceased individuals and positives individuals divided by age groups.} 

%\vspace{-0.8cm}

\begin{figure}[t!]
\centering
\includegraphics[width=1\textwidth]{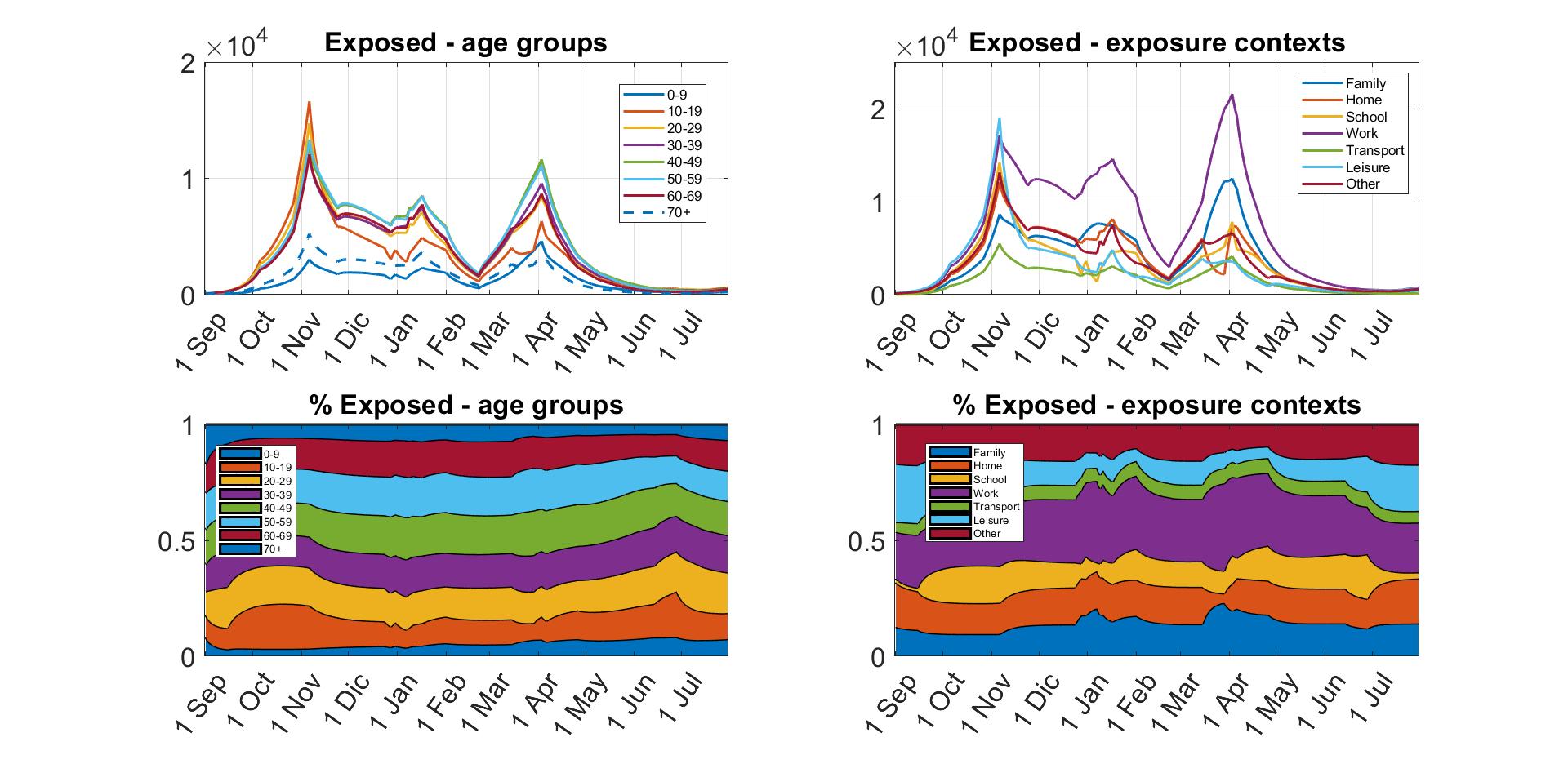}
\caption{Exposed in Lazio: at the top, evolution of the exposed individuals divided by age group (a) and context of exposition (b); at the bottom, distribution of age groups (c) and contexts of exposition (d) in the exposed compartment. }
\label{fig:exposed_lazio}
\end{figure}

In Figure \ref{fig:exposed_lazio} {the amount of exposed is represented divided by age groups and context of exposition, together with} how age groups and contexts of exposition are distributed in the exposed compartment.
The curves in Figure \ref{fig:exposed_lazio} are different both from the national curves of exposed (Figure \ref{fig:exposed}) and the Lombardy curves (Figure \ref{fig:exposed_lombardy}).

\subsection{September 2020 - December 2021: Italy}
\label{subsec:natGP}
{In this section, we take into account for the Green Pass restrictions as introduced in Section \ref{greenpass}.}
To set the scenarios, until the end of July we used the same setting {imposed in Section \ref{calibration_sept_july}}, while from August onwards we took into account the new rules relating to the Green Pass described below:
\begin{itemize}
    \item From August 6 to September 12: Green Pass requirement for indoor restaurants, indoor pubs, discos, indoor sports and various events (concerts, museums, cinemas, stadium, shows, theaters);
    \item From September 13 to October 14: Green Pass requirement also extended to school staff and university students;
    \item From October 15 onward: Green Pass requirement also extended to workers.
\end{itemize}
{For the sake of} simplicity we considered as Green Pass holders only vaccinated individuals{, since our model does not  explicitly monitor swabs}. We {consider} the same initial condition of the first calibration seen in (\ref{calibration_sept_july}), using a non-zero rate of end of immunity for recovered individual since we are considering a time period of over a year. 

\begin{figure}[t!]
\centering
\includegraphics[width=0.4\textwidth]{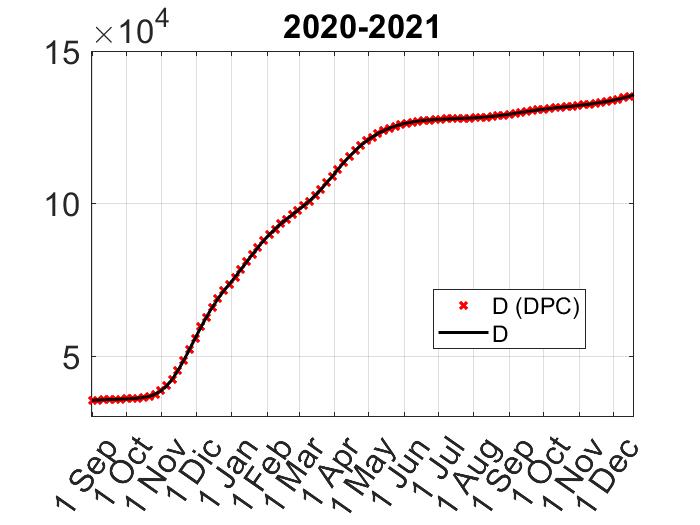}
\includegraphics[width=0.4\textwidth]{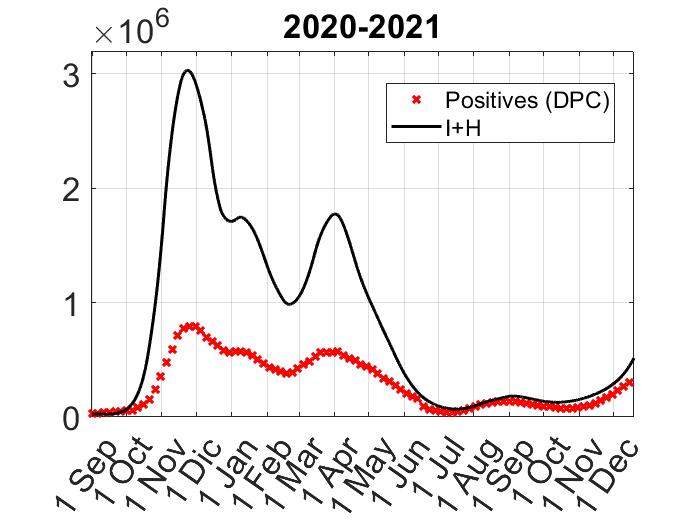}
\caption{Calibration results in Italy: the calibration on reported deceased at the left, the sum of Infectious and Healing estimated by the model compared with reported positives at the right, introducing GP containment measures. }
\label{fig:calibration_results_Italy_GP}
\end{figure}

In Figure \ref{fig:calibration_results_Italy_GP} {we show the trend of positive individuals in the GP case, and observe that it matches available data}. {Since the gap between the curve and the data is smaller from July, we can deduce that a better monitoring policy is induced by the introduction of Green Pass obligation} (notice that the Green Pass can also be granted if swabs are negative). The curve catches also the rise in cases that began in mid-October.

\begin{figure}[t!]
\centering
\includegraphics[width=0.24\textwidth]{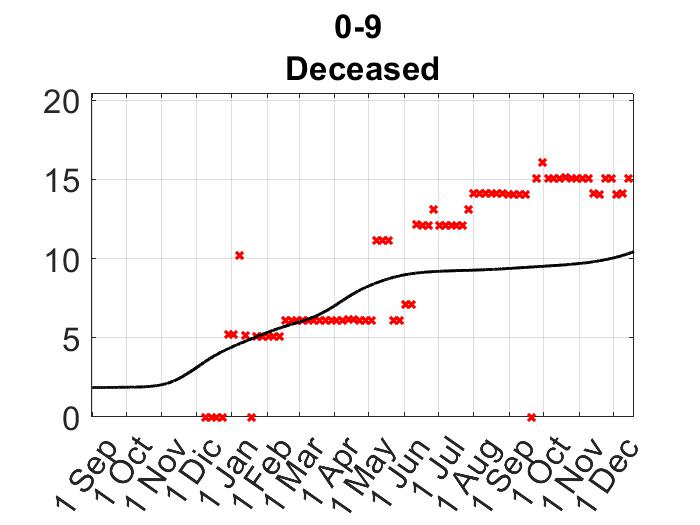}
\includegraphics[width=0.24\textwidth]{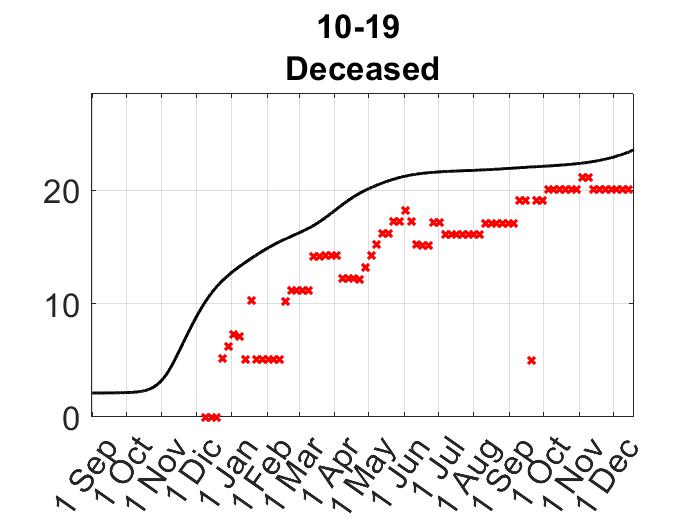}
\includegraphics[width=0.24\textwidth]{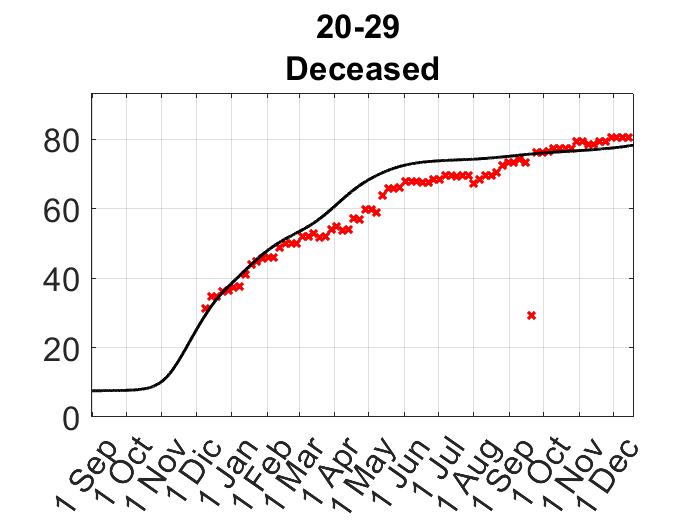}
\includegraphics[width=0.24\textwidth]{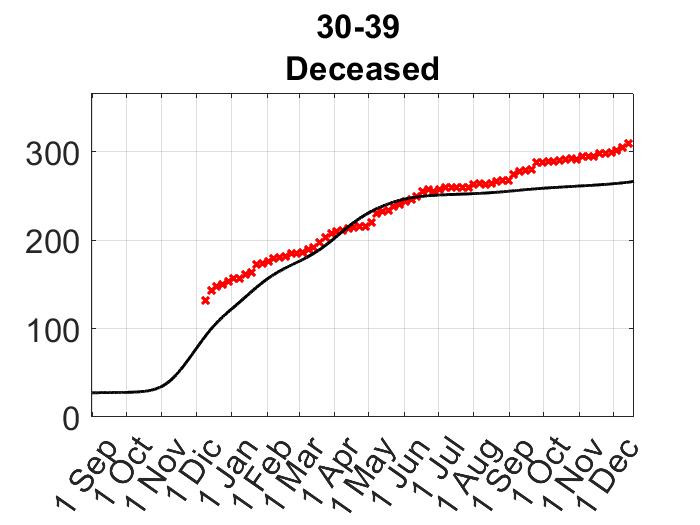}
\includegraphics[width=0.24\textwidth]{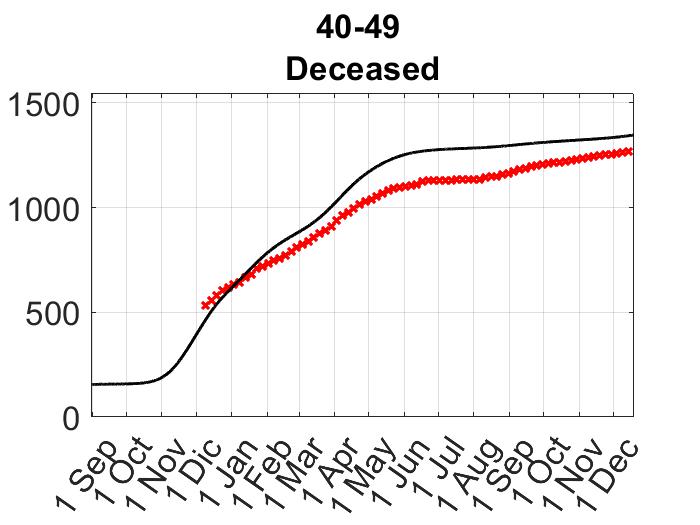}
\includegraphics[width=0.24\textwidth]{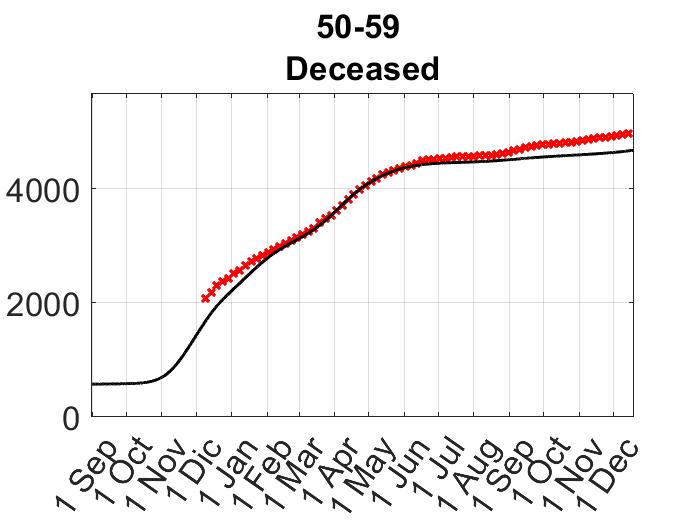}
\includegraphics[width=0.24\textwidth]{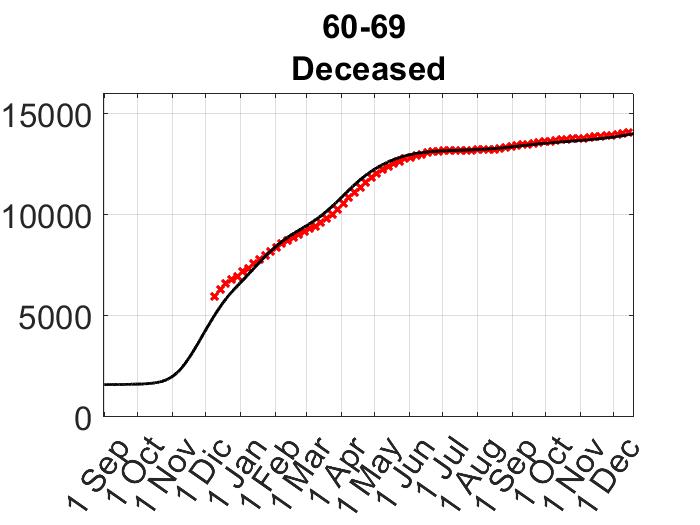}
\includegraphics[width=0.24\textwidth]{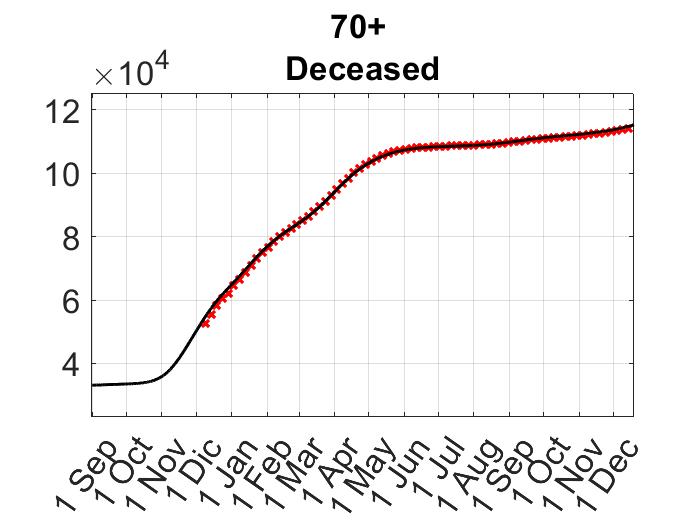}
\caption{Deceased in Italy estimated by SEIHRDV multi-age/multi-context model divided by age group, introducing GP containment measures.}
\label{fig:deceased_age_Italy_GP}
\end{figure}

\begin{figure}[t!]
\centering
\includegraphics[width=0.24\textwidth]{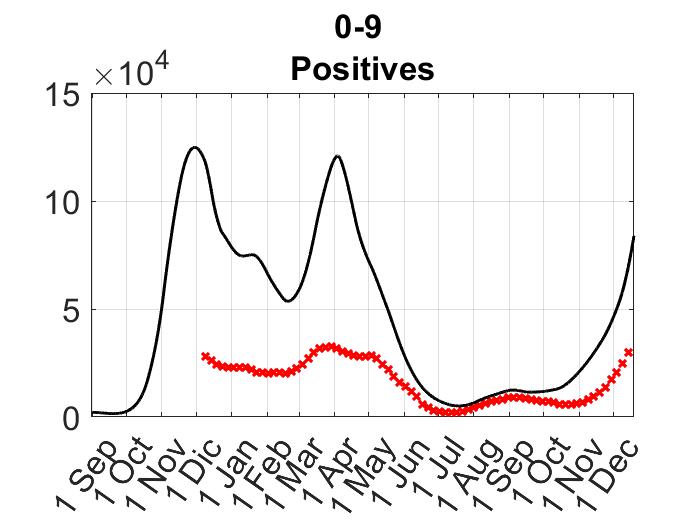}
\includegraphics[width=0.24\textwidth]{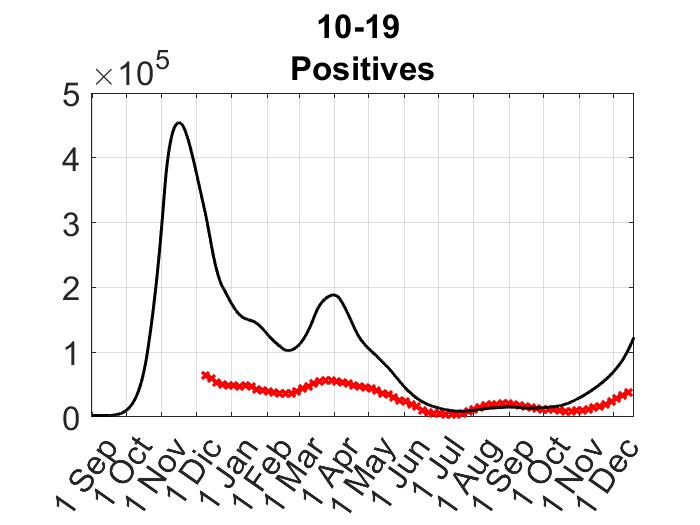}
\includegraphics[width=0.24\textwidth]{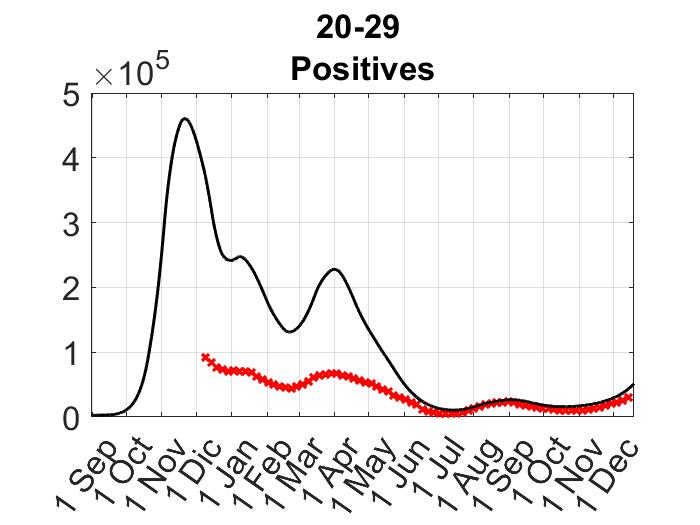}
\includegraphics[width=0.24\textwidth]{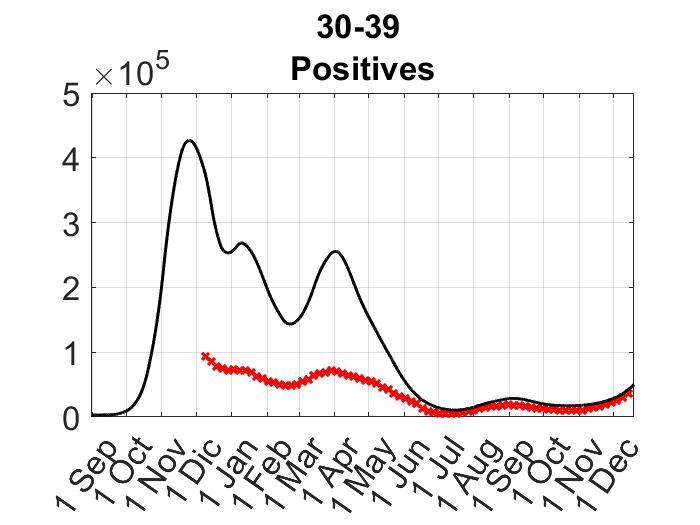}
\includegraphics[width=0.24\textwidth]{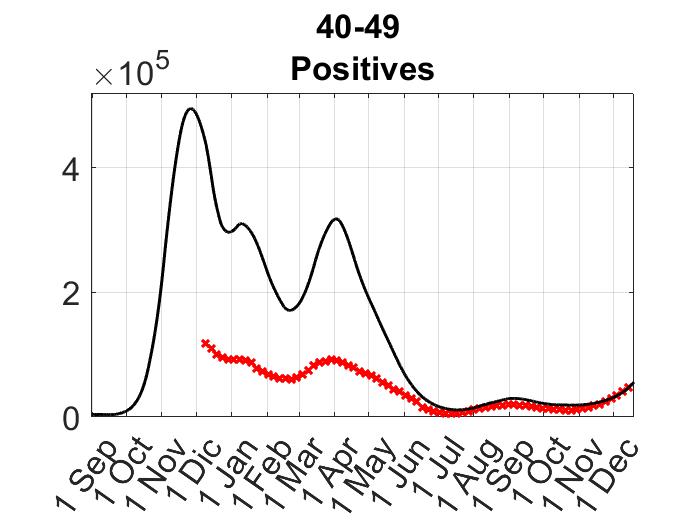}
\includegraphics[width=0.24\textwidth]{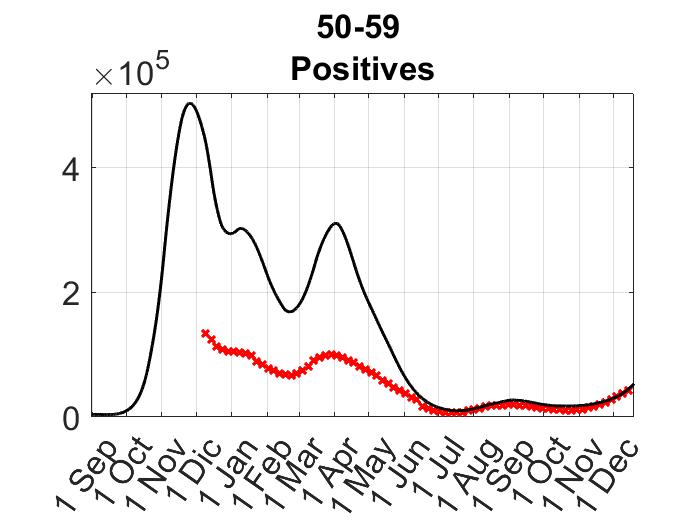}
\includegraphics[width=0.24\textwidth]{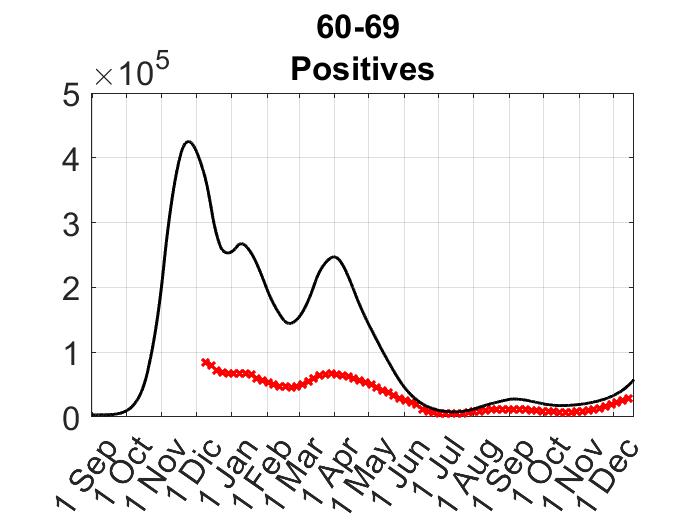}
\includegraphics[width=0.24\textwidth]{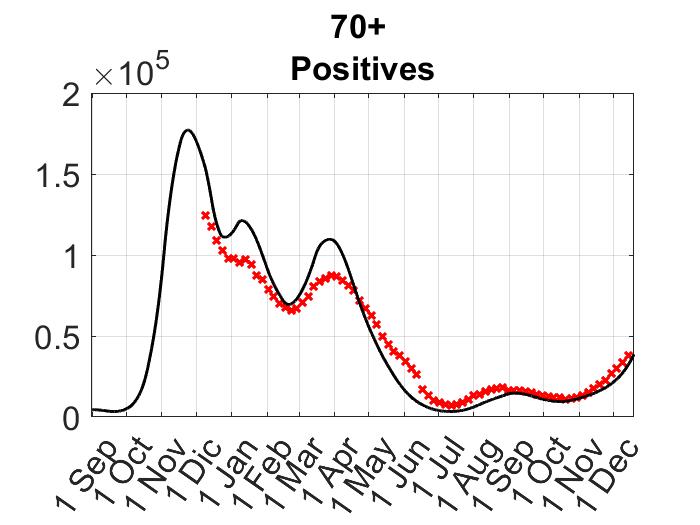}
\caption{Positives ($I+H$) in Italy estimated by SEIHRDV multi-age/multi-context model divided by age group, introducing GP containment measures.}
\label{fig:positives_age_Italy_GP}
\end{figure}

\begin{figure}[t!]
\centering
\includegraphics[width=1\textwidth]{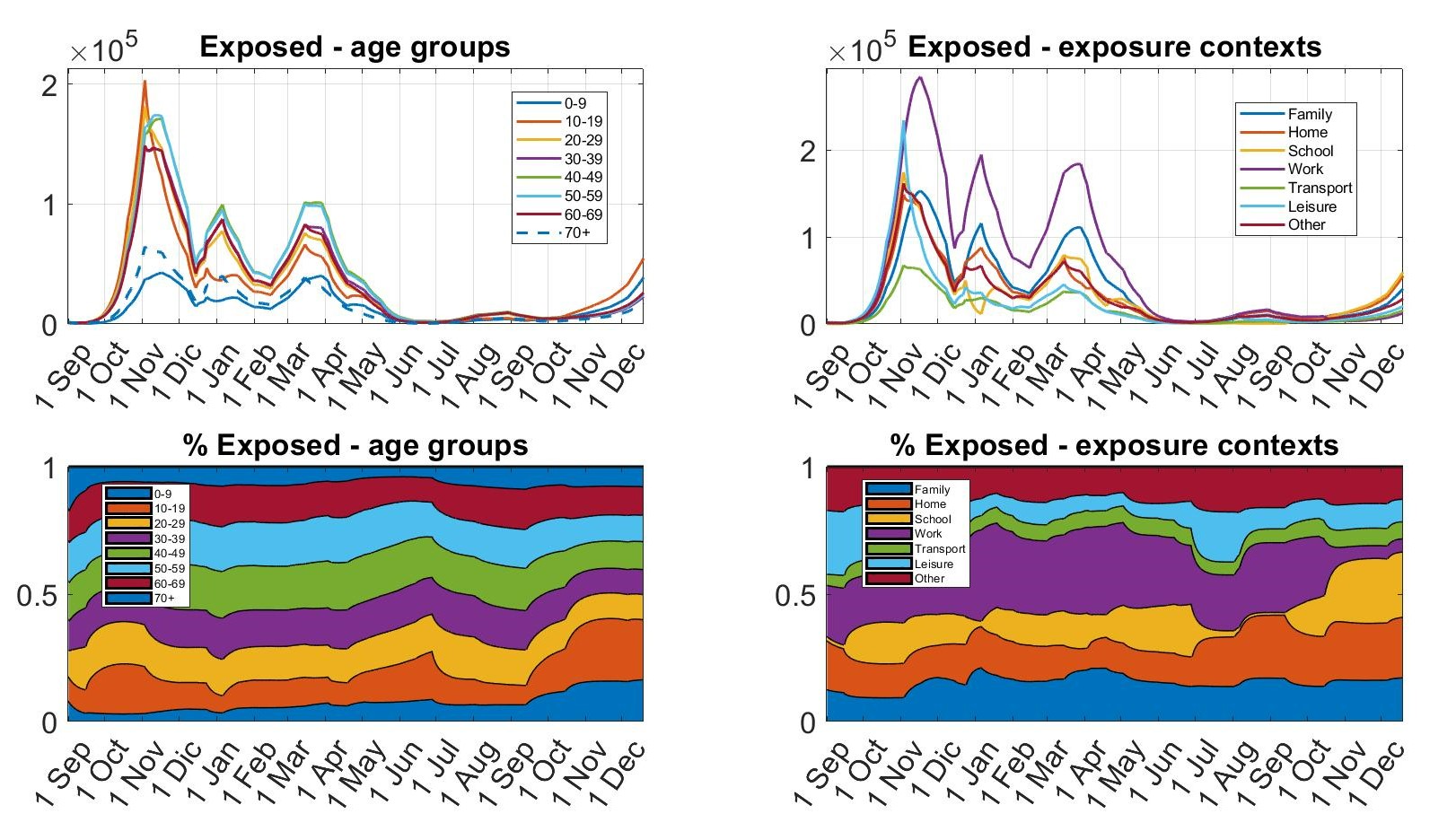}
\caption{Exposed in Italy with GP containment measures: at the top, evolution of the exposed individuals divided by age group (a) and context of exposition (b); at the bottom, distribution of age groups (c) and contexts of exposition (d) in the exposed compartment. }
\label{fig:exposed_italy_GP}
\end{figure}

In Figure \ref{fig:positives_age_Italy_GP} it is possible to observe differences between the age groups in this last time period. In particular from SEIHRDV multi-age/multi-context model it is possible to see a significant growth in 0-9 and 10-19, maybe due to the school reopening. Indeed in July and August schools were closed, as shown in the distribution of exposed in the different contexts of exposition (Figure \ref{fig:exposed_italy_GP}): starting on September the exposed in \virgolette{school} context started growing again. In particular, the age groups most affected {by school reopening are indeed} those aging 0-9 and 10-19 (as can be seen in Figure \ref{fig:exposed_italy_GP}), which are also the ones less vaccinated (the vaccination campaign concerns the over 12s).

{Finally, we conclude that the Green Pass NPIs} impact more in the context of exposition of \textit{work}, and in a lighter way in \textit{leisure}, as we can see in Figure \ref{fig:exposed_italy_GP}. {On the other hand}, an increase in the rate of exposure to the virus is observed in contexts such as \textit{school} and \textit{home}.  

\section{Conclusions and perspectives }\label{conclusions}

We introduced a new mathematical model to describe COVID-19 epidemic in Italy during the years 2020 and 2021. This is a multi-age and multi-context SEIHRDV model, {constituted by} seven compartments, susceptible individuals $S$, Exposed individuals $E$, infectious individuals $I$, healing individuals $H$, recovered individuals $R$, dead individuals $D$, vaccinated individuals $V$.
{This model is able to track the infections patterns among age groups happening in different exposition contexts}. {In this work, we exploited this model to reconstruct COVID-19 spread in different scenarios between} September 2020 and July 2021 in Italy, in Lombardy and in Lazio, and taking into account the vaccination campaign and Green Pass Interventions. 
{This model allows to assess the impact of different NPIs that can be implemented in various contexts heterogeneously.}
Indeed, we obtained numerical results differentiated for age and context: this was useful to better understand which age groups and contexts of exposition were the most risky.

{The model has been validated using data from the Dipartimento di Protezione Civile, Italy, which were also employed to calibrate the model parameters through a standard Least Squares optimization scheme. In this calibration process, we relied on information on deceased individuals, for sure those most reliable and least affected by monitoring uncertainties.}

The SEIHRDV multiage and multicontext model provides a valuable framework for analyzing the differing behaviors observed between age groups {and understanding how infections spread within various exposure contexts. Nonetheless, further adjustments are needed, such as introducing an additional compartment to account for fully vaccinated individuals or tracking undetected cases, as done in other studies. Despite these limitations, we are confident that this model represents a fundamental tool for policymakers, enabling reliable scenario analyses during epidemic outbreaks.}

\section*{Acknowledgments}
L.D., N.P., and G.Z. acknowledge their membership to INdAM group GNCS -- Gruppo Nazionale per il Calcolo Scientifico (National Group for Scientific Computing, Italy). The present reasearch is part of
the activities of ”Dipartimento di Eccellenza 2023-2027”, MUR, Italy, Department of Mathematics, Politecnico di Milano.

% Bibliografia
%\nocite{*}
%\bibliographystyle{abbrv}
%\nocite{*}

\bibliographystyle{unsrt}
\bibliography{biblio}

\end{document}